\documentclass[11pt, a4paper]{article}
\PassOptionsToPackage{unicode}{hyperref}
\usepackage{jheppub}
\let\U\relax
\let\C\relax

\usepackage{tikz}
\definecolor{maroon}{RGB}{128,0,0}
\definecolor{midnightblue}{RGB}{0,51,102}
\definecolor{firebrick}{RGB}{178,34,34}

\usepackage{relsize}
\usepackage[caption=false,labelformat=simple]{subfig}

\usetikzlibrary{calc, positioning, arrows.meta,
  decorations.pathreplacing} \tikzset{
  baseline={([yshift=-0.75ex]current bounding box.center)},
  zerosep/.style={inner sep=0pt, outer sep=0pt, minimum size=0pt},
  node distance=8pt, align at top/.style={baseline=(current bounding
    box.north)}, align at bottom/.style={baseline=(current bounding
    box.south)}, }

\usetikzlibrary{decorations,decorations.markings, decorations.pathreplacing}
\tikzset{
  strike out/.style={
    postaction=decorate,
    decoration={
      markings,
      mark=at position 0.5 with {
        \draw[-] (-2pt, -3pt) -- (2pt, 3pt);
      }
    }
  }
}

\tikzset{
vev/.style={strike out},
pdvev/.style={snake=snake},
}

\newlength{\qsep}
\usetikzlibrary{decorations.markings, arrows, intersections}
\tikzset{
  x=\qsep, y=\qsep,  font=\smaller,
  ->-/.style={decoration={
      markings, mark=at position #1 with
      {\arrow{>}}},postaction={decorate}},
  -<-/.style={decoration={
      markings, mark=at position #1 with
      {\arrow{<}}},postaction={decorate}},
  node/.style={draw, fill=white, shape=circle, minimum size=7pt, inner
    sep=0pt},
  gnode/.style={node},
  dgnode/.style={node, densely dashed},
  ggnode/.style={node, double},
  fnode/.style={node, shape=rectangle},
  tnode/.style={fnode, double, minimum size=12pt},
  q-/.style={-},
  q->/.style={->,  >=stealth, shorten >=1pt, font=\smaller[2]},
  q<-/.style={q->, <-, shorten >=0pt, shorten <=1pt},
  eq-/.style={double, double distance=2pt},
}

\newcommand{\RM}{\check{R}}

\tikzset{
  r-/.style={-, >={Classical TikZ Rightarrow[length=2pt]}, thick},
  r->/.style={r-, ->},
  r<-/.style={r-, <-},
  r->-/.style={->-=#1, thick},
}

\tikzset{
  z-/.style={-, >={Classical TikZ Rightarrow[length=2pt]}, thick},
  z->/.style={z-, ->},
  z<-/.style={z->, <-},
  dr-/.style={-, densely dashed, line width=0.6pt, >={Classical TikZ
      Rightarrow[length=2.5pt]}},
  dr->/.style={dr-, ->},
  dr<-/.style={dr-, <-},
  dtr-/.style={-, double, double distance=1pt, >={Classical TikZ
      Rightarrow[length=3pt]}},
  dtr->/.style={dtr-, ->},
  dtr<-/.style={dtr-, <-},
  tr-/.style={-, line width=1pt, >={Classical TikZ
      Rightarrow[length=3pt]}},
  tr->/.style={tr-, ->},
  tr<-/.style={tr-, <-},
  wz-/.style={z-, densely dotted}, 
  wz->/.style={wz-, ->},
  wz<-/.style={wz-, <-},
  shaded/.style={fill=black!20},
  dshaded/.style={fill=midnightblue!40},
  lshaded/.style={fill=firebrick!40},
  frame/.style={draw=olive},
  ws/.style={fill=olive!5},
  boundary/.style={draw=maroon},
  fboundary/.style={boundary, double},
}

\tikzset{
  minp/.style={draw, shape=circle, minimum size=3pt, inner 
    sep=0pt, font=\tiny},
  maxp/.style={minp, double, double distance=1pt, fill=white, minimum
    size=5pt},
}

\usepackage{upgreek}
\newcommand{\uplambdab}{\bar{\uplambda}}
\newcommand{\upmub}{\bar{\upmu}}
\newcommand{\upomegab}{\bar{\upomega}}

\newcommand{\gf}{\mathfrak{g}}

\newcommand{\tf}{\mathfrak{t}}

\newcommand{\del}{\partial}
\newcommand{\delb}{{\bar\partial}}

\newcommand{\id}{\mathop{\mathrm{id}}\nolimits}

\newcommand{\vol}{\mathrm{vol}}

\newcommand{\biggvev}[1]{\biggl\langle #1 \biggr\rangle}

\newcommand{\diag}{\mathop{\mathrm{diag}}\nolimits}

\newcommand{\ad}{\mathop{\mathrm{ad}}\nolimits}

\renewcommand{\Im}{\mathop{\mathrm{Im}}\nolimits}
\renewcommand{\Re}{\mathop{\mathrm{Re}}\nolimits}

\newcommand{\Tr}{\mathop{\mathrm{Tr}}\nolimits}
\newcommand{\End}{\mathop{\mathrm{End}}\nolimits}

\newcommand{\Lie}{\mathop{\mathrm{Lie}}\nolimits}

\newcommand{\crit}{\mathop{\mathrm{Crit}}}

\newcommand{\SU}{\mathrm{SU}}
\newcommand{\PSU}{\mathrm{PSU}}

\newcommand{\Spin}{\mathrm{Spin}}

\newcommand{\GL}{\mathrm{GL}}

\newcommand{\slf}{\mathfrak{sl}}

\newcommand{\U}{\mathrm{U}}

\newcommand{\Gr}{\mathrm{Gr}}

\newcommand{\iso}{\cong}

\newcommand{\Z}{\mathbb{Z}}

\newcommand{\R}{\mathbb{R}}
\newcommand{\C}{\mathbb{C}}

\newcommand{\T}{\mathbb{T}}

\let\nc\newcommand
\let\renc\renewcommand
\nc{\wbar}{\overline}
\let\td\tilde
\let\wtd\widetilde
\let\wht\widehat
\let\mcl\mathcal

\nc{\ab}{{\bar{a}}} \nc{\at}{\tilde{a}} \nc{\ah}{\hat{a}}
\nc{\bb}{{\bar{b}}} \nc{\bt}{\tilde{b}} \nc{\bh}{\hat{b}}
\nc{\cb}{{\bar{c}}} \nc{\ct}{\tilde{c}} 
\nc{\db}{{\bar{d}}} \nc{\dt}{\tilde{d}} \renc{\dh}{\hat{d}}
\nc{\eb}{{\bar{e}}} \nc{\et}{\tilde{e}} \nc{\eh}{\hat{e}}
\nc{\fb}{{\bar{f}}} \nc{\ft}{\tilde{f}} \nc{\fh}{\hat{f}}
\nc{\gb}{{\bar{g}}} \nc{\gt}{\tilde{g}} \nc{\gh}{\hat{g}}
\nc{\hb}{{\bar{h}}} \nc{\hh}{\hat{h}} 
\nc{\ib}{{\bar{\imath}}} \nc{\ih}{\hat{\imath}} 
\nc{\jb}{{\bar{\jmath}}} \nc{\jt}{\tilde{\jmath}} \nc{\jh}{\hat{\jmath}}
\nc{\kb}{{\bar{k}}} \nc{\kt}{\tilde{k}} \nc{\kh}{\hat{k}}
\nc{\lb}{{\bar{l}}} \nc{\lt}{\tilde{l}} \nc{\lh}{\hat{l}}
\nc{\mb}{{\bar{m}}} \nc{\mt}{\tilde{m}} \nc{\mh}{\hat{m}}
\nc{\nb}{{\bar{n}}} \nc{\nt}{\tilde{n}} \nc{\nh}{\hat{n}}
\nc{\ob}{{\bar{o}}} \nc{\ot}{\tilde{o}} \nc{\oh}{\hat{o}}
\nc{\pb}{{\bar{p}}} \nc{\pt}{\tilde{p}} \nc{\ph}{\hat{p}}
\nc{\qb}{{\bar{q}}} \nc{\qt}{\tilde{q}} \nc{\qh}{\hat{q}}
\nc{\rb}{{\bar{r}}} \nc{\rt}{\tilde{r}} \nc{\rh}{{\hat{r}}}
\renc{\sb}{{\bar{s}}} \nc{\st}{\tilde{s}} \nc{\sh}{\hat{s}}
\nc{\tb}{{\bar{t}}} \renc{\th}{\hat{t}} 
\nc{\ub}{{\bar{u}}} \nc{\ut}{\tilde{u}} \nc{\uh}{\hat{u}}
\nc{\vb}{{\bar{v}}} \nc{\vt}{\tilde{v}} \nc{\vh}{\hat{v}}
\nc{\wb}{{\bar{w}}} \nc{\wt}{\tilde{w}} \nc{\wh}{\hat{w}}
\nc{\xb}{{\bar{x}}} \nc{\xt}{\tilde{x}} \nc{\xh}{\hat{x}}
\nc{\yb}{{\bar{y}}} \nc{\yt}{\tilde{y}} \nc{\yh}{\hat{y}}
\nc{\zb}{{\bar{z}}} \nc{\zt}{\tilde{z}} \nc{\zh}{\hat{z}}

\nc{\Ab}{{\wbar{A}}} \nc{\At}{{\wtd{A}}} \nc{\Ah}{{\wht{A}}}
\nc{\Bb}{{\wbar{B}}} \nc{\Bt}{{\wtd{B}}} \nc{\Bh}{{\wht{B}}}
\nc{\Cb}{{\wbar{C}}} \nc{\Ct}{{\wtd{C}}} \nc{\Ch}{{\wht{C}}}
\nc{\Db}{{\wbar{D}}} \nc{\Dt}{{\wtd{D}}} \nc{\Dh}{{\wht{D}}}
\nc{\Eb}{{\wbar{E}}} \nc{\Et}{{\wtd{E}}} \nc{\Eh}{{\wht{E}}}
\nc{\Fb}{{\wbar{F}}} \nc{\Ft}{{\wtd{F}}} \nc{\Fh}{{\wht{F}}}
\nc{\Gb}{{\wbar{G}}} \nc{\Gt}{{\wtd{G}}} \nc{\Gh}{{\wht{G}}}
\nc{\Hb}{{\wbar{H}}} \nc{\Ht}{{\wtd{H}}} \nc{\Hh}{{\wht{H}}}
\nc{\Ib}{{\bar{I}}} \nc{\It}{{\wtd{I}}} \nc{\Ih}{{\wht{I}}}
\nc{\Jb}{{\bar{J}}} \nc{\Jt}{{\wtd{J}}} \nc{\Jh}{{\wht{J}}}
\nc{\Kb}{{\wbar{K}}} \nc{\Kt}{{\wtd{K}}} \nc{\Kh}{{\wht{K}}}
\nc{\Lb}{{\wbar{L}}} \nc{\Lt}{{\wtd{L}}} \nc{\Lh}{{\wht{L}}}
\nc{\Mb}{{\wbar{M}}} \nc{\Mt}{{\wtd{M}}} \nc{\Mh}{{\wht{M}}}
\nc{\Nb}{{\wbar{N}}} \nc{\Nt}{{\wtd{N}}} \nc{\Nh}{{\wht{N}}}
\nc{\Ob}{{\wbar{O}}} \nc{\Ot}{{\wtd{O}}} \nc{\Oh}{{\wht{O}}}
\nc{\Pb}{{\wbar{P}}} \nc{\Pt}{{\wtd{P}}} \nc{\Ph}{{\wht{P}}}
\nc{\Qb}{{\wbar{Q}}} \nc{\Qt}{{\wtd{Q}}} \nc{\Qh}{{\wht{Q}}}
\nc{\Rb}{{\wbar{R}}} \nc{\Rt}{{\wtd{R}}} \nc{\Rh}{{\wht{R}}}
\nc{\Sb}{{\wbar{S}}} \nc{\St}{{\wtd{S}}} \nc{\Sh}{{\wht{S}}}
\nc{\Tb}{{\wbar{T}}} \nc{\Tt}{{\wtd{T}}} \nc{\Th}{{\wht{T}}}
\nc{\Ub}{{\wbar{U}}} \nc{\Ut}{{\wtd{U}}} \nc{\Uh}{{\wht{U}}}
\nc{\Vb}{{\wbar{V}}} \nc{\Vt}{{\wtd{V}}} \nc{\Vh}{{\wht{V}}}
\nc{\Wb}{{\wbar{W}}} \nc{\Wt}{{\wtd{W}}} \nc{\Wh}{{\wht{W}}}
\nc{\Xb}{{\wbar{X}}} \nc{\Xt}{{\wtd{X}}} \nc{\Xh}{{\wht{X}}}
\nc{\Yb}{{\wbar{Y}}} \nc{\Yt}{{\wtd{Y}}} \nc{\Yh}{{\wht{Y}}}
\nc{\Zb}{{\wbar{Z}}} \nc{\Zt}{{\wtd{Z}}} \nc{\Zh}{{\wht{Z}}}

\nc{\CA}{{\mcl{A}}} \nc{\CAb}{{\wbar{\CA}}} \nc{\CAt}{{\wtd{\CA}}} \nc{\CAh}{{\wht{\CA}}}
\nc{\CB}{{\mcl{B}}} \nc{\CBb}{{\wbar{\CB}}} \nc{\CBt}{{\wtd{\CB}}} \nc{\CBh}{{\wht{\CB}}}
\nc{\CC}{{\mcl{C}}} \nc{\CCb}{{\wbar{\CC}}} \nc{\CCt}{{\wtd{\CC}}} \nc{\CCh}{{\wht{\CC}}}
\nc{\cD}{{\mcl{D}}} \nc{\cDb}{{\wbar{\cD}}} \nc{\cDt}{{\wtd{\cC}}} \nc{\cDh}{{\wht{\cD}}}
\nc{\CE}{{\mcl{E}}} \nc{\CEb}{{\wbar{\CE}}} \nc{\CEt}{{\wtd{\CE}}} \nc{\CEh}{{\wht{\CE}}}
\nc{\CF}{{\mcl{F}}} \nc{\CFb}{{\wbar{\CF}}} \nc{\CFt}{{\wtd{\CF}}} \nc{\CFh}{{\wht{\CF}}}
\nc{\CG}{{\mcl{G}}} \nc{\CGb}{{\wbar{\CG}}} \nc{\CGt}{{\wtd{\CG}}} \nc{\CGh}{{\wht{\CG}}}
\nc{\CH}{{\mcl{H}}} \nc{\CHb}{{\wbar{\CH}}} \nc{\CHt}{{\wtd{\CH}}} \nc{\CHh}{{\wht{\CH}}}
\nc{\CI}{{\mcl{I}}} \nc{\CIb}{{\wbar{\CI}}} \nc{\CIt}{{\wtd{\CI}}} \nc{\CIh}{{\wht{\CI}}}
\nc{\CJ}{{\mcl{J}}} \nc{\CJb}{{\wbar{\CJ}}} \nc{\CJt}{{\wtd{\CJ}}} \nc{\CJh}{{\wht{\CJ}}}
\nc{\CK}{{\mcl{K}}} \nc{\CKb}{{\wbar{\CK}}} \nc{\CKt}{{\wtd{\CK}}} \nc{\CKh}{{\wht{\CK}}}
\nc{\CL}{{\mcl{L}}} \nc{\CLb}{{\wbar{\CL}}} \nc{\CLt}{{\wtd{\CL}}} \nc{\CLh}{{\wht{\CL}}}
\nc{\CM}{{\mcl{M}}} \nc{\CMb}{{\wbar{\CM}}} \nc{\CMt}{{\wtd{\CM}}} \nc{\CMh}{{\wht{\CM}}}
\nc{\CN}{{\mcl{N}}} \nc{\CNb}{{\wbar{\CN}}} \nc{\CNt}{{\wtd{\CN}}} \nc{\CNh}{{\wht{\CN}}}
\nc{\CO}{{\mcl{O}}} \nc{\COb}{{\wbar{\CO}}} \nc{\COt}{{\wtd{\CO}}} \nc{\COh}{{\wht{\CO}}}
\nc{\CP}{{\mcl{P}}} \nc{\CPb}{{\wbar{\CP}}} \nc{\CPt}{{\wtd{\CP}}} \nc{\CPh}{{\wht{\CP}}}
\nc{\CQ}{{\mcl{Q}}} \nc{\CQb}{{\wbar{\CQ}}} \nc{\CQt}{{\wtd{\CQ}}} \nc{\CQh}{{\wht{\CQ}}}
\nc{\CR}{{\mcl{R}}} \nc{\CRb}{{\wbar{\CR}}} \nc{\CRt}{{\wtd{\CR}}} \nc{\CRh}{{\wht{\CR}}}
\nc{\CS}{{\mcl{S}}} \nc{\CSb}{{\wbar{\CS}}} \nc{\CSt}{{\wtd{\CS}}} \nc{\CSh}{{\wht{\CS}}}
\nc{\CT}{{\mcl{T}}} \nc{\CTb}{{\wbar{\CT}}} \nc{\CTt}{{\wtd{\CT}}} \nc{\CTh}{{\wht{\CT}}}
\nc{\CU}{{\mcl{U}}} \nc{\CUb}{{\wbar{\CU}}} \nc{\CUt}{{\wtd{\CU}}} \nc{\CUh}{{\wht{\CU}}}
\nc{\CV}{{\mcl{V}}} \nc{\CVb}{{\wbar{\CV}}} \nc{\CVt}{{\wtd{\CV}}} \nc{\CVh}{{\wht{\CV}}}
\nc{\CW}{{\mcl{W}}} \nc{\CWb}{{\wbar{\CW}}} \nc{\CWt}{{\wtd{\CW}}} \nc{\CWh}{{\wht{\CW}}}
\nc{\CX}{{\mcl{X}}} \nc{\CXb}{{\wbar{\CX}}} \nc{\CXt}{{\wtd{\CX}}} \nc{\CXh}{{\wht{\CX}}}
\nc{\CY}{{\mcl{Y}}} \nc{\CYb}{{\wbar{\CY}}} \nc{\CYt}{{\wtd{\CY}}} \nc{\CYh}{{\wht{\CY}}}
\nc{\CZ}{{\mcl{Z}}} \nc{\CZb}{{\wbar{\CZ}}} \nc{\CZt}{{\wtd{\CZ}}} \nc{\CZh}{{\wht{\CZ}}}

\let\eps\epsilon
\let\ups\upsilon
\let\veps\varepsilon
\let\vtht\vartheta
\let\vtheta\vtht
\let\vsgm\varsigma
\let\vphi\varphi
\let\vrho\varrho

\nc{\alphab}{{\bar{\alpha}}} \nc{\alphat}{{\td{\alpha}}} \nc{\alphah}{{\hat{\alpha}}}
\nc{\betab}{{\bar{\beta}}}   \nc{\betat}{{\td{\beta}}}   \nc{\betah}{{\hat{\beta}}} 
\nc{\gammab}{{\bar{\gamma}}} \nc{\gammat}{{\td{\gamma}}} \nc{\gammah}{{\hat{\gamma}}} 
\nc{\deltab}{{\bar{\delta}}} \nc{\deltat}{{\td{\delta}}} \nc{\deltah}{{\hat{\delta}}} 
\nc{\epsilonb}{{\bar{\eps}}} \nc{\epsilont}{{\td{\eps}}} \nc{\epsilonh}{{\hat{\eps}}} 
\nc{\vepsb}{{\bar{\veps}}}   \nc{\vepst}{{\td{\veps}}}   \nc{\vepsh}{{\hat{\veps}}} 
\nc{\zetab}{{\bar{\zeta}}}   \nc{\zetat}{{\td{\zeta}}}   \nc{\zetah}{{\hat{\zeta}}} 
\nc{\etab}{{\bar{\eta}}}     \nc{\etat}{{\td{\eta}}}     \nc{\etah}{{\hat{\eta}}} 
\nc{\thetab}{{\bar{\theta}}} \nc{\thetat}{{\td{\theta}}} \nc{\thetah}{{\hat{\theta}}} 
\nc{\vthetab}{{\bar{\vtht}}} \nc{\vthetat}{{\td{\vtht}}} \nc{\vthetah}{{\hat{\vtht}}} 
\nc{\lambdab}{{\bar{\lambda}}} \nc{\lambdat}{{\td{\lambda}}} \nc{\lambdah}{{\hat{\lambda}}} 
\nc{\iotab}{{\bar{\iota}}}   \nc{\iotat}{{\td{\iota}}}   \nc{\iotah}{{\hat{\iota}}} 
\nc{\kappab}{{\bar{\kappa}}} \nc{\kappat}{{\td{\kappa}}} \nc{\kappah}{{\hat{\kappa}}} 
\nc{\lmdb}{{\bar{\lmd}}}     \nc{\lmdt}{{\td{\lmd}}}     \nc{\lmdh}{{\hat{\lmd}}} 
\nc{\mub}{{\bar{\mu}}}       \nc{\mut}{{\td{\mu}}}       \nc{\muh}{{\hat{\mu}}} 
\nc{\nub}{{\bar{\nu}}}       \nc{\nut}{{\td{\nu}}}       \nc{\nuh}{{\hat{\nu}}} 
\nc{\xib}{{\bar{\xi}}}       \nc{\xit}{{\td{\xi}}}       \nc{\xih}{{\hat{\xi}}} 
\nc{\pib}{{\bar{\pi}}}       \nc{\pit}{{\td{\pi}}}       \nc{\pih}{{\hat{\pi}}} 
\nc{\vpib}{{\bar{\vpi}}}     \nc{\vpit}{{\td{\vpi}}}     \nc{\vpih}{{\hat{\vpi}}} 
\nc{\rhob}{{\bar{\rho}}}     \nc{\rhot}{{\td{\rho}}}     \nc{\rhoh}{{\hat{\rho}}} 
\nc{\vrhob}{{\bar{\vrho}}}   \nc{\vrhot}{{\td{\vrho}}}   \nc{\vrhoh}{{\hat{\vrho}}} 
\nc{\sigmab}{{\bar{\sigma}}} \nc{\sigmat}{{\td{\sigma}}} \nc{\sigmah}{{\hat{\sigma}}} 
\nc{\vsigmab}{{\bar{\vsgm}}} \nc{\vsigmat}{{\td{\vsgm}}} \nc{\vsigmah}{{\hat{\vsgm}}} 
\nc{\taub}{{\bar{\tau}}}     \nc{\taut}{{\td{\tau}}}     \nc{\tauh}{{\hat{\tau}}} 
\nc{\upsb}{{\bar{\ups}}} \nc{\upst}{{\td{\ups}}} \nc{\upsh}{{\hat{\ups}}} 
\nc{\phib}{{\bar{\phi}}}     \nc{\phit}{{\td{\phi}}}     \nc{\phih}{{\hat{\phi}}} 
\nc{\varphib}{{\bar{\vphi}}}   \nc{\varphit}{{\td{\vphi}}}   \nc{\varphih}{{\hat{\vphi}}} 
\nc{\chib}{{\bar{\chi}}}     \nc{\chit}{{\td{\chi}}}     \nc{\chih}{{\hat{\chi}}} 
\nc{\psib}{{\bar{\psi}}}     \nc{\psit}{{\td{\psi}}}     \nc{\psih}{{\hat{\psi}}} 
\nc{\omegab}{{\bar{\omega}}} \nc{\omegat}{{\td{\omega}}} \nc{\omegah}{{\hat{\omega}}} 

\nc{\Gammab}{{\wbar{\Gamma}}}     \nc{\Gammat}{{\wtd{\Gamma}}}     \nc{\Gammah}{{\wht{\Gamma}}}
\nc{\Deltab}{{\wbar{\Delta}}}     \nc{\Deltat}{{\wtd{\Delta}}}     \nc{\Deltah}{{\wht{\Delta}}}
\nc{\Thetab}{{\wbar{\Theta}}}     \nc{\Thetat}{{\wtd{\Theta}}}     \nc{\Thetah}{{\wht{\Theta}}}
\nc{\Lambdab}{{\wbar{\Lambda}}}   \nc{\Lambdat}{{\wtd{\Lambda}}}   \nc{\Lambdah}{{\wht{\Lambda}}}
\nc{\Xib}{{\wbar{\Xi}}}           \nc{\Xit}{{\wtd{\Xi}}}           \nc{\Xih}{{\wht{\Xi}}}
\nc{\Pib}{{\wbar{\Pi}}}           \nc{\Pit}{{\wtd{\Pi}}}           \nc{\Pih}{{\wht{\Pi}}}
\nc{\Sigmab}{{\wbar{\Sigma}}}     \nc{\Sigmat}{{\wtd{\Sigma}}}     \nc{\Sigmah}{{\wht{\Sigma}}}
\nc{\Upsilonb}{{\wbar{\Upsilon}}} \nc{\Upsilont}{{\wtd{\Upsilon}}} \nc{\Upsilonh}{{\wht{\Upsilon}}}
\nc{\Phib}{{\wbar{\Phi}}} \nc{\Phit}{{\wtd{\Phi}}} \nc{\Phih}{{\wht{\Phi}}}
\nc{\Psib}{{\wbar{\Psi}}}         \nc{\Psit}{{\wtd{\Psi}}}         \nc{\Psih}{{\wht{\Psi}}}
\nc{\Omegab}{{\wbar{\Omega}}}     \nc{\Omegat}{{\wtd{\Omega}}}     \nc{\Omegah}{{\wht{\Omega}}}

\newcommand{\rmd}{\mathrm{d}}
\newcommand{\epsb}{\epsilonb}
\newcommand{\ChS}{\mathrm{CS}}

\title{Unification of integrability in supersymmetric gauge theories}

\author[a]{Kevin Costello}
\author[a, b]{and Junya Yagi}

\affiliation[a]{Perimeter Institute for Theoretical Physics, Waterloo, ON
  N2L 2Y5 Canada}

\affiliation[b]{Yau Mathematical Sciences Center, Tsinghua University,
  Beijing 100084 China}

\abstract{A four-dimensional analog of Chern--Simons theory produces
  integrable lattice models from Wilson lines and surface operators.
  We show that this theory describes a quasi-topological sector of
  maximally supersymmetric Yang--Mills theory in six dimensions,
  topologically twisted and subjected to an $\Omega$-deformation.  By
  realizing the six-dimensional theory in string theory and applying
  dualities, we unify various phenomena in which the eight-vertex
  model and the XYZ spin chain, as well as variants thereof, emerge
  from supersymmetric gauge theories.}


\arxivnumber{}

\newcommand{\auxP}{\mathsf{P}}
\newcommand{\auxF}{\mathsf{F}}
\newcommand{\auxFb}{\overline{\auxF}}
\newcommand{\auxD}{\mathsf{D}}
\newcommand{\auxB}{\mathsf{B}}

\newcommand{\iu}{\mathrm{i}}

\let\starx\star
\let\star\relax
\newcommand{\star}{\mathop{\starx}\nolimits}

\newcommand{\SV}{S_{\text{V}}}
\newcommand{\SC}{S_{\text{C}}}
\newcommand{\SW}{S_W}
\newcommand{\GF}{\text{GF}}
\newcommand{\SGF}{S_{\text{GF}}}
\newcommand{\BRST}{\text{B}}

\newcommand{\mm}{\upmu} 

\newcommand{\V}{\mathbb{V}}

\newcommand{\RF}{R^{\text{F}}}

\newcommand{\RB}{R^{\text{B}}}

\usepackage[mathscr]{euscript}

\newcommand{\Spinor}{S}
\newcommand{\PB}{P}
\newcommand{\KB}{\overline{K}}
\newcommand{\SG}{\mathscr{G}}
\newcommand{\SP}{\mathscr{P}}
\newcommand{\ST}{\mathscr{T}}
\newcommand{\STb}{\overline{\mathscr{T}}}
\newcommand{\Mod}{\mathscr{M}}
\newcommand{\Lag}{\mathscr{L}}

\usetikzlibrary{arrows.meta}

\DeclareRobustCommand{\uarrow}{{\hspace{0.13pt} \raisebox{-1.3pt}{\tikz{\draw[thin, round cap->, shorten >= 0.5pt, >={Classical TikZ Rightarrow[length=2pt]}] (0,0) -- (0,7pt);}}}}

\DeclareRobustCommand{\darrow}{{\hspace{0.13pt} \raisebox{-1.3pt}{\tikz{\draw[thin, round cap->, shorten >= 0.5pt, >={Classical TikZ Rightarrow[length=2pt]}] (0,7pt) -- (0,0);}}}}

\DeclareRobustCommand{\dashrarrow}{{\hspace{0.26pt} \raisebox{-1.53pt}{\tikz{\draw[thin, densely dashed, dash pattern=on 1.2pt off 1pt, round cap->,shorten >= 0.5pt, >={Classical TikZ Rightarrow[length=2pt]}] (0,0) -- (7pt,0);}}}}

\DeclareRobustCommand{\dashuarrow}{{\hspace{0.13pt} \raisebox{-1.3pt}{\tikz{\draw[thin, densely dashed, dash pattern=on 1.2pt off 1pt, round cap->, shorten >= 0.5pt, >={Classical TikZ Rightarrow[length=2pt]}] (0,0) -- (0,7pt);}}}}

\DeclareRobustCommand{\suarrow}{{\hspace{0.31pt} \raisebox{-1.4pt}{\tikz{\draw[thin, round cap->, shorten >= 0.5/1.2pt, >={Classical TikZ Rightarrow[length=2/1.2pt]}] (0,0) -- (0,7/1.2pt);}}}}

\DeclareRobustCommand{\sdarrow}{{\hspace{0.31pt} \raisebox{-1.85pt}{\tikz{\draw[thin, round cap->, shorten >= 0.5/1.2pt, >={Classical TikZ Rightarrow[length=2/1.2pt]}] (0,7/1.2pt) -- (0,0);}}}}

\DeclareRobustCommand{\sdashrarrow}{{\hspace{0.35pt} \raisebox{-0.1pt}{\tikz{\draw[thin, densely dashed, dash pattern=on 1.2/1.2pt off 1/1.2pt, round cap->, shorten >= 0.5/1.2pt, >={Classical TikZ Rightarrow[length=2/1.2pt]}] (0,0) -- (7/1.2pt,0);}}}}

\DeclareRobustCommand{\sdashuarrow}{{\hspace{0.31pt} \raisebox{-1.4pt}{\tikz{\draw[thin, densely dashed, dash pattern=on 1.2/1.2pt off 1/1.2pt, round cap->, shorten >= 0.5/1.2pt, >={Classical TikZ Rightarrow[length=2/1.2pt]}] (0,0) -- (0,7/1.2pt);}}}}

\begin{document}
\maketitle

\section{Introduction}

Over the past decade there has been considerable progress in our
understanding of connections between quantum field theories and
quantum integrable systems.  Many phenomena have been discovered in
which structures of integrable quantum spin chains and integrable
lattice models emerge from quantum field theories in diverse spacetime
dimensions, in most cases supersymmetric ones.

Among these phenomena, there are several instances where the same
family of integrable systems appears.  The most notable example is the
XXX spin chain and its generalizations the XXZ and XYZ spin chains, or
equivalently, the six- and eight-vertex models~\cite{Baxter:1982zz}.
These spin chains and lattice models have been found to arise in two-,
three- and four-dimensional supersymmetric gauge theories with four
supercharges~\cite{Nekrasov:2009uh, Nekrasov:2009ui}, four-, five- and
six-dimensional supersymmetric gauge theories with eight
supercharges~\cite{Nekrasov:2009rc, Dorey:2011pa, Chen:2011sj,
  Nekrasov:2012xe, Nekrasov:2013xda}, three-dimensional $\CN = 4$
supersymmetric gauge theories~\cite{Bullimore:2015lsa,
  Braverman:2016pwk}, four-dimensional supersymmetric gauge theories
in the presence of surface operators~\cite{Gaiotto:2012xa,
  Gadde:2013ftv, Alday:2013kda, Gaiotto:2015usa, Maruyoshi:2016caf,
  Yagi:2017hmj, Ito:2016fpl}, and a four-dimensional analog of
Chern--Simons theory~\cite{Costello:2013zra, Costello:2013sla,
  Costello:2017dso, Costello:2018gyb}.

Then a question comes to mind: why does a single family of integrable
systems makes appearances in multiple contexts?

In this paper we provide an answer to this question.  We argue that
these field theory setups are actually different descriptions of one
and the same physical system, all related by dualities in string
theory.

Another, closely related, aim of the paper is to better understand
four-dimensional Chern--Simons theory.  This bosonic theory has a
fairly direct connection with integrable lattice models, which can be
elegantly deduced solely from its topological--holomorphic nature.
Yet, this is by far the strangest of the theories listed above.  For
one thing, it can only be defined on a product $\Sigma \times C$ of
two surfaces, with $C$ being either the complex plane~$\C$, the
punctured complex plane $\C^\times = \C \setminus \{0\}$ or an
elliptic curve $E = \C/(\Z + \tau\Z)$; the three choices correspond to
the three levels of the rational--trigonometric--elliptic hierarchy of
integrable systems.  Moreover, it has a complex gauge group and a
complex action functional.

One of the main results of this paper is that four-dimensional
Chern--Simons theory in fact has an origin in six dimensions: it
describes a six-dimensional topological--holomorphic theory, subjected
to a so-called ``$\Omega$-deformation''~\cite{Nekrasov:2002qd,
  Nekrasov:2003rj, Nekrasov:2010ka}.  This six-dimensional theory is a
partial topological twist~\cite{Witten:1988ze} of maximally
supersymmetric Yang--Mills theory, and the restriction on the choice
of $C$ comes from the requirement for unbroken supercharges.  The
complex gauge group and the complex action functional naturally
arise when the path integral is partially carried out to yield a
four-dimensional description.

In turn, the six-dimensional construction allows us to realize
four-dimensional Chern--Simons theory and its observables using branes
in string theory.  Various chains of dualities then relate the brane
configuration thus obtained to different but physically equivalent
configurations which realize the other relevant theories, thereby
unifying the connections between quantum field theories and the
eight-vertex model mentioned above.

Since this paper is somewhat long and at times technical, let us give
a brief overview here before proceeding to detailed discussions.

We begin in section~\ref{sec:6dTHT} by formulating the six-dimensional
topological--holomorphic theory.  The theory is defined on a product
$M \times C$, and is topological on the four-manifold~$M$ and
holomorphic on $C$.  For $M = D \times \Sigma$, we may regard the
theory as a B-twisted gauge theory~\cite{Witten:1991zz, Hori:2003ic}
on the surface~$D$, with an infinite-dimensional gauge group and
infinite-dimensional matter representations.  It turns out that in
this two-dimensional description, the theory has a superpotential
which coincides with the action of four-dimensional Chern--Simons
theory.

In section~\ref{sec:Omega-def}, we turn to general B-twisted gauge
theories and explain how to introduce $\Omega$-deformations to these
theories~\cite{Yagi:2014toa, Luo:2014sva}.  By localization of the
path integral, we show that when the spacetime $D$ is $\R^2$, the
quasi-topological sector of an $\Omega$-deformed B-twisted gauge
theory is equivalent to a zero-dimensional gauge theory with complex
gauge group, whose action is given by the superpotential of the
two-dimensional theory~\cite{Yagi:2014toa, Luo:2014sva,
  Nekrasov:2018pqq}.  The integration domain of the path integral for
this zero-dimensional theory consists of the gradient flow
trajectories generated by the superpotential, terminating on a
Lagrangian submanifold chosen in a relevant moduli space.

Then we apply this result to the six-dimensional
topological--holomorphic theory, viewing it as a B-twisted gauge
theory.  We are immediately led to the conclusion that the
topological--holomorphic theory on $\R^2 \times \Sigma \times C$, with
an $\Omega$-deformation on $\R^2$, is equivalent to four-dimensional
Chern--Simons theory on $\Sigma \times C$.  This is done in
section~\ref{sec:4dCS-from-6d}.

Section~\ref{sec:ILM-4dCS} is devoted to discussions on the relations
between four-dimensional Chern--Simons theory and integrable lattice
models in the case when $C = E$.  We explain how a lattice
model~\cite{MR908997, Jimbo:1987mu} whose Boltzmann weights are given
by Felder's dynamical R-matrix~\cite{Felder:1994pb, Felder:1994be}
arises from a lattice of Wilson lines, and how certain surface
operators transform this R-matrix to the Baxter--Belavin
R-matrix~\cite{Baxter:1971cr, Baxter:1972hz, Belavin:1981ix} for the
eight-vertex model and its $\slf_N$ generalization.  We also use these
surface operators to define two kinds of L-operators, which may be
thought of as R-matrices associated with a pair of finite- and
infinite-dimensional representations.  The section ends with some
discussions on framing anomaly and junctions of Wilson lines; these
lie outside the main line of argument and are not strictly necessary
for understanding of the rest of the paper.

In section~\ref{sec:brane-construction}, we present a string theory
realization of the $\Omega$-deformed six-dimensional
topological--holomorphic theory.  This realization involves a stack of
D5-branes placed in a background with a nontrivial Ramond--Ramond (RR)
two-form field.  Wilson lines are created by fundamental strings
ending on the D5-branes, whereas surface operators are produced by
D3-branes forming bound states with the D5-branes.  Applying string
dualities, we map this brane configuration to those realizing brane
tiling~\cite{Hanany:2005ve, Franco:2005rj} and
class-$\CS_k$~\cite{Gaiotto:2009we, Gaiotto:2009hg, Gaiotto:2015usa}
theories, linear quiver theories~\cite{Witten:1997sc}, and theories
related to the cotangent bundles of partial flag manifolds.  In each
dual picture we identify how the structures of lattice models and spin
chains arise.

There are many directions for future research.  One important question
which we hope this paper will shed some light on is the origin of the
chiral Potts model and its higher genus curve for spectral parameters.
The mysterious coincidence between the chiral Potts model and magnetic
monopoles, pointed out by Atiyah~\cite{MR1117746} in 1990, hints that
we are on the right track.  Indeed, we have necessary ingredients in
our construction: monopoles create surface operators, and crossings of
surface operators produce~\cite{Spiridonov:2010em, Yamazaki:2012cp,
  Yagi:2015lha} a variant of the Bazhanov--Sergeev
model~\cite{Bazhanov:2010kz, Bazhanov:2011mz} which is known to reduce
to the chiral Potts model in a special limit.  It is plausible that
the higher genus curve emerges in low energy physics as a geometric
object, in a way similar to how the Seiberg--Witten curve does when a
D4--NS5 brane configuration is lifted to
M-theory~\cite{Witten:1997sc}.

Finally, we remark that another string theory construction of
four-dimensional Chern--Simons theory has been proposed recently
in~\cite{Ashwinkumar:2018tmm}.  Their construction appears to be
related to a T-dual version of ours.  Also, a string theory
realization was discussed by Nikita Nekrasov in his talk on his joint
work with Samson Shatashvili and Mina Aganagic at the conference
String--Math 2017, where we also announced our results.

\section{Six-dimensional topological--holomorphic theory}
\label{sec:6dTHT}

In this section we formulate the six-dimensional
topological--holomorphic theory, from which four-dimensional
Chern--Simons theory arises via an $\Omega$-deformation.  After
explaining the construction, we reformulate this theory as a
two-dimensional gauge theory, in a form more suited for the
application of the $\Omega$-deformation.

\subsection{\texorpdfstring{$\CN = (1,1)$ super Yang--Mills theory}{N = (1,1) super Yang--Mills theory}}

The topological--holomorphic theory is defined as a topological twist
of $\CN = (1,1)$ super Yang--Mills theory in six dimensions, which in
turn can be constructed from super Yang--Mills theory in ten
dimensions by dimensional reduction.  So let us quickly review these
super Yang--Mills theories.  We mainly follow the convention of
\cite{Kapustin:2006pk}.
 
To describe spinors in ten dimensions, we use the gamma matrices
$\Gamma_I$, $I = 0$, $\dotsc$, $9$, obeying the anticommutation
relation
\begin{equation}
  \{\Gamma_I, \Gamma_J\} = 2\eta_{IJ} \,,
\end{equation}
where $\eta = -(\rmd x^0)^2 + (\rmd x^1)^2 + \dotsb + (\rmd x^9)^2$ is
the ten-dimensional Minkowski metric.  They can be chosen to be real
$32 \times 32$ matrices.  We let $\Gamma_{I_1 \dots I_k}$ be the
matrix that equals $\Gamma_{I_1} \dotsm \Gamma_{I_k}$ if $I_1$,
$\dotsc$, $I_k$ are all different and vanishes otherwise.

The generators of the Lorentz group $\Spin(9,1)$ are represented on
$\R^{32}$ by the matrices~$\Gamma_{IJ}$.  The chirality operator
$\Gamma_{0123456789}$ squares to $1$ and anticommutes with $\Gamma_I$.
Its eigenspaces therefore furnish irreducible spinor representations
of $\Spin(9,1)$ on which the chirality operator acts as multiplication
by $+1$ or $-1$; let $\Spinor^+$ and $\Spinor^-$ denote the space of
positive chirality spinors and that of negative chirality spinors,
respectively.  There is a charge conjugation matrix $C$ such that
$C \Gamma_{IJ} C^{-1} = -\Gamma_{IJ}^T$, and the map
\begin{equation}
  \alpha \mapsto \alphab = \alpha^T C
\end{equation}
sends $\alpha \in \Spinor^\pm$ to its dual
$\alphab \in (\Spinor^\pm)^*$.  For $\alpha$, $\beta \in \Spinor^\pm$,
the product $\alphab \Gamma_{I_1 \dots I_k} \beta$ transforms under
$\Spin(9,1)$ like the corresponding component of a $k$-form.

The fields of ten-dimensional super Yang--Mills theory with gauge
group $G$ are the gauge field $A$ and a fermionic field $\Psi$, the
latter being a positive chirality spinor in the adjoint
representation.  More precisely, $A$ is a connection of a principal
$G$-bundle $\PB$ over Minkowski spacetime $\R^{9,1}$, and $\Psi$ is a
section of $\Spinor^+ \otimes \ad(\PB)$, where $\ad(\PB)$ is the adjoint
bundle of $\PB$.  The theory is governed by the action
\begin{equation}
  \label{eq:10d-SYM}
  -\frac{1}{e^2} \int \! \rmd^{10} x \,
  \Tr\biggl(\frac12 F^{IJ} F_{IJ} - \iu \Psib \Gamma^I D_I \Psi\biggr)
  \,.
\end{equation}
Here $e$ is the gauge coupling, $F = \rmd A + A \wedge A$ is the field
strength and $D = \rmd + A$ is the covariant derivative.  The symbol
$\Tr$ denotes an invariant symmetric bilinear form on the Lie algebra
$\gf$ of the compact Lie group $G$.  We have chosen $A$ in such a way
that it is antihermitian in a unitary representation of $G$, and $\Tr$
to be negative definite.  We pick generators $T_a$, $a = 1$, $\dotsc$,
$\dim G$, of $\gf$ such that $\Tr(T_a T_b) = -\delta_{ab}$ so that we
can write, for example, $A_I = \sum_{a=1}^{\dim G} A_I^a T_a$ with
real coefficients $A_I^a$.

The action \eqref{eq:10d-SYM} is invariant under the supersymmetry
variation
\begin{equation}
  \label{eq:10d-SUSY}
  \delta_\epsilon A_I = \iu \epsilonb \Gamma_I \Psi \,,
  \qquad
  \delta_\epsilon \Psi = \frac12 F_{IJ} \Gamma^{IJ} \epsilon \,,
\end{equation}
whose parameter $\epsilon$ is a constant spinor of positive chirality.
Hence, the theory has sixteen supercharges, which is the maximum
amount of supercharges for theories without gravity.  Let $Q_\epsilon$
be the supercharge generating the above transformation.  Up to
equations of motion, the supercharges obey the anticommutation
relation
\begin{equation}
  \{Q_\epsilon, Q_\eta\} = \epsilonb\Gamma^I\eta P_I \,,
\end{equation}
where the momentum $P_I$ generates translations in the
$x^I$-direction.

Now, let us demand the fields to be independent of the coordinates
$x^6$, $x^7$, $x^8$, $x^9$.  Then we obtain a six-dimensional gauge
theory, which is $\CN = (1,1)$ super Yang--Mills theory.

Under the splitting of $\R^{9,1}$ into $\R^{5,1}$ and $\R^4$, the
ten-dimensional Lorentz group $\Spin(9,1)$ decomposes into the product
$\Spin(5,1) \times \Spin(4)_R$.  The first factor is the
six-dimensional Lorentz group, while the second is the R-symmetry
group of the six-dimensional theory.  The ten-dimensional gauge field
$\sum_{I=0}^9 A_I \rmd x^I$ descends to a gauge field
$\sum_{I=0}^5 A_I \rmd x^I$ and four adjoint scalar fields
\begin{equation}
  \phi_\mu \,, \quad \text{$\mu = 0$, $\dotsc$, $3$} \,,
\end{equation}
in the six-dimensional theory, where the latter come from the
components $A_{\mu+6}$ and transform in the vector representation
$\mathbf{4}$ of $\Spin(4)_R$.  Upon the dimensional reduction the
bosonic part of the action becomes
\begin{equation}
  \label{eq:bos-action}
  -\frac{1}{e^2} \int \! \rmd^6 x
  \Tr\Biggl(
  \frac12 \sum_{I,J=0}^5 F^{IJ} F_{IJ}
  + \sum_{I=0}^5 \sum_{\mu=0}^3   D^I\phi^\mu D_I\phi_\mu
  + \frac12 \sum_{\mu=0}^3 [\phi^\mu, \phi^\nu] [\phi_\mu, \phi_\nu]
  \Biggr) \,.
\end{equation}

To understand what the fermion $\Psi$ becomes in six dimensions, we
recall that $\Spin(4)$ is isomorphic to $\SU(2) \times \SU(2)$.  In
the case of $\Spin(4)_R$, we can take the first $\SU(2)$ factor to be
generated by
\begin{equation}
  \frac12 (\Gamma_{67} + \Gamma_{89}) \,,
  \qquad
  \frac12 (\Gamma_{68} + \Gamma_{97}) \,,
  \qquad
  -\frac12 (\Gamma_{69} + \Gamma_{78})
\end{equation}
and the second to be generated by
\begin{equation}
  \frac12 (\Gamma_{67} - \Gamma_{89}) \,,
  \qquad
  \frac12 (\Gamma_{68} - \Gamma_{97}) \,,
  \qquad
  \frac12 (\Gamma_{69} - \Gamma_{78}) \,.
\end{equation}
Acting on these generators with the chirality operator $\Gamma_{6789}$
for $\Spin(4)_R$, we see that the irreducible spinor representations
of $\Spin(4)_R$ of positive and negative chirality are the
representations $(\mathbf{1},\mathbf{2})$ and
$(\mathbf{2},\mathbf{1})$ of $\SU(2) \times \SU(2)$, respectively.
The chirality operator for $\Spin(5,1)$ is $\Gamma_{012345}$, so
$\Spinor^+$ of $\Spin(9,1)$ decomposes with respect to
$\Spin(5,1) \times \Spin(4)_R$ as
\begin{equation}
  \bigl(\mathbf{4}_+, (\mathbf{1},\mathbf{2})\bigr)
  \oplus \bigl(\mathbf{4}_-, (\mathbf{2},\mathbf{1})\bigr)
  \,,
\end{equation}
where $\mathbf{4}_\pm$ are the spinor representations of $\Spin(5,1)$
with the chirality indicated by the subscripts.  Thus, in six
dimensions, $\Psi$ becomes two sets of spinors which have opposite
chirality and are doublets of different $\SU(2)$ factors of
$\Spin(4)_R$.

The six-dimensional super Yang--Mills theory inherits the sixteen
supercharges from ten dimensions.  Since the parameter $\epsilon$ of
supersymmetry variations transforms in the same way as $\Psi$ does,
the theory has two $\SU(2)$ doublets of supercharges with opposite
chirality, generating $\CN = (1,1)$ supersymmetry in six dimensions.

\subsection{Topological--holomorphic theory}

The topological--holomorphic theory we are going to construct is a
topological twist of the Euclidean version of six-dimensional
$\CN = (1,1)$ super Yang--Mills theory, and can be defined on a
product $M \times C$, with $M$ being a four-manifold and $C$ either
$\C$, $\C^\times$ or $\C/(\Z + \tau\Z)$.  The relevant topological
twist is essentially the GL-twist of $\CN = 4$ super Yang--Mills
theory in four dimensions, which plays an important role in a gauge
theoretic approach to the geometric Langlands
duality~\cite{Kapustin:2006pk}.

The Euclidean theory is obtained from the Lorentzian one by the Wick
rotation $x^0 \mapsto -ix^0$.  Correspondingly, we get gamma matrices
in Euclidean signature by making the replacement
$\Gamma_0 \mapsto \iu\Gamma_0$ in those in Lorentzian signature.

For a moment, suppose that $M$ is a spin manifold.  The structure group
of the spinor bundle over $M \times C$ is
$\Spin(4)_M \times \Spin(2)_C$.  To implement the topological twist in
question, we turn on a background gauge field for $\Spin(4)_R$ whose
value is equal to the spin connection of $M$, and interpret the
diagonal subgroup $\Spin(4)_M'$ of $\Spin(4)_M \times \Spin(4)_R$ as a
new rotation group on $M$.
Under $\Spin(4)_M'$, the scalars $\phi_\mu$ transform as the
components of a one-form
\begin{equation}
  \phi = \phi_\mu \rmd x^\mu
\end{equation}
since they originated from the components of the ten-dimensional gauge
field along the directions rotated by $\Spin(4)_R$.

The transformation properties of the fermions can be identified as
follows.  In Minkowski signature, we can take the chirality operators
for $\Spin(3,1)$, $\Spin(2)$ and $\Spin(5,1)$ to be
$-\iu\Gamma_{0123}$, $-\iu\Gamma_{45}$ and
$\Gamma_{012345} = -(-\iu\Gamma_{0123})(-\iu\Gamma_{45})$,
respectively.  Then, $\mathbf{4}_+$ of $\Spin(5,1)$ transforms under
the subgroup $\Spin(3,1) \times \Spin(2)$ as
$\mathbf{2}_+^{-1} \oplus \mathbf{2}_-^{1}$, while $\mathbf{4}_-$
transforms as $\mathbf{2}_+^{1} \oplus \mathbf{2}_-^{-1}$.  Here the
superscripts indicate the charges under $\Spin(2) \iso \U(1)$,
measured by $-\iu\Gamma_{45}$.  In Euclidean signature, $\Spin(3,1)$
is replaced with $\Spin(4) \iso \SU(2) \times \SU(2)$, and
$\mathbf{2}_+$ becomes $(\mathbf{1},\mathbf{2})$ and $\mathbf{2}_-$
becomes $(\mathbf{2},\mathbf{1})$.  Using the decomposition
$\mathbf{2} \otimes \mathbf{2} = \mathbf{1} \oplus \mathbf{3}$ of
$\SU(2)$, we find that the fermions transform under
$\Spin(4)_M' \times \Spin(2)_C$ as
\begin{equation}
  2(\mathbf{1},\mathbf{1})^{-1}
  \oplus (\mathbf{1},\mathbf{3})^{-1}
  \oplus (\mathbf{3},\mathbf{1})^{-1}
  \oplus 2(\mathbf{2},\mathbf{2})^{1}
  \,.
\end{equation}
The first three summands represent two scalars and one two-form on $M$,
transforming as negative chirality spinors on $C$:
\begin{equation}
  \xi,\, \xi'
  \in \Gamma\bigl(\Lambda_M^0 \otimes \KB_C^{-1/2} \otimes \ad(\PB)\bigr) \,,
  \qquad
  \chi
  \in \Gamma\bigl(\Lambda_M^2 \otimes \KB_C^{-1/2} \otimes \ad(\PB)\bigr) \,.
\end{equation}
The last summand gives two one-forms on $M$ which are positive
chirality spinors on $C$:
\begin{equation}
  \psi\,, \psi'
  \in \Gamma\bigl(\Lambda_M^1 \otimes \KB_C^{1/2} \otimes \ad(\PB)\bigr) \,.
\end{equation}
Here $\Lambda_M^p$ is the bundle of $p$-forms on $M$ and
$\KB_C^{\pm 1/2}$ are the bundles of spinors on $C$ with positive and
negative chirality, all pulled back to $M \times C$.  We have used the
same symbol $\PB$ as in the ten-dimensional case to denote the gauge
bundle.

Since the twisted theory does not contain any spinors on $M$, at this
point we can relax the assumption that $M$ is spin.  The twisted
theory can be defined on any four-manifold~$M$.

Looking at the transformation properties of the fermions, we see that
the twisted theory has two supercharges that are invariant under
$\Spin(4)_M'$.  They generate supersymmetry transformations whose
parameters are constant scalars on~$M$ and constant spinors on $C$,
and as such are present for any choice of $M$.  In contrast, the other
supercharges are broken unless $M$ admits covariantly constant
one-forms or two-forms.

Let us describe the supercharges that are scalars on $M$ more
explicitly.  The relevant supersymmetry parameters are annihilated by
the generators of $\Spin(4)'_M$:
\begin{equation}
  \label{eq:scalar-SUSY}
  (\Gamma_{\mu\nu} + \Gamma_{\mu+6,\nu+6}) \epsilon = 0
  \,,\quad
  \text{$\mu$, $\nu = 0$, $\dotsc$, $3$}
  \,.
\end{equation}
These equations can be rewritten as
\begin{equation}
  \label{eq:scalar-SUSY-mod}
  \epsilon = \Gamma_{\mu\nu \mu+6,\nu+6} \epsilon
  \,,
\end{equation}
and impose three independent constraints on $\epsilon$.  Each of them
reduces the dimension of the parameter space by half, so there are
$16 \times (1/2)^3 = 2$ independent solutions, as expected.

We can single out a supercharge by further demanding
\begin{equation}
  \label{eq:cond-Q}
  \eps = -\iu\Gamma_{\mu,\mu+6} \eps
  \,.
\end{equation}
These equations are compatible with the
condition~\eqref{eq:scalar-SUSY-mod} and the chirality condition
\begin{equation}
  \label{eq:chirality+}
  \iu \Gamma_{0123456789} \epsilon = \epsilon
\end{equation}
in Euclidean signature since $\Gamma_{\mu,\mu+6}$ commute with
$\Gamma_{0123456789}$ and $\Gamma_{\mu\nu \mu+6,\nu+6}$.  We are left
with a unique solution up to rescaling, and call the corresponding
supercharge $Q$.  Similarly, imposing the condition
\begin{equation}
  \label{eq:cond-Q'}
  \eps = \iu\Gamma_{\mu,\mu+6} \eps
\end{equation}
we obtain another supercharge $Q'$.

An important property of the supercharges thus defined is that they
square to zero:
\begin{equation}
  Q^2 = (Q')^2 = 0 \,.
\end{equation}
It is clear that $P_\mu$ cannot appear in $Q^2$ or $(Q')^2$ because
these supercharges are $\Spin(4)_M'$-invariant.  To see that $P_I$ for
any $I = 0$, $\dotsc$, $9$ makes no appearance either, say in $Q^2$,
we pick $\mu$ such that $I \neq \mu$, $\mu + 6$ and note
$\iu\epsilonb\Gamma_{\mu,\mu+6} = -\iu\epsilon^T \Gamma_{\mu,\mu+6}^T
C = \epsilonb$.  So we have
\begin{equation}
  \epsilonb \Gamma_I \eps 
  = \epsilonb \Gamma_I (-\iu\Gamma_{\mu,\mu+6} \eps)
  = -\iu\epsilonb\Gamma_{\mu,\mu+6} \Gamma_I \eps
  = -\epsilonb \Gamma_I \eps
\end{equation}
and $Q^2 = \epsilonb \Gamma^I \eps P_I = 0$.

From the constraints~\eqref{eq:scalar-SUSY-mod} and the chirality
condition~\eqref{eq:chirality+}, it follows
\begin{equation}
  \eps
  = \Gamma_{0167} \Gamma_{2389} \eps
  = \iu\Gamma_{45} \eps \,.
\end{equation}
Since the parameters $\epsilon$ under consideration have charge $-1$
with respect to the $\U(1)$ symmetry generated by $-\iu\Gamma_{45}$,
the corresponding supercharges have charge $+1$.  As can be seen from
the transformation property of $\alphab(\Gamma_4 - \iu\Gamma_5)\beta$,
the linear combination $P_4 - \iu P_5$ has charge~$2$ and is the only
translation generator with that charge.  Hence,
$\{Q,Q'\} \propto P_4 - \iu P_5$.  Introducing the complex coordinate
\begin{equation}
  \label{eq:z}
  z = \frac12(x^4 - \iu x^5) \,,
\end{equation}
we normalize the supercharges in such a way that
\begin{equation}
  \label{eq:QQ'}
  \{Q,Q'\} = P_\zb \,.
\end{equation}

Comparing the equation $(\Gamma_4 - i\Gamma_5) \eps = 0$ with the
formula~\eqref{eq:10d-SUSY} for supersymmetry variations, we see that
$A_\zb$ is invariant under the action of $Q$ and $Q'$. The extra
condition~\eqref{eq:cond-Q} says
$(\Gamma_\mu + \iu\Gamma_{\mu+6}) \epsilon = 0$, meaning that
$A_\mu + \iu A_{\mu+6}$ is annihilated by $Q$.  The twisted theory
therefore has the $Q$-invariant partial connection
\begin{equation}
  \CA = \CA_\mu \rmd x^\mu + A_\zb \rmd \zb \,,
  \qquad
  \CA_\mu = A_\mu + \iu\phi_\mu \,.
\end{equation}
By the same token, if we define
\begin{equation}
  \CAb =  (A_\mu - \iu\phi_\mu) \rmd x^\mu + A_z \rmd z
  \,,
\end{equation}
then $\CAb_\mu \rmd x^\mu + A_\zb \rmd \zb$ is $Q'$-invariant.  We
write $\cD$ and $\CF$ for the covariant derivative and curvature of
$\CA + A_z \rmd z$, and $\cDb$ and $\CFb$ for those of
$\CAb + A_\zb \rmd\zb$.

We can readily write down the action of $Q$ on the rest of the fields.
From the $Q$-invariance of~$\CA$, the transformation properties of the
fields under $\Spin(4)_M' \times \Spin(2)_C$, and the fact that the
supersymmetry variation of a fermion is a linear combination of the
field strength~$F_{IJ}$ in ten dimensions, we deduce that the
$Q$-action can be written as
\begin{equation}
  \label{eq:SUSY-6d}
  \begin{alignedat}{2}
    \delta\CA_\mu &= 0 \,,
    &
    \delta\CAb_\mu &= \psi_\mu \,,
    \\
    \delta A_z &= \xi \,,
    &
    \delta A_\zb &= 0 \,,
    \\
    \delta\xi &= 0 \,,
    &
    \delta\xi' &= \auxP \,,
    \\
    \delta\psi_\mu &= 0 \,,
    &
    \delta\psi'_\mu &= \CF_{\mu \zb} \,,
    \\
    \delta\auxP &= 0 \,,
    &\qquad
    \delta\chi_{\mu\nu} &= \CF_{\mu\nu} \,.
  \end{alignedat}
\end{equation}
We have introduced an auxiliary bosonic scalar $\auxP$ in order to
realize the correct supersymmetry algebra off-shell.

The variation $\delta'$ under the action of $Q'$ can be identified
with the help of relation~\eqref{eq:QQ'}.  Setting
$\{\delta, \delta'\} \CA_\mu = \CF_{\zb\mu}$ leads to
$\delta \delta'\CA_\mu = -\delta\psi'_\mu$, so we have
$\delta'\CA_\mu = -\psi'_\mu$ up to $Q$-invariant terms which we can
absorb in the definition of $\psi'_\mu$.  Likewise, the relation
$\{\delta, \delta'\} \CAb_\mu = \CFb_{\zb\mu}$ implies
$\delta'\psi_\mu = -\CFb_{\mu\zb}$.  Redefining $\xi'$ and $\auxP$ if
necessary, we can set $\delta' A_z = -\xi'$.  Then,
$\{\delta, \delta'\} A_z = F_{\zb z}$ gives
$\delta'\xi = \auxP - F_{z\zb}$.

To have $\{\delta, \delta'\} \chi = D_\zb\chi$, we need to use the
equation of motion for~$\chi$.  Let us postulate that the part of the
action that contains $\chi$ is given by
\begin{equation}
  \label{eq:6d-S1}
  \begin{split}
    S_1
    &=
    \frac{1}{e^2} \int_{M \times C} \! \rmd^2 z
    \Tr\biggl(
    -\delta\bigl(\chi \wedge \star_M \CFb\bigr)
    + \chi \wedge \cD \psi' 
    + \frac12\chi \wedge D_\zb \chi\biggr)
    \\
    &=
    \frac{1}{e^2} \int_{M \times C} \! \rmd^2 z
    \Tr\biggl(
    -\CF \wedge \star_M \CFb
    + \chi \wedge \star_M \cDb\psi
    + \chi \wedge \cD \psi' 
    + \frac12\chi \wedge D_\zb \chi\biggr) \,,
  \end{split}
\end{equation}
where
$\rmd^2 z = -2\iu \, \rmd z \wedge \rmd\zb = \rmd x^4 \wedge \rmd
x^5$.  Note that we have chosen a metric $g_M$ on $M$ to define the
Hodge star $\star_M$ on $M$.  The above expression is $Q$-invariant
thanks to the Bianchi identity $\cD\CF = 0$, provided that boundary
terms do not arise in the $Q$-variation.
The equation of motion for $\chi$ derived from this action is
\begin{equation}
  D_\zb \chi = -\delta\bigl(\star_M\CFb\bigr) + \delta'\delta\chi \,.
\end{equation}
Equating the left-hand side with $\{\delta,\delta'\}\chi$, we find
$\delta'\chi = -\star_M\CFb$ up to $Q$-invariant terms.
Because we did not redefine $\chi$, its supersymmetry variation is
still given by a linear combination of $F_{IJ}$, and the $Q$-invariant
terms, if exist, must be constructed from $\CF$.  Such terms are not
compatible with $\delta'^2\chi = 0$ and should be absent (unless we
are willing to use other equations of motion).  The $Q'$-variations of
the remaining fields can be fixed by the requirement $\delta'^2 = 0$.

We have thus obtained the following formula for the supersymmetry
transformation generated by $Q'$:
\begin{equation}
  \begin{alignedat}{2}
    \delta'\CA_\mu &= -\psi'_\mu \,,
    &
    \delta'\CAb_\mu &= 0\,,
    \\
    \delta' A_z &= -\xi' \,,
    &
    \delta' A_\zb &= 0 \,,
    \\
    \delta'\xi &= \auxP - F_{z\zb} \,,
    &
    \delta'\xi' &= 0\,,
    \\
    \delta'\psi_\mu &= -\CFb_{\mu \zb}  \,,
    &\qquad
    \delta'\psi'_\mu &= 0\,,
    \\
    \delta'\auxP &= D_\zb\xi' \,,
    &
    \delta'\chi_{\mu\nu} &= -(\star_M\CFb)_{\mu\nu} \,.
  \end{alignedat}
\end{equation}
With these transformation rules, we can write $S_1$ as
\begin{equation}
  S_1
  =
  \frac{1}{e^2} \int_{M \times C} \! \rmd^2 z
  \Tr\biggl(
  \delta'(\CF \wedge \chi)
  +
  \chi \wedge \star_M \cDb \psi
  + \frac12 \chi \wedge D_\zb \chi\biggr)
  \,.
\end{equation}
This is $Q'$-invariant, again up to boundary contributions.  Hence,
$S_1$ is invariant under both $Q$ and $Q'$.

The rest of the action of the twisted theory is
\begin{equation}
  \label{eq:6d-S2}
  \begin{split}
    S_2
    &=
    \frac{1}{e^2}
    \int_{M \times C} \! \sqrt{g} \, \rmd^6 x \,
    \delta\delta'
    \Tr(-\xi\xi' + 2\iu F_{\mu z} \phi^\mu)
    \\
    &= 
    \frac{1}{e^2}
    \int_{M \times C} \! \sqrt{g} \, \rmd^6 x \,
    \delta
    \Tr\bigl((-\auxP + F_{z\zb} + 2\iu D^\mu\phi_\mu) \xi'
    - \CFb_{\mu z} \psi'^\mu\bigr)
    \,,
  \end{split}
\end{equation}
where we have ignored boundary terms in going to the last expression.
To define the volume form $\sqrt{g} \, \rmd^6 x$ we have endowed $C$
with the metric $g_C = (\rmd x^4)^2 + (\rmd x^5)^2$; the total metric
on $M \times C$ is $g = g_M \oplus g_C$ and we have
$g_{z\zb} = g_{\zb z} = 2$.  This action is manifestly $Q$- and
$Q'$-invariant.  Explicitly, we have
\begin{multline}
  S_2
  =
  \frac{1}{e^2}
  \int_{M \times C} \! \sqrt{g} \, \rmd^6 x
  \Tr\bigl(
  -\auxP (\auxP - F_{z\zb} - 2\iu D^\mu\phi_\mu)
  - 2 \CFb^{\mu\zb} \CF_{\mu\zb}
  \\
  + \xi' D_\zb \xi 
  +  \xi' \cD_\mu\psi^\mu
  + \psi'_\mu \cDb^\mu \xi
  + D_z \psi^\mu \psi'_\mu
  \bigr)
  \,.
\end{multline}

The bosonic part of the full action $S_1 + S_2$ is
\begin{multline}
  \frac{1}{e^2}
  \int_{M \times C} \! \sqrt{g} \, \rmd^6 x
  \Tr\biggl(
  -\frac12 \CFb^{\mu\nu} \CF_{\mu\nu}
  -\auxP(\auxP - F_{z\zb} - 2\iu D^\mu\phi_\mu)
  - 2\CFb^{\mu\zb} \CF_{\mu\zb}\biggr)
  \\
  =
  \frac{1}{e^2}
  \int_{M \times C} \! \sqrt{g} \, \rmd^6 x
  \Tr\biggl(
  -\biggl(\auxP - \frac12 F_{z\zb} - \iu D^\mu\phi_\mu\biggr)^2
  -\frac12 F^{\mu\nu} F_{\mu\nu}
  -2F^{\mu z} F_{\mu z}
  - F^{z\zb} F_{z\zb}
  \\
  - D^\mu\phi^\nu D_\mu\phi_\nu
 - 2D^z\phi^\mu D_z\phi_\mu
  - \frac12 [\phi^\mu, \phi^\nu] [\phi_\mu, \phi_\nu]
  - R_{\mu\nu} \phi^\mu \phi^\nu 
  \biggr) \,,
\end{multline}
where $R_{\mu\nu}$ is the Ricci curvature of $g$.  For $M = \R^4$, it
reproduces the bosonic part~\eqref{eq:bos-action} of the action for
$\CN = (1,1)$ super Yang--Mills theory.

Now that we have constructed a theory with two supercharges that
square to zero, let us pick one of them, say $Q$, and consider the
$Q$-invariant sector of the theory.  The correlation function of a
$Q$-exact operator vanishes because in the path integral
representation it is the integral of a ``total derivative'' over an
infinite-dimensional supermanifold.  Therefore, the correlation
function of a $Q$-invariant operator depends only on the
$Q$-cohomology class of that operator.  We can also define the
$Q$-cohomology of states.  This is a module over the $Q$-cohomology of
operators, and the partition function with $Q$-closed states specified
on the boundary components of spacetime depends only on the
$Q$-cohomology classes of those states.  (More generally, correlation
functions of $Q$-closed operators with $Q$-closed states have a
similar property.)

Since the dependence of the action on the metric of $M$ is completely
buried in the $Q$-exact part, the theory becomes topological on $M$
once we restrict it to the $Q$-invariant sector.  Similarly, the
$Q$-invariant sector of the theory depends on the complex structure of
$C$ but not on the metric.  The anticommutation
relation~\eqref{eq:QQ'} shows that $P_\zb = 0$ in the $Q$-cohomology,
so correlation functions of $Q$-closed operators supported at points
on $C$ vary holomorphically on $C$.  In this sense, the twisted theory
is a topological--holomorphic theory on $M \times C$.

An example of a $Q$-closed operator is a Wilson line constructed from
$\CA$, supported along a closed curve $K \subset M$ and a point
$z \in C$:
\begin{equation}
  \Tr_V P\exp\biggl(\oint_{K \times \{z\}} \! \CA\biggr) \,.
\end{equation}
The trace is taken in some representation $G_\C \to \GL(V)$ of the
complexification~$G_\C$ of~$G$.  Such Wilson lines form one of the two
classes of observables from which we construct integrable lattice
models.

\subsection{Two-dimensional formulation}

Suppose that $C$ is an elliptic curve.  If we make $C$ very small and
discard the Kaluza--Klein modes, the topological--holomorphic theory
on $M \times C$ reduces to a topological theory on~$M$.  This is the
GL-twisted $\CN = 4$ super Yang--Mills theory \cite{Kapustin:2006pk}.
We can further take $M$ to be the product of a surface $D$ and a
torus, and make the torus very small.  Then, after discarding the
Kaluza--Klein modes, we obtain a topologically twisted $\CN = (8,8)$
super Yang--Mills theory on $D$.

In this series of reduction from six to two dimensions, we could as
well keep all Kaluza--Klein modes.  If we chose to do so, we would end
up with a formulation of the six-dimensional topological--holomorphic
theory as a two-dimensional gauge theory.  Let us describe this
two-dimensional formulation concretely, as we will use it later when
we introduce an $\Omega$-deformation of the theory.

\subsubsection{Two-dimensional supersymmetry}

Recall that the topological twist of the six-dimensional theory
replaces the generators $\Gamma_{\mu\nu}$ of the rotation group
$\Spin(6)$ with $\Gamma_{\mu\nu} + \Gamma_{\mu+6,\nu+6}$.  The
supercharges $Q$, $Q'$ are characterized by three conditions on the
parameter $\epsilon$ of supersymmetry transformation:
\begin{equation}
  (\Gamma_{01} + \Gamma_{67})\epsilon
  = (\Gamma_{12} + \Gamma_{78})\epsilon
  = (\Gamma_{23} + \Gamma_{89})\epsilon
  = 0 \,.
\end{equation}
Requiring the additional condition
\begin{equation}
 \epsilon = -\iu\Gamma_{39}\epsilon
\end{equation}
then picks out the supercharge $Q$ used to define the
topological--holomorphic theory.  To describe this procedure in
two-dimensional terms, let us impose the above conditions in a
different order.

First, we demand
$(\Gamma_{23} + \Gamma_{89})\epsilon = (1 + \iu\Gamma_{39})\epsilon =
0$.  These equations have four independent solutions that are
eigenvectors of the two-dimensional chirality operator
$-\iu\Gamma_{01}$.  The action of $\Gamma_{06}$ leaves the space of
solutions invariant but changes chirality, so there are equal number
of positive and negative chirality solutions.  The corresponding
supercharges generate $\CN = (2,2)$ supersymmetry in two dimensions.

Next, we impose $(\Gamma_{01} + \Gamma_{67})\epsilon = 0$, which
reduces the number of independent solutions to two.  There are two
$\U(1)$ R-symmetries that rotate the supercharges of $\CN = (2,2)$
supersymmetry, $\U(1)_V$ generated by $-\iu\Gamma_{45}$ and $\U(1)_A$
generated by $-\iu\Gamma_{67}$.  This condition means that we twist
the two-dimensional rotation group $\U(1)_D$ by replacing it with the
diagonal subgroup $\U(1)_D'$ of $\U(1)_D \times \U(1)_A$, and keep
only those supercharges that are scalars under $\U(1)_D'$.  The
R-symmetry $\U(1)_A$ used in this twisting acts on the scalars
$\phi_6$, $\phi_7$, which are part of the $Q$-invariant connection
$\sum_{i=0}^1 \CA_i \rmd x^i$ in two dimensions and belong to the
vector multiplet of $\CN = (2,2)$ supersymmetry.  It is known as the
axial $\U(1)$ R-symmetry, and the topological twist with respect to it
is called the \emph{B-twist}~\cite{Witten:1991zz, Hori:2003ic}.  Since
$\eps = \iu\Gamma_{45} \eps$, the scalar supercharges have charge $1$
under the other R-symmetry $\U(1)_V$, referred to as the vector
$\U(1)$ R-symmetry. The topological twist using $\U(1)_V$ is called
the A-twist.

Finally, the condition $(\Gamma_{12} + \Gamma_{78})\epsilon = 0$ picks
out a particular linear combination of the scalar supercharges, which
we have been calling $Q$.  We can choose another scalar
supercharge~$\Qt$ (which is different from $Q'$) such that $Q$ and
$\Qt$ obey the relations
\begin{equation}
  Q^2 = \Qt^2 = \{Q,\Qt\} = 0
\end{equation}
up to central charges.  These supercharges do not have definite
chirality since the last condition is not compatible with the
chirality condition $\eps = \pm\iu\Gamma_{01} \eps$.

\subsubsection{B-twisted gauge theory}

Our task is therefore to describe the six-dimensional
topological--holomorphic theory as a B-twisted gauge theory in two
dimensions.  To this purpose we briefly review the construction of the
latter theory.

We write $\SG$ for the gauge group of a B-twisted gauge theory to
distinguish it from the gauge group of the six-dimensional theory.  We
pick generators $\ST_a$, $a=1$, $\dotsc$, $\dim\SG$, of the Lie
algebra $\Lie(\SG)$ of $\SG$ that are orthonormal with respect to the
minus of an invariant symmetric bilinear form $\Tr$.  The spacetime of
the theory is a surface $D$, and the gauge bundle is a principal
$\SG$-bundle $\SP \to D$.

The basic ingredients of a B-twisted gauge theory are vector
multiplets and chiral multiplets.  A vector multiplet consists of a
gauge field $A$ of $\SP$, bosonic fields
\begin{equation}
  \sigma \in \Gamma\bigl(\Lambda_D^1 \otimes \ad(\SP)\bigr) \,,
  \qquad
  \auxD \in \Gamma\bigl(\Lambda_D^0 \otimes \ad(\SP)\bigr) \,,
\end{equation}
and fermionic fields
\begin{equation}
  \alpha \in \Gamma\bigl(\Lambda_D^0 \otimes \ad(\SP)\bigr) \,,
  \qquad
  \lambda \in \Gamma\bigl(\Lambda_D^1\otimes \ad(\SP)\bigr) \,,
  \qquad
  \zeta \in \Gamma\bigl(\Lambda_D^2 \otimes \ad(\SP)\bigr) \,.
\end{equation}
A chiral multiplet is valued in a unitary representation $R$ of $\SG$.
It consists of bosonic fields
\begin{equation}
  \varphi \in \Gamma\bigl(\Lambda_D^0 \otimes R(\SP)\bigr) \,,
  \qquad
  \auxF \in \Gamma\bigl(\Lambda_D^2\otimes R(\SP)\bigr) \,,
\end{equation}
and fermionic fields
\begin{equation}
  \etab \in \Gamma\bigl(\Lambda_D^0 \otimes \Rb(\SP)\bigr) \,,
  \qquad
  \rho \in \Gamma\bigl(\Lambda_D^1\otimes R(\SP)\bigr) \,,
  \qquad
  \mub \in \Gamma\bigl(\Lambda_D^2 \otimes \Rb(\SP)\bigr) \,,
\end{equation}
where $R(\SP)$ denotes the vector bundle associated to~$\SP$
constructed from $R$, and $\Rb$ is the complex conjugate of $R$ which
we also regard as the dual of $R$ by a hermitian form on the
representation space.

As mentioned already, a B-twisted $\CN = (2,2)$ supersymmetric theory
has two scalar supercharges, $Q$ and $\Qt$.  Under the action of $Q$,
the vector multiplet transforms as
\begin{equation}
  \label{eq:Q-VM}
  \begin{alignedat}{2}
    \delta\CA &= 0 \,,
    &
    \delta\CAb &= \lambda \,,
    \\
    \delta\lambda &= 0 \,,
    &\qquad
    \delta\alpha &= \auxD \,,
    \\
    \delta\auxD &= 0 \,,
    &
    \delta\zeta &= \CF \,,
  \end{alignedat}
\end{equation}
while the chiral multiplet transforms as
\begin{equation}
  \label{eq:Q-CM}
  \begin{alignedat}{2}
    \delta\varphi &= 0 \,,
    &
    \delta\varphib &= \etab \,,
    \\
    \delta\rho &= \cD\varphi \,,
    &
    \delta\etab &= 0 \,,
    \\
    \delta\auxF &= \cD\rho - \zeta\varphi \,,
    &\qquad
    \delta\auxFb &= 0 \,,
    \\
    &&
    \delta\mub &= \auxFb \,.
  \end{alignedat}
\end{equation}
Here we have introduced the notation
\begin{equation}
  \CA = A + \iu\sigma \,,
  \qquad
  \CAb = A - \iu\sigma \,.
\end{equation}
The fields $\varphib$ and $\auxFb$ are the hermitian conjugates of
$\varphi$ and $\auxF$.

The other supercharge $\Qt$ depends on a choice of a metric on $D$.
It acts on the vector multiplet by
\begin{equation}
  \begin{alignedat}{2}
    \deltat\CA &= \star\lambda \,,
    &
    \deltat\CAb &= 0 \,,
    \\ 
   \deltat\lambda &= 0 \,,
    &\qquad
    \deltat\alpha &= \star\CFb \,,
    \\
    \deltat\auxD &= -\star\cDb\lambda \,,
    &
    \deltat\zeta &= -\star\auxD + 2\iu D\star\sigma
  \end{alignedat}
\end{equation}
and on the chiral multiplet by
\begin{equation}
  \begin{alignedat}{2}
    \deltat\varphi &= 0 \,,
    &
    \deltat\varphib &= -\star\mub \,,
    \\
    \deltat\rho &= -\star\cDb\varphi \,,
    &\qquad
    \deltat\etab &= \star\auxFb \,,
    \\
    \deltat\auxF &= \cDb\star\rho - \star\alpha\varphi \,,
    &
    \deltat\auxFb &= 0 \,,
    \\
    &&
    \deltat\mub &= 0 \,.
  \end{alignedat}
\end{equation}

The main part of the action is exact with respect to both $Q$ and
$\Qt$.  The action governing the dynamics of the vector multiplet is
\begin{equation}
  \begin{split}
    \label{eq:SV}
    \SV
    &=
    \int_D \delta\deltat \Tr(\zeta\alpha)
    \\
    &=
    \int_D \delta \Tr\bigl(
    (-\star\auxD + 2\iu D\star\sigma) \alpha
    - \zeta \star\CFb\bigr)
    \\
    &=
    \int_D \Tr\bigl(
    -\CFb \star\CF
    - \auxD \star(\auxD - 2\iu \star D \star \sigma)
    + \alpha \cD\star\lambda
    + \zeta \star\cDb\lambda
    \bigr)
    \,.
  \end{split}
\end{equation}
The action for the chiral multiplet is
\begin{equation}
  \label{eq:SC}
  \begin{split}
    \SC
    &=
    \int_D \delta\deltat(-\varphib\auxF)
    \\
    &=
    \int_D \delta \bigl(
    -\varphib (\cDb\star\rho - \star\alpha \varphi)
    + \mub \star\auxF\bigr)
    \\
    &=
    \int_D \bigl(
    -\varphib \cDb\star\cD\varphi
    + \star\varphib\auxD \varphi
    + \auxFb \star\auxF
    \\
    & \qquad \qquad \qquad
    - \etab \cDb\star\rho
    - \mub \star\cD\rho
    + \star\etab\alpha\varphi
    - \varphib \lambda \wedge \star\rho
    + \mub \star\zeta\varphi
    \bigr)
    \,.
  \end{split}
\end{equation}

In addition, we can turn on a superpotential $W$, which is a gauge
invariant holomorphic function of the chiral multiplet scalars.  It
generates the interaction terms given by
\begin{equation}
  \begin{split}
    \SW
    &=
    \int_D \biggl(
    \auxF \frac{\del W}{\del\varphi}
    + \frac12 \rho \wedge \rho \frac{\del^2 W}{\del\varphi\del\varphi}
    + \star\delta\deltat\Wb
    \biggr)
    \\
    &=
    \int_D \biggl(
    \auxF \frac{\del W}{\del\varphi}
    + \frac12 \rho \wedge \rho \frac{\del^2 W}{\del\varphi\del\varphi}
    - \auxFb \frac{\del\Wb}{\del\varphib}
    - \etab \mub \frac{\del^2\Wb}{\del\varphib \del\varphib}
    \biggr)
    \,.
  \end{split}
\end{equation}

Unlike $\SV$ and $\SC$, this is neither $Q$-exact nor $\Qt$-exact.
Furthermore, it is not automatically invariant under $Q$ or $\Qt$ if
$D$ has a boundary.  The $Q$-invariance requires
\begin{equation}
  \int_{\del D} \rho \frac{\del W}{\del\varphi} = 0
  \,,
\end{equation}
while for the $\Qt$-invariance we need
\begin{equation}
  \int_{\del D} \star\rho \frac{\del W}{\del\varphi} = 0 \,.
\end{equation}
Appropriate boundary conditions must be imposed for the supercharges
to be unbroken.

\subsubsection{Topological--holomorphic theory as a B-twisted gauge
  theory}

Now we take $M = D \times \Sigma$ and describe the six-dimensional
topological--holomorphic theory on $D \times \Sigma \times C$ as a
B-twisted gauge theory on $D$.  We use letters $i$, $j$, $\dotsc$ for
indices for $D$ and $m$, $n$, $\dotsc$ for those for $\Sigma$.  For
simplicity we assume $D \times \Sigma \times C$ has no boundary (or
impose appropriate boundary conditions so that all boundary terms
arising from integration by parts vanish).

We need to organize the fields of the six-dimensional theory into
supermultiplets of B-twisted gauge theory.  Clearly, the theory has a
single vector multiplet whose gauge field $A_i \rmd x^i$ is part of
the six-dimensional gauge field.  Comparing the transformation
rules~\eqref{eq:Q-VM} and \eqref{eq:SUSY-6d} in two and six
dimensions, we identify the other fields in the vector multiplet as
\begin{equation}
  \sigma_i = \phi_i \,,
  \qquad
  \auxD = \auxP \,,
  \qquad
  \alpha = \xi' \,,
  \qquad
  \lambda_i = \psi_i \,,
  \qquad
  \zeta_{ij} = \chi_{ij} \,.
\end{equation}

In order to lift the vector multiplet action~\eqref{eq:SV} to six
dimensions, we interpret the bilinear form~$\Tr$ on~$\Lie(\SG)$ as
\begin{equation}
  \frac{1}{e^2} \int_{\Sigma \times C} \star_{\Sigma \times C} \Tr \,,
\end{equation}
where $\Tr$ in the integrand stands for the bilinear form on $\gf$.
This gives
\begin{equation}
  \SV
  =
  \frac{1}{e^2} \int_{D \times \Sigma \times C} \! \sqrt{g} \, \rmd^6 x
  \, \delta \Tr\biggl(
  (-\auxP + 2\iu D^i\phi_i) \xi'
  - \frac12 \chi^{ij} \CFb_{ij}\biggr)
  \,.
\end{equation}

As can be seen from the bosonic fields annihilated by $Q$, we have
three chiral multiplets in the adjoint representation, whose scalar
components are $\CA_m$ and $A_\zb$.  We name the fields of these
multiplets as $(\CA_m, \auxF_m, \etab_m, \rho_m, \mub_m)$ and
$(A_\zb, \auxF_\zb, \etab_z, \rho_\zb, \mub_z)$.

To lift formula~\eqref{eq:Q-CM} to six dimensions, what we have to do
is essentially to replace the scalar fields with the corresponding
covariant derivatives so that when we perform dimensional reduction on
$\Sigma \times C$, we get back to the same formula.  In this way we
obtain
\begin{equation}
  \begin{alignedat}{2}
    \delta\CA_m &= 0 \,,
    &
    \delta\CAb_m &= \etab_m \,,
    \\
    \delta\rho_{mi} &= \CF_{im} \,,
    &
    \delta\etab_m &= 0 \,,
    \\
    \delta\auxF_{mij} &= \cD_i \rho_{mj} - \cD_j \rho_{mi} + \cD_m \zeta_{ij} \,,
    &\qquad
    \delta\auxFb_{mij} &= 0 \,,
    \\
    &&
    \delta\mub_{mij} &= \auxFb_{mij}
  \end{alignedat}
\end{equation}
and
\begin{equation}
  \begin{alignedat}{2}
    \delta A_\zb &= 0 \,,
    &
    \delta A_z &= \etab_z \,,
    \\
    \delta\rho_{\zb i} &= \CF_{i\zb} \,,
    &
    \delta\etab_z &= 0 \,,
    \\
    \delta\auxF_{\zb ij} &= \cD_i \rho_{\zb j} - \cD_j \rho_{\zb i} + D_\zb \zeta_{ij} \,,
    &\qquad
    \delta\auxFb_{z ij} &= 0 \,,
    \\
    &&
    \delta\mub_{z ij} &= \auxFb_{z ij} \,.
  \end{alignedat}
\end{equation}

From the $Q$-variations involving the gauge field, we see
\begin{equation}
  \etab_m = \psi_m \,,
  \qquad
  \rho_{mi} = \chi_{im} \,,
  \qquad
  \etab_z = \xi \,,
  \qquad
  \rho_{\zb i} = \psi'_i \,.
\end{equation}
With this identification, we can write
$\delta\auxF_{mij} = (\cD\chi)_{mij}$ and
$\delta\auxF_{\zb ij} = (\cD\psi')_{ij} + D_\zb\chi_{ij}$.  On the
other hand, from the six-dimensional action we derive the equations of
motion
\begin{equation}
  \cD_M\chi
  = -\star_M D_z\psi + \star_M\cD_M\xi
  = -\star_M \iota_{\del_z} \delta\CFb
\end{equation}
and
\begin{equation}
  \cD_M\psi' + D_\zb\chi
  = -\star_M\cDb_M\psi
  = -\star_M\delta\CFb
  \,,
\end{equation}
with $\cD_M = \cD_\mu \rmd x^\mu$.  Combining these equations we
deduce the on-shell relations
\begin{equation}
  \label{eq:EOM-auxF}
  \auxF_{mij} = -\bigl(\star_M \iota_{\del_z} \CFb\bigr)_{mij} \,,
  \qquad
  \auxF_{\zb ij} = -\bigl(\star_M\CFb\bigr)_{ij} \,.
\end{equation}
Then, we have $\auxFb_{mij} = -(\star_M \iota_{\del_\zb} \CF)_{mij} $
and $\auxFb_{zab} = -(\star_M\CF)_{ij}$ on shell (note the sign; the
on-shell value of $\auxFb$ is minus the hermitian conjugate of
$\auxF$) and
\begin{equation}
  \label{eq:mub}
  \mub_{mij} = (\star_M\psi')_{mij}
  \,,
  \qquad
  \mub_{zij} = -(\star_M\chi)_{ij} \,.
\end{equation}

Lifting the chiral multiplet action \eqref{eq:SC} to six dimensions is
also straightforward.  For example, the term
$\delta(-\varphib \cDb\star\rho)$ in the integrand on the second line
can be converted to $\delta(\cDb\varphib \wedge \star\rho)$ by
integration by parts and lifted to
$\delta\Tr( - \CFb^{im} \chi_{im} - \CFb_{iz} \psi'^i)$.  Also, since
$\varphi$ is in the adjoint representation,
$\delta(\varphib\star\alpha\varphi)$ can be written as
$\delta\Tr(\star[\varphi,\varphib]\alpha)$ and is lifted to
$\delta \Tr((2\iu D^m\phi_m + F_{z\zb})\xi')$.
For comparison with the six-dimensional description, it is useful to
express the chiral multiplet action as
\begin{multline}
  \label{eq:SC-6d}
  \SC
  =
  \frac{1}{e^2} \int_{D \times \Sigma \times C} \! \sqrt{g} \, \rmd^6 x
  \Tr\biggl(
  \delta\bigl(
  - \CFb^{im} \chi_{im}
  - \CFb_{iz} \psi'^i
  + (2\iu D^m\phi_m + F_{z\zb})\xi' 
  \bigr)
  \\
  + \frac12 \auxFb^{mij} \auxF_{mij}
  + \frac12 \auxFb_z^{ij} \auxF_{\zb ij}
  \biggr)
  \\
  +
  \frac{1}{e^2} \int_{D \times \Sigma \times C} \! \rmd^2 z
  \Tr\bigl(
  - \cD\chi \wedge \psi'_\Sigma
  + \chi_\Sigma \wedge (\cD\psi' + D_\zb\chi)
  \bigr)
  \,.
\end{multline}
Here we have defined
\begin{equation}
  \chi_\Sigma = \frac12 \chi_{mn} \rmd x^m \wedge \rmd x^n \,,
  \qquad
  \psi'_\Sigma = \psi'_m \rmd x^m \,.
\end{equation}

We also need to determine the superpotential.  In order to reproduce
the equations of motion \eqref{eq:EOM-auxF}, the superpotential must
be
\begin{equation}
  \label{eq:W-6d}
  W = -\frac{\iu}{e^2} \int_{\Sigma \times C} \rmd z \wedge \ChS(\CA) \,,
\end{equation}
where
\begin{equation}
  \ChS(\CA)
  = \Tr\biggl(\CA \wedge \rmd\CA + \frac23 \CA \wedge \CA \wedge \CA\biggr)
\end{equation}
is the Chern--Simons three-form constructed from $\CA$.  The
corresponding superpotential terms are
\begin{multline}
  \label{eq:SW-6d}
  \SW
  =
  \frac{1}{e^2} \int_{D \times \Sigma \times C} \! \sqrt{g} \, \rmd^6 x
  \Tr\biggl(
  \frac12 \auxF^{mij} (\star_M\iota_{\del_\zb}\CF)_{mij}
  + \frac12 \auxF_\zb^{ij} (\star_M \CF)_{ij} 
  \\
  + \frac12 \auxFb^{mij} (\star_M\iota_{\del_z}\CFb)_{mij}
  + \frac12 \auxFb_z^{ij} (\star_M \CFb)_{ij} 
  - \delta\CFb_{mz} \psi'^m
  + \frac12 \chi^{mn} \delta\CFb_{mn}\biggr)
  \\
  +
  \frac{1}{e^2} \int_{D \times \Sigma \times C} \! \rmd^2 z
  \Tr\biggl(
  \frac12 \chi_{D|\Sigma} \wedge D_\zb \chi
  + \chi_{D|\Sigma} \wedge \cD \psi'_D
  \biggr) \,,
\end{multline}
where
\begin{equation}
  \chi_{D|\Sigma} = \chi_{im} \, \rmd x^i \wedge \rmd x^m \,,
  \qquad
  \psi'_D = \psi'_i \, \rmd x^i \,.
\end{equation}
The superpotential~\eqref{eq:W-6d} is not quite gauge invariant, but
this is not a problem because the resulting action is gauge invariant.

While the sum $\SV + \SC + \SW$ reproduces the fermionic part of the
six-dimensional action, they lack the terms
\begin{equation}
  \frac{1}{e^2} \int_{D \times \Sigma \times C} \! \sqrt{g} \, \rmd^6 x
  \Tr\biggl(
  -\frac12 \CFb^{mn} \CF_{mn}
  - 2 \CFb^{m\zb} \CF_{m\zb}
  \biggr)
\end{equation}
from the bosonic part.  These missing terms are supplied when the
auxiliary fields are integrated out.  Thus, we have obtained a
two-dimensional formulation of the six-dimensional
topological--holomorphic theory, which is applicable for
$M = D \times \Sigma$.

\section{\texorpdfstring{$\Omega$-deformation of B-twisted gauge theories}{Ω-deformation of B-twisted gauge theories}}
\label{sec:Omega-def}

Once we reformulate the six-dimensional topological--holomorphic
theory as a two-dimensional B-twisted gauge theory, we can subject it
to an $\Omega$-deformation~\cite{Nekrasov:2002qd, Nekrasov:2003rj,
  Nekrasov:2010ka} following the construction of~\cite{Luo:2014sva}.
Via localization of the path integral, the $\Omega$-deformation
reduces the topological sector of the B-twisted gauge theory to a
zero-dimensional gauge theory with complex gauge
group~\cite{Yagi:2014toa, Luo:2014sva, Nekrasov:2018pqq}.  In this
section we discuss this localization mechanism for a general B-twisted
gauge theory.

\subsection{\texorpdfstring{$\Omega$-deformation}{Ω-deformation}}

As we have seen above, a B-twisted gauge theory has two scalar
supercharges $Q$ and~$\Qt$.  If the spacetime $D$ is flat, the theory
additionally has a one-form supercharge $G = G_i \rmd x^i$, satisfying
$\{Q, G_i\} = P_i$ and $\{G_i, G_j\} = 0$ in some coordinates.  More
generally, if $V$ is a parallel vector field on $D$ (which may or may
not be curved), there is an associated fermionic symmetry and hence
the corresponding supercharge $\iota_V G$.  The linear combination
$Q + \iota_V G$ is then a supercharge which squares to $V$.

If $V$ is not covariantly constant, $\iota_V G$ does not exist in
general.  Nevertheless, if $V$ is a Killing vector field generating an
isometry of $D$, we can construct a deformation of the theory such
that it has a supercharge $Q_V$ that squares to the generator of the
isometry and reduces to $Q$ for $V = 0$.  This deformation is what we
call an $\Omega$-deformation of the B-twisted gauge theory.

Specifically, the deformed supercharge $Q_V$ acts on the vector
multiplet by
\begin{equation}
  \begin{alignedat}{2}
    \delta_V\CA &= \iota_V\zeta \,,
    &
    \delta_V\CAb &= \lambda - \iota_V\zeta \,,
    \\
    \delta_V\lambda &= 2\iota_V F - 2\iu  D\iota_V\sigma \,,
    &\qquad
    \delta_V\zeta &= \CF \,,
    \\
    \delta_V\alpha &= \auxD \,,
    &
    \delta_V \auxD &= \iota_V\cD\alpha \,,
  \end{alignedat}
\end{equation}
and on the chiral multiplet by
\begin{equation}
  \begin{alignedat}{2}
    \delta_V\varphi &= \iota_V\rho \,,
    &
    \delta_V\varphib &= \etab \,,
    \\
    \delta_V\rho &= \cD\varphi + \iota_V\auxF \,,
    &\qquad
    \delta_V\etab &= \iota_V \cD \varphib \,,
    \\
    \delta_V\auxF &= \cD\rho - \zeta\varphi \,,
    &
    \delta_V\auxFb &= \cD\iota_V \mub \,,
    \\
    &&
    \delta_V\mub &= \auxFb \,.
  \end{alignedat}
\end{equation}
Its square is essentially the Lie derivative
$\CL_V = \rmd \iota_V + \iota_V \rmd$, but made covariant with respect
to the complexified gauge symmetry:
\begin{equation}
  Q_V^2
  = \cD \iota_V + \iota_V \cD
  \,.
\end{equation}
The right-hand side equals $\CL_V$ plus the infinitesimal gauge
transformation generated by~$\iota_V\CA$.

Being a generator of an isometry, $V$ is a real vector field.  More
generally, we allow $V$ to be a complex Killing vector field that
commutes with its complex conjugate $\Vb$.  Also, $V|_{\del D}$ must
be tangent to $\del D$ so that the isometry preserves the boundary.

The action of the $\Omega$-deformed theory is again of the form
$\SV + \SC + \SW$, each term being a $Q_V$-invariant deformation of
the corresponding term in the undeformed action.  As in the undeformed
case, we can take $\SV$ and $\SC$ to be $Q_V$-exact.  A minimal choice
is
\begin{equation}
  \label{eq:SV-Omega}
  \begin{split}
    \SV
    &=
    \delta_V \int_D \Tr\bigl(
    (-\star\auxD + 2\iu D\star\sigma) \alpha
    - \zeta \star\CFb\bigr)
    \\
    &=
    \int_D \Tr\bigl(
    -\CFb \star\CF
    - \auxD \star(\auxD - 2\iu \star D \star \sigma)
    + \alpha \cD\star\lambdat
    + \zeta \star\cDb\lambdat
    + \alpha \rmd V^\flat \star\zeta
    \bigr)
    \\
    &\qquad\qquad\qquad
    +
    \int_{\del D} \Tr(\star\zeta\star\iota_V\star\alpha)
  \end{split}
\end{equation}
and
\begin{equation}
  \label{eq:SC-Omega}
  \begin{split}
    \SC
    &=
    \delta_V\int_D \bigl(
    \bigl(\cDb\varphib + \iota_\Vb\auxFb\bigr) \wedge \star\rho
    + \star\varphib\alpha \varphi
    + \mub \star\auxF
    \bigr)
    \\
    &=
    \int_D \bigl(
    \bigl(\cDb\varphib + \iota_\Vb\auxFb\bigr)
    \wedge \star \bigl(\cD\varphi + \iota_V\auxF\bigr)
    + \star\varphib\auxD \varphi
    + \auxFb \star\auxF
    \\
    & \qquad \qquad \qquad
    + \cDb\etab \star\rho
    - \mub \star\cD\rho
    + \star\etab\alpha\varphi
    - \varphib \lambdat \wedge \star\rho
    + \mub \star\zeta\varphi
    + \iota_\Vb \cD\iota_V\mub \star\rho
    \bigr)
    \,,
  \end{split}
\end{equation}
where $V^\flat$ is the one-form dual to $V$ with respect to the metric
on $D$ and
\begin{equation}
  \lambdat = \lambda - \iota_V\zeta - \star\iota_V\star\alpha \,.
\end{equation}
It is important here that $V$ is a Killing vector field and
$[V, \Vb] = 0$.  The former property means that $\CL_V$ annihilates
the metric and commutes with $\star$, while the latter implies
$[\CL_V, \iota_\Vb] = 0$.  Together with the identity
$[\CL_V, \iota_V] = 0$, these properties ensure the $Q_V$-invariance
of $\SV$ and $\SC$.

Remarkably, the $\Omega$-deformation allows $\SW$ to be
$Q_V$-invariant without resorting to any boundary conditions.
Suppose, for simplicity, that there is only one boundary component in
$ D$, and parametrize this boundary circle by an angular coordinate
$\theta$.  Then
\begin{equation}
  \label{eq:SW-Omega}
  \begin{split}
    \SW
    &=
    \int_D \biggl(
    \auxF \frac{\del W}{\del\varphi}
    + \frac12 \rho \wedge \rho \frac{\del^2 W}{\del\varphi\del\varphi}
    - \delta_V \biggl(\mub \frac{\del\Wb}{\del\varphib}\biggr)
    \biggr)
    - \int_{\del D} W \frac{\rmd\theta}{V^\theta}
    \\
    &=
    \int_D \biggl(
    \auxF \frac{\del W}{\del\varphi}
    + \frac12 \rho \wedge \rho \frac{\del^2 W}{\del\varphi\del\varphi}
    - \auxFb \frac{\del\Wb}{\del\varphib}
    - \etab \mub \frac{\del^2\Wb}{\del\varphib \del\varphib}
    \biggr)
    - \int_{\del D} W  \frac{\rmd\theta}{V^\theta}
  \end{split}
\end{equation}
is a $Q_V$-invariant superpotential action.

Since $Q_V$ squares to zero on operators and states that are invariant
under the gauge symmetry and the isometry, we can define its
cohomology in the spaces of such states and operators.  Unlike the
undeformed case, the $Q_V$-invariant sector of the theory is not quite
topological: it is invariant under deformations of the metric only if
$V$ remains as a Killing vector field.  For this reason, we refer to
the $\Omega$-deformed B-twisted gauge theory as a quasi-topological
theory.

\subsection{Localization on a disk}

As we have just seen, an $\Omega$-deformation can be applied to a
B-twisted gauge theory whenever the spacetime $D$ has an isometry.  A
basic example is when $D$ is a disk of finite radius, equipped with a
rotation invariant metric, and $V$ is a generator of rotations.  We
now show that for a suitable boundary condition, the quasi-topological
sector of the $\Omega$-deformed theory is in this case equivalent to a
zero-dimensional theory, whose path integral is performed over a
domain specified by the boundary condition.

To be concrete, we endow $D$ with the metric of a hemisphere of unit
area.  In terms of polar coordinates $(r,\theta)$, the metric takes
the form
\begin{equation}
  g(r,\theta) = g_{rr}(r) \rmd r^2 + g_{\theta\theta}(r) \rmd\theta^2
\end{equation}
and we have
\begin{equation}
  V = \epsilon\del_\theta
\end{equation}
for some $\epsilon \in \C$.  We use hatted indices $(\rh, \thetah)$ to
denote components of tensors with respect to the orthonormal vectors
$\del_\thetah = \sqrt{g^{\theta\theta}} \, \del_\theta$,
$\del_\rh = \sqrt{g^{rr}} \, \del_r$ and their duals
$\rmd\thetah = \sqrt{g_{\theta\theta}} \, \rmd\theta$,
$\rmd\rh = \sqrt{g_{rr}} \, \rmd\rh$.  For example,
$V^\thetah = \sqrt{g_{\theta\theta}} \, V^\theta$ and $|V^\thetah|$
equals the norm $\|V\|$ of $V$.

\subsubsection{Boundary conditions}

To begin with, let us figure out what sort of boundary condition
should be imposed.  In general, a good boundary condition ensures that
the boundary terms vanish in the variation of the action so that the
classical equations of motion are obtained from the variational
principle.  In our case, we moreover want the boundary condition to be
$Q_V$-invariant.

We start with the vector multiplet.  Integrating $\auxD$ out and
varying the gauge field, we see that the boundary terms in the
variation of the action vanish if either the Dirichlet condition
$\delta A_\theta = 0$ or the Neumann condition $F_{r\theta} = 0$ is
satisfied on the boundary.  The former breaks the gauge symmetry on
the boundary.  For our applications we look for a boundary condition
that preserves the gauge symmetry, so we pick the Neumann condition.
We can also choose a gauge in which
\begin{equation}
  \label{eq:Ar=0}
  A_r = 0
\end{equation}
on the boundary.  Then, the Neumann condition reads
\begin{equation}
  \label{eq:del_r-A_theta=0}
  \del_r A_\theta = 0 \,.
\end{equation}

Next, varying $\sigma$, we find that each component $\sigma_i$ of
$\sigma$ should obey either the Dirichlet condition
$\delta\sigma_i = 0$ or the Neumann condition $\del_r\sigma_i = 0$.
In view of the fact that the gauge field appears in $Q_V^2$ through
the combination $\CA = A + \iu\sigma$, it is natural to choose
\begin{equation}
  \sigma_r = \del_r\sigma_\theta = 0 \,.
\end{equation}

Letting $Q_V$ act on the boundary conditions we have so far, we get
\begin{equation}
  \zeta_{r\theta} = \lambda_r = \del_r\lambda_\theta = 0 \,.
\end{equation}
These conditions already ensure that the boundary terms vanish under
variations of the fermions.

Since $\lambda_\theta$ does not vanish on the boundary, the action
should have a term that contains the boundary value of
$\lambda_\theta$.  The only term that may not vanish on the boundary
is $\alpha \cD^\theta \lambda_\theta$.  So we require $\alpha$ to obey
the Neumann condition
\begin{equation}
  \del_r\alpha = 0 \,,
\end{equation}
just as $\lambda_\theta$ does.  Then we must have
\begin{equation}
  \del_r\auxD = 0
\end{equation}
for $Q_V$-invariance.

The set of boundary conditions for the vector multiplet thus obtained
is $Q_V$-invariant.  Repeated action of $Q_V$ does not lead to any
further conditions since $Q_V^2$ just generates translations on the
boundary.

On the chiral multiplet, we impose a boundary condition of brane type.
The target space $X$ for the chiral multiplet is the representation
space of the representation $R$.  We choose a submanifold $\gamma$ in
$X$, and demand the boundary value of the scalar field to lie
in~$\gamma$:
\begin{equation}
  \varphi \in \gamma \,.
\end{equation}
The $Q_V$-action on this condition yields
\begin{equation}
  \label{eq:BC-rho-eta}
  (\iota_V\rho, \etab) \in T_\varphi\gamma \otimes \C
  \,.
\end{equation}
We require $\gamma$ to be $G$-invariant so that the gauge symmetry is
preserved.  Furthermore, we assume that $\Re(W/\epsilon)$ is bounded
above on $\gamma$ so that the boundary term in the superpotential
action \eqref{eq:SW-Omega} does not render the path integral
divergent.

Varying the fermions we get the boundary terms
\begin{equation}
  \int_{\del D} \! \rmd\thetah \,
  \biggl(
  - \frac{1}{\|V\|^2} \delta(\iota_V \rho) (\iota_\Vb \mub)_\rh
  + \delta\etab \rho_\rh
  \biggr)
  \,.
\end{equation}
For these terms to vanish, we should have
\begin{equation}
  \label{eq:BC-rho-mu}
  \bigl(\rho_\rh, -\|V\|^{-2} (\iota_\Vb\mub)_\rh\bigr)
  \in N_\varphi\gamma \otimes \C
  \,,
\end{equation}
where $N\gamma$ is the normal bundle of $\gamma$ with respect to the
K\"ahler metric
\begin{equation}
  \label{eq:gX}
  g_X = \Re(\rmd\varphi \otimes \rmd\varphib)
\end{equation}
of $X$.  The $Q_V$-variation of this condition, together with the
gauge condition~\eqref{eq:Ar=0}, gives
\begin{equation}
  \label{eq:BC-normal}
  \bigl(\del_\rh\varphi + (\iota_V \auxF)_\rh,
  - \|V\|^{-2} (\iota_\Vb \auxFb)_\rh,\bigr)
  \in N_\varphi\gamma \otimes \C
  \,,
\end{equation}
which completes a $Q_V$-invariant set of boundary conditions on the
chiral multiplet.

The equations of motion for $\auxF$ and $\auxFb$ are
\begin{equation}
  \label{eq:F-Fb}
    \auxF_{\rh\thetah}
    =
    \frac{1}{1 + \|V\|^2}
    \biggl(\Vb^\thetah \cD_\rh\varphi
    + \frac{\del\Wb}{\del\varphib}\biggr)
    \,,
    \qquad
    \auxFb_{\rh\thetah}
    =
    \frac{1}{1 + \|V\|^2}
    \biggl(V^\thetah \cDb_\rh\varphib
    - \frac{\del W}{\del\varphi}\biggr)
    \,.
\end{equation}
Plugging these equations into the boundary
condition~\eqref{eq:BC-normal}, we get
\begin{equation}
  \label{eq:BC-normal-onshell}
  \biggl(\del_\rh\varphi
  - V^\thetah \frac{\del\Wb}{\del\varphib},
  \del_\rh\varphib
  - \frac{1}{V^\thetah} \frac{\del W}{\del\varphi}\biggr)
  \in N_\varphi\gamma \otimes \C
  \,.
\end{equation}
As a check, let us verify that boundary terms are absent in the
variation of the action under this boundary condition.  After $\auxF$
and $\auxFb$ are integrated out, the bosonic terms in $\SC + \SW$ are
given by
\begin{multline}
  \label{eq:boson-SC+SW}
  \int_D \! \rmd\rh \, \rmd\thetah
  \biggl(
  \frac{1}{1 + \|V\|^2}
  \biggl(\cDb_\rh\varphib \cD_\rh\varphi
  + \Vb^\thetah \del_\rh W - V^\thetah \del_\rh\Wb
  + \frac{\del W}{\del\varphi} \frac{\del\Wb}{\del\varphib}\biggr)
  + \cDb_\thetah\varphib \cD_\thetah\varphi
  + \varphib\auxD \varphi
  \biggr)
  \\
  - \int_{\del D} W \frac{\rmd\thetah}{V^\thetah}
  \,.
\end{multline}
Varying the scalars, we see that the boundary terms indeed vanish.

In the undeformed case $\epsilon = 0$, the boundary
condition~\eqref{eq:BC-normal-onshell} implies that $W$ is locally
constant on $\gamma$.  The same condition then requires
$(\del_r\varphi, \del_r\varphib) \in N_\varphi\gamma$ on the boundary.
If $\gamma$ is a complex submanifold, this is (part of) a ``B-brane''
boundary condition for a B-twisted Landau--Ginzburg
model~\cite{Hori:2003ic}, which preserves half of $\CN = (2,2)$
supersymmetry.  For our application, however, we will actually take
$\gamma$ to be, roughly speaking, a \emph{Lagrangian} submanifold.

We remark that the boundary condition described here depends on
$\|V\|^2$.  As a consequence, the presence of boundary mildly breaks
the quasi-topological invariance of the theory.  We are still allowed
to deform the metric as long as we continue to impose the same
boundary condition defined with respect to the original metric.

\subsubsection{Gauge fixing}

Performing the path integral requires gauge fixing.  We do this by
adapting the BRST gauge fixing procedure to the present setting.

We enlarge the set of fields with additional fermionic fields $b$, $c$
and auxiliary bosonic field $\auxB$, all transforming in the adjoint
representation:
\begin{equation}
  b,\, c, \, \auxB \in \Gamma\bigl(\ad(\SP)\bigr) \,.
\end{equation}
Then we introduce the BRST symmetry that acts on these fields by
\begin{equation}
  \delta_\BRST b = \auxB
  \,,
  \qquad
  \delta_\BRST \auxB = 0
  \,,
  \qquad
  \delta_\BRST c = \frac12 \{c,c\}
  \,.
\end{equation}
On the other fields the BRST symmetry acts by the gauge transformation
generated by $c$; for instance, $\delta_\BRST\varphi = c\varphi$.
Since the action of the theory is gauge invariant, it is invariant
under the BRST symmetry.

The conserved charge $Q_\BRST$ for the BRST symmetry squares to zero.
In the standard BRST gauge fixing, one adds $Q_\BRST$-exact
gauge fixing terms to the action and considers the
$Q_\BRST$-cohomology.  However, such terms will not be $Q_V$-invariant
and breaks the quasi-topological invariance of the theory.

To remedy this problem we combine $Q_\BRST$ with $Q_V$.  Let us
postulate that $Q_V$ acts on $b$, $c$ and $\auxB$ by
\begin{equation}
  \delta_V b = 0
  \,,
  \qquad
  \delta_V \auxB = \iota_V\rmd b
  \,,
  \qquad
  \delta_V c = -\iota_V\CA
  \,.
\end{equation}
With this definition of the action of $Q_V$, the combined charge
$Q_{V + \BRST} = Q_V + Q_\BRST$ satisfies
\begin{equation}
  Q_{V+\BRST}^2 = \iota_V \rmd + \rmd \iota_V
  \,.
\end{equation}
The right-hand side is the ordinary Lie derivative instead of a gauge
covariant one, so we can define the cohomology with respect to the
action of $Q_{V+\BRST}$ on rotation invariant states and operators
which are not necessarily gauge invariant.

After gauge fixing, therefore, what we should consider is not the
$Q_V$-cohomology, but the $Q_{V+\BRST}$-cohomology in the spaces of
rotation invariant states and operators.  Since $\SV$ and $\SC$ are
$Q_V$-commutators of gauge invariant expressions, they are
automatically exact with respect to $Q_{V+\BRST}$.  The
quasi-topological invariance of the theory is thus maintained.

Now we can perform gauge fixing as in the usual BRST procedure,
treating $Q_{V+\BRST}$ as the BRST operator.  We pick a suitable
$\Lie(\SG)$-valued function $f_\GF$ constructed from the original set of
fields, and add to the action the $Q_{V+\BRST}$-exact term
\begin{equation}
  \delta_{V + \BRST} \int_D \star\Tr(2\iu b f_\GF)
  =
  \int_D \star \Tr(2\iu \auxB f_\GF
  - 2\iu b \, \delta_{V + \BRST} f_\GF)
  \,.
\end{equation}
Integrating over $\auxB$ produces a delta function which imposes the
gauge fixing condition
\begin{equation}
  f_\GF = 0 \,.
\end{equation}
The fermionic part can be written as
\begin{equation}
  2\iu \int_D \star (\ST_b \cdot f_\GF)^a b^a \ct^b
\end{equation}
for some fermion
\begin{equation}
  \ct = c + \dotsb \,,
\end{equation}
where $\ST_b \cdot f_\GF$ denotes the infinitesimal gauge
transformation of $f_\GF$ by $\ST_b \in \Lie(\SG)$.  This is possible
because for a sensible choice of $f_\GF$, the matrix-valued function
$((\ST_b \cdot f_\GF)^a)_{a,b = 1}^{\dim\SG}$ represents an invertible
operator on the field space.  The integration over the fermions
produces the Faddeev--Popov determinant for the gauge fixing.

For the convenience of computation, we actually make another choice of
gauge fixing terms:
\begin{equation}
  \begin{split}
    \SGF
    &=
    \delta_{V + \BRST} \int_D \star\Tr(-b\auxB + 2\iu b f_\GF)
    \\
    &=
    \int_D \star \Tr\bigl(-\auxB^2 + b \iota_V \rmd b
    + 2\iu \auxB f_\GF
    - 2\iu b \, \delta_{V + \BRST} f_\GF\bigr)
    \,.
  \end{split}
\end{equation}
This is a $Q_{V+\BRST}$-exact deformation of the previous gauge
fixing action, so it leads to the same result.  With this choice,
integrating $\auxB$ out yields the potential term $\Tr(-f_\GF^2)$
rather than setting $f_\GF = 0$.

In the present situation where $D$ is endowed with a hemisphere
metric, the gauge field on $D$ can be obtained from one on the sphere
$S^2$ by restriction.  Then, Uhlenbeck's theorem~\cite{MR648355}
guarantees that there exists a representative satisfying the Lorenz
gauge condition
\begin{equation}
  \nabla^i A_i = 0
\end{equation}
in each gauge equivalence class (at least when the field strength is
sufficiently small), and it is unique up to constant gauge
transformations.  Further, the Lorenz gauge is compatible with the
boundary conditions~\eqref{eq:Ar=0} and \eqref{eq:del_r-A_theta=0} on
the equator $\del D$.%
\footnote{Let $(\theta,\phi)$ be the spherical coordinates on $S^2$,
  and $A$ a gauge field on $D$ obeying the boundary condition
  $A_\phi = \del_\phi A_\theta = 0$ at $\phi = \pi/2$.  We extend $A$
  to $S^2$ by setting
  $A_\theta(\theta,\phi) = A_\theta(\theta,\pi - \phi)$ and
  $A_\phi(\theta,\phi) = -A_\phi(\theta,\pi - \phi)$ for
  $\phi \geq \pi/2$.  Suppose that a gauge transformed connection
  $A^g = gA g^{-1} - g^{-1} \rmd g$ satisfies the Lorenz gauge
  condition, and let $\gt(\theta,\phi) = g(\theta,\pi - \phi)$.  Then,
  $A^{\gt}_\theta(\theta,\phi) = A^g_\theta(\theta,\pi - \phi)$ and
  $A^{\gt}_\phi(\theta,\phi) = -A^g_\phi(\theta,\pi - \phi)$, and
  $A^{\gt}$ also satisfies the Lorenz gauge condition.  By the
  uniqueness property we have $\gt = g_0 g$ for some constant
  element~$g_0$.  This implies that at $\phi = \pi/2$, we have
  $-g^{-1} \del_\phi g = -\gt^{-1} \del_\phi\gt = +g^{-1} \del_\phi g$
  and hence $A^g_\phi = -g^{-1} \del_\phi g = 0$ and
  $\del_\phi A^g_\theta = g F_{\phi\theta} g^{-1} = 0$.}

We can remove the residual gauge freedom by imposing an additional
gauge fixing condition:
\begin{equation}
  f_{\GF,0}(\varphi_0,\varphib_0)= 0 \,.
\end{equation}
Here $f_{\GF,0}$ is an appropriate $\Lie(\SG)$-valued function on $X$,
and $\varphi_0$ is the constant part of~$\varphi$, defined as the
average of $\varphi$ with respect to the volume form of~$D$:
\begin{equation}
  \varphi_0 = \int_D \star\varphi \,.
\end{equation}
Hence, we take our gauge fixing function to be
\begin{equation}
  f_\GF = \nabla^i A_i + f_{\GF,0}(\varphi_0,\varphib_0)
  \,.
\end{equation}
The corresponding gauge fixing action is
\begin{multline}
    \SGF
    =
    \int_D \star \Tr\bigl(
    - (\nabla^i A_i)^2
    + b \iota_V \rmd b
    + 2\iu b \nabla^i D_i c
    - \iu b \nabla^i \lambda_i
    \bigr)
    \\
    -
    \Tr\bigl(f_{\GF,0}(\varphi_0,\varphib_0)^2
    + 2\iu b_0 \, \delta_{V + \BRST} f_{\GF,0}(\varphi_0,\varphib_0)\bigr)
    \,,
\end{multline}
where $b_0$ is the constant part of $b$.

Before proceeding, we need to specify the boundary conditions for $b$,
$c$ and $\auxB$.  For the $Q_{V+\BRST}$-action to preserve the
boundary gauge condition~\eqref{eq:Ar=0}, $c$ must obey the Neumann
condition.  Then, for the nondegeneracy of $\SGF$ on the boundary $b$
should also obey the Neumann condition, and so should $\auxB$ for
$Q_{V+\BRST}$-invariance.  Thus, we impose the boundary condition
\begin{equation}
  \del_r b
  = \del_r c 
  = \del_r \auxB
  = 0
  \,.
\end{equation}
The set of all boundary conditions is then invariant under $Q_\BRST$
as well as $Q_V$.

\subsubsection{Localization}

We are ready to demonstrate the equivalence between the
$Q_{V+\BRST}$-invariant sector of the $\Omega$-deformed B-twisted
gauge theory and a zero-dimensional theory.  To this end we take
advantage of the invariance of the theory under $Q_{V+\BRST}$-exact
deformations and rescale the kinetic terms by large factors.  Such a
rescaling makes the oscillating modes of fields very massive and
yields an effective description in terms of constant modes.

There are various ways of doing this by a $Q_{V+\BRST}$-exact
deformation, but perhaps the most transparent is to rescale the metric
as
\begin{equation}
  g \to t^{-2} g
\end{equation}
and send $t$ to a large value.  (A disadvantage of this choice of
deformation is that the metric appearing in the action no longer
matches the one that enters the boundary condition.)  In other words,
we shrink the spacetime $D$ by a large factor so that the excited
modes get large masses.  To cancel the accompanied rescaling of the
volume form, at the same time we also rescale the $Q_{V+\BRST}$-exact
part of the action by a factor of $t^{2}$:
\begin{equation}
  \label{eq:rescaling}
  \SV + \SC + \SGF + \SW
  \to t^2(\SV + \SC + \SGF) + \SW \,.
\end{equation}
We want to show that in the limit $t \to \infty$, the path integral
with respect to this deformed action reduces to the path integral for
a zero-dimensional theory.

After this rescaling, the bosonic part of the action, with the
auxiliary fields integrated out, becomes
\begin{multline}
  \label{eq:bosonic-terms}
  \int_D \! \rmd\rh \, \rmd\thetah \biggl( t^4 \Tr\biggl(
  - F_{\rh\thetah}^2 - D^i\sigma^j D_i\sigma_j
  - R_{ij} \sigma^i \sigma^j
  - [\sigma_\rh, \sigma_\thetah]^2
  - (\nabla^i A_i)^2
  \biggr)
  \\
  + t^2 \biggl( \frac{t^2 - \|V\|^2}{t^2 + \|V\|^2}
  \bigl(|D_\rh\varphi|^2 + |\sigma_\rh \varphi|^2\bigr)
  + \frac{\|V\|^2}{t^2 + \|V\|^2}
    |\cDb_\rh\varphi|^2
    + |D_\thetah\varphi|^2
    + |\sigma_\thetah \varphi|^2 \biggr)
  \\
  - \frac{1}{4} (\varphib \ST_a \varphi)^2
  + \frac{2\iu}{t^2 + \|V\|^2}
    \Im\bigl(\Vb^\thetah \del_\rh W\bigr)
  + \frac{t^{-2}}{t^2 + \|V\|^2}
    \biggl|\frac{\del W}{\del\varphi}\biggr|^2 \biggr)
  \\
  - \frac{1}{\epsilon} \int_{\del D} W \rmd\theta
  - \Tr\bigl(f_{\GF,0}(\varphi_0,\varphib_0)^2\bigr) \,,
\end{multline}
where $R_{ij}$ is the Ricci curvature of $g$.  The metric used in this
expression is the original one before the rescaling.

The real part of the integrand of the bulk integral is a sum of
squares, while the boundary integral is bounded below by the
assumption on the boundary condition for~$\varphi$.  Looking at the
terms multiplied by positive powers of~$t$, we find that as
$t \to \infty$, the action diverges away from the field configurations
such that
\begin{equation}
  F_{ij}
  = \nabla^i A_i
  = D_i\sigma_j
  = [\sigma_i, \sigma_j]
  = \Tr (R_{ij} \sigma^i \sigma^j)
  = D_i\varphi
  = \sigma_i\varphi
  = 0
  \,.
\end{equation}
The path integral therefore localizes in this limit to the locus of
the field space defined by these equations.%
\footnote{It is crucial here that $D$ is compact.  If $D$ were
  noncompact, $D_r\varphi$, for example, could vanish as $t^{-1}$ in
  the limit $t \to \infty$ but $\varphi$ could still vary by a finite
  amount over $D$.}

A general solution $(A_0,\sigma_0,\varphi_0)$ of the localization
equations can be easily identified.  The obvious solution of
$F_{ij} = 0$ is the vanishing gauge field, and any flat connection on
$D$ is gauge equivalent to it under the Neumann boundary condition
which preserves the gauge symmetry.  This solution also satisfies the
Lorenz gauge condition $\nabla^i A_i = 0$, so we have
\begin{equation}
  A_0 = 0 \,.
\end{equation}
Also, as we have equipped $D$ with a hemisphere metric which has a
positive Ricci curvature, $\Tr(R_{ij} \sigma^i \sigma^j) = 0$ implies
\begin{equation}
  \sigma_0 = 0
  \,.
\end{equation}
Given the vanishing of the gauge field, $D_i\varphi = 0$ simply means
that $\varphi_0$ is a constant, which by the boundary condition must
belong to the submanifold $\gamma$ of $X$:
\begin{equation}
  \varphi_0 \in \gamma
  \,.
\end{equation}

We can evaluate the path integral by perturbation theory around these
localization configurations.  To facilitate the calculation we write
\begin{equation}
  A = A_0 + t^{-2} A' \,,
  \qquad
  \sigma = \sigma_0 + t^{-2} \sigma' \,,
  \qquad
  \varphi = \varphi_0 + t^{-1} \varphi' \,,
\end{equation}
and rescale the fermions by the usual scale transformation:
\begin{equation}
  (\alpha,\lambda,\zeta) \to  (\alpha, t^{-1} \lambda, t^{-2} \zeta) \,,
  \qquad
  (\etab,\rho,\mub) \to (\etab, t^{-1} \rho, t^{-2} \mub) \,.
\end{equation}
The rescaling suppresses the fermionic terms that contain $V$.

Further, for each fermion $\Psi$, let $\Psi_0$ be the part of $\Psi$
that is a zero mode of the Laplace--de Rham operator
$\Delta_\rmd = (\rmd - \star\rmd\star)^2$ and satisfies the relevant
boundary condition, and write
\begin{equation}
  \Psi
  =
  \begin{cases}
    \Psi_0 + t^{-1/2} \Psi' & (\Psi \notin \{b, c\}) \,; \\ 
    \Psi_0 + t^{-1} \Psi' & (\Psi \in \{b, c\})  \,.
  \end{cases}
\end{equation}
There are no zero modes for $\lambda$ and $\rho$ since there are no
harmonic one-forms on a hemisphere.  A zero mode for $\zeta$ would be
proportional to the volume form but this is killed by the boundary
condition $\zeta_0 = 0$.  Thus we have
\begin{equation}
  \lambda_0 = \zeta_0 = \rho_0 = 0 \,.
\end{equation}
Then, $\eta_0$ and $\star\mu_0$ are constants satisfying the boundary
conditions
\begin{equation}
  \label{eq:BC-zero}
  (0, \etab_0) \in T_{\varphi_0}\gamma \otimes \C \,,
  \qquad
  (0, \star\mub_0) \in N_{\varphi_0}\gamma \otimes \C \,.
\end{equation}
The zero modes $b_0$, $c_0$ of $b$, $c$ are constant and not affected
by the boundary conditions $\del_r b_0 = \del_r c_0 = 0$.

In terms of the field variables introduced above, the action reads, to
the zeroth order in $t$,
\begin{multline}
  \int_D \biggl(\Tr\bigl(
  - A' \star\Delta_\rmd A'
  - \sigma' \star\Delta_\rmd \sigma'
  + \lambda'
    \wedge \star(\rmd - \star\rmd\star)(\alpha' - \zeta')\bigr)
  \\
  +
  \frac12(\varphib' \star\Delta_\rmd\varphi'
  + \Delta_\rmd\varphi' \star\varphib')
  - \rho' \wedge \star(\rmd - \star\rmd\star) (\etab' - \mub')
  - 2\iu b' \star\Delta_\rmd c'\biggr)
  \\
  + \int_{\del D} \rmd\thetah
  \biggl(\frac12(\varphib' \del_{\rh}\varphi'
  + \del_{\rh}\varphib' \varphi')
  - \mub'_{\rh\thetah} \rho'_\thetah
  \biggr)
  + S_0
  \,.
\end{multline}
The last term contains only $\varphi_0$, $\varphib_0$ and the fermion
zero modes:
\begin{equation}
  \label{eq:S0}
  S_0
  =
  - \frac{2\pi}{\epsilon} W(\varphi_0)
  + (f_{\GF,0}^a)^2
  + 2\iu (\ST_b \cdot f_{\GF,0}^a) b_0^a \ct_0^b
  - \frac{1}{4} (\varphib_0 \ST_a \varphi_0)^2
  + \etab_0 \alpha_0\varphi_0
  \,.
\end{equation}
Here $\ct_0 = c_0 + \dotsb$ is the constant part of the fermion $\ct$
defined earlier.  The contributions from the higher order terms vanish
in the limit $t \to \infty$.

Thus, to the order relevant in the limit we are interested in, the
bosonic and fermionic nonzero modes (the primed variables) enter the
action quadratically and can be integrated out exactly.  In general,
the one-loop determinant
$\Delta_{\text{1-loop}}(\varphi_0,\varphib_0)$ produced by this
integration is a function on $\gamma$: even though the quadratic terms
are independent of the point $\varphi_0 \in \gamma$ around which we
are expanding $\varphi$, the boundary conditions do depend on it.

Now that the nonzero modes have been integrated out, we are only left
with the integration over the zero modes.  This step can be expressed
schematically as
\begin{equation}
  \label{eq:loc}
  \int_\gamma \vol_\gamma
  \int \!
  \rmd\alpha_0 \, \rmd\mub_0 \, \rmd\etab_0 \,
  \rmd b_0 \, \rmd\ct_0 \,
  \Delta_{\text{1-loop}}(\varphi_0,\varphib_0)
  e^{-S_0} \,,
\end{equation}
where $\vol_\gamma$ is a volume form of $\gamma$.  The final
expression may be thought of as the path integral for a
zero-dimensional theory, which is what we wanted to obtain.

\subsubsection{Lagrangian branes and complex gauge symmetry}

The first thing to notice about the integral \eqref{eq:loc} is that
$\mub_0$ is absent from the action~\eqref{eq:S0}; as $\zeta$ has no
zero modes, the interaction term $\mub \star\zeta\varphi$ dropped out
in the localization process.  For the integral to be nonvanishing,
then, $\mub$ should have no zero modes either.  In the same way, the
number of zero modes for $\etab$ should equal that of $\alpha$ or the
integral vanishes.

The numbers of zero modes for $\etab$ and $\mub$ depend on the
boundary conditions~\eqref{eq:BC-zero}.  If we wish to have a
nontrivial result, we should choose the submanifold $\gamma$ for the
support of the brane appropriately so that both of the above
requirements are satisfied.

First, suppose that the gauge symmetry is trivial.  Then, the theory
has no vector multiplet, and we want the boundary conditions to kill
$\etab_0$ and $\mub_0$ completely.  This is achieved if we take
$\gamma$ to be a Lagrangian submanifold of the target space $X$, with
respect to the K\"ahler form
\begin{equation}
  \omega_X = \frac{\iu}{2} \rmd\varphi \wedge \rmd\varphib \,.
\end{equation}
An interesting property of a Lagrangian submanifold of a K\"ahler
manifold is that the action of the complex structure $J$ interchanges
the tangent and normal bundles.  It follows that
$(0, -\iu\etab_0) \in N_{\varphi_0}\gamma \otimes \C$ and
$(0, -\iu\star\mub_0) \in T_{\varphi_0}\gamma \otimes \C$, hence
$\etab_0 = \mub_0 = 0$, as desired.

Now suppose that the theory has a nontrivial gauge symmetry.  In this
case, the action contains the potential
\begin{equation}
  (f_{\GF,0}^a)^2 - \frac14 (\varphib_0 \ST_a \varphi_0)^2 \,.
\end{equation}
Actually, we can rescale this potential by an arbitrarily large factor
without affecting the localization argument; we just have to rescale
the $Q_{V+\BRST}$-exact part of the action by that factor.  Hence, for
a nontrivial result, $\gamma$ must intersect with the zero locus of
the potential.

The zero locus is characterized by the equations $f_{\GF,0} = 0$ and
$\iu \varphib_0 \ST_a \varphi_0/2 = 0$.  The former is the gauge
fixing condition, so we can drop it and instead undo the gauge fixing.
This puts us in a situation where we have the K\"ahler manifold $X$,
endowed with a $\SG$-action and the $\SG$-invariant K\"ahler form
$\omega_X$.  The quantities
\begin{equation}
  \mm_a(\varphi_0, \varphib_0)
  = \frac{\iu}{2} \varphib_0 \ST_a \varphi_0
\end{equation}
which we are setting to zero are the moment map
$\mm\colon X \to \Lie(\SG)^*$ for the $\SG$-action evaluated on
$\ST_a$.  By $\mm$ being the moment map, we mean that
$\rmd\mm_a = \iota_{v_a} \omega_X$, where
$v_a = \ST_a \varphi_0 \del_{\varphi_0} - \varphib_0 \STb_a
\del_{\varphib_0}$ is the vector field on $X$ generated by $\ST_a$.

As $\mm$ is $\SG$-equivariant (that is,
$\langle\mm(g\cdot x), \ST_a\rangle = \langle\mm(x), g^{-1} \ST_a
g\rangle$), the level set $\mm^{-1}(0)$ is $\SG$-invariant.  The zero
locus of the potential is homeomorphic to the quotient
\begin{equation}
  \Mod = \mm^{-1}(0)/\SG \,.
\end{equation}
This is the symplectic reduction of $X$ by the $\SG$-action and itself
a symplectic manifold.  The symplectic form of $\Mod$ is naturally
induced from $\omega_X$ since $\omega_X(v_a, v) = v(\mm_a) = 0$ for
any vector field $v$ tangent to $\mm^{-1}(0)$.

The equation $\mm = 0$, like the other equation $f_{\GF,0} = 0$, can
be regarded as a gauge fixing condition, albeit for a complex gauge
symmetry.  The $\SG$-action on $X$ naturally extends to a holomorphic
action of the complexified gauge group~$\SG_\C$, whose Lie algebra
$\Lie(\SG_\C)$ is spanned by $\{\ST_a, \iu \ST_a\}$.  The vector
fields $Jv_a$ generated by $\iu \ST_a$ are normal to $\mm^{-1}(0)$:
for $v \in \Gamma(T\mm^{-1}(0))$, we have
$g_X(J v_a, v) = \omega_X(v_a, v) = 0$.  Hence, the $\SG_\C$-orbit
$\SG_\C \cdot x$ of a point $x \in \mm^{-1}(0)$ intersects the
$\SG$-orbit $\SG \cdot x \subset \mm^{-1}(0)$ orthogonally.  Moreover,
it can be shown that every $\SG_\C$-orbit contains in its closure at
most a single $\SG$-orbit inside $\mm^{-1}(0)$.

A point of $X$ such that the closure of its $\SG_\C$-orbit has a
nonempty intersection with $\mm^{-1}(0)$ is said to be
\emph{semistable}.  The fact just mentioned implies that $\Mod$ is
homeomorphic to the quotient of the set $X^{\text{ss}}$ of semistable
points by the $\SG_\C$-action:
\begin{equation}
  \Mod \simeq X^{\text{ss}}/\SG_\C \,.
\end{equation}
Put differently, imposing the condition $\mm = 0$, roughly speaking,
gauge fixes the noncompact part of the complex gauge symmetry
generated by $\{\iu \ST_a\}$.  Being a quotient by a holomorphic
$\SG_\C$-action, $\Mod$ is complex, hence K\"ahler.

At low energies, the theory effectively becomes one without gauge
symmetry whose target space is the curved K\"ahler manifold $\Mod$.
Then, an argument similar to what we have given for flat target spaces
would show that $\etab_0$ and $\mub_0$ should vanish when pushed
forward by the projection $\pi\colon X^{\text{ss}} \to \Mod$.  Thus, we
take $\gamma$ to be the preimage of a Lagrangian submanifold $\Lag$ of
$\Mod$:
\begin{equation}
  \gamma = \pi^{-1}(\Lag) \,.
\end{equation}
This is indeed a good choice.  The kernel of $\pi_*$ is spanned by the
vectors $(\ST_a \varphi_0, 0)$ and $(0, \varphib_0 \ST_a)$.  These lie
in $T_{\varphi_0}\gamma \otimes \C$, so we still have $\mub_0 = 0$.
However, $\etab_0$ no longer needs to vanish and can be anything of
the form
\begin{equation}
  \etab_0 = \varphib_0 \beta_0 \,,
\end{equation}
with $\beta_0 \in \Lie(\SG)$.  The number of zero modes for $\etab_0$
is therefore $\dim\SG$, just as for $\alpha_0$.

We call a boundary condition of the type described above a
\emph{Lagrangian brane}.  Its support $\gamma$ gives a Lagrangian
submanifold in the symplectic reduction $\Mod$ of $X$.

Putting together what we have found, we conclude that the localized
path integral is given by
\begin{equation}
  \label{eq:LPI}
  \int_\gamma \vol_\gamma
  \int \! \prod_a \bigl(\rmd\alpha_0^a \, \rmd\beta_0^a \,
  \rmd b_0^a \, \rmd\ct_0^a\bigr)
  \Delta_{\text{1-loop}}
  \exp\biggl(\frac{2\pi}{\epsilon} W - S_0'\biggr)
  \,,
\end{equation}
with
\begin{equation}
  S_0'
  =
  (f_{\GF,0}^a)^2
  + 2\iu (\ST_b \cdot f_{\GF,0}^a) b_0^a \ct_0^b
  + \mm_a^2
  + (\varphib_0 \ST_a \ST_b \varphi_0) \beta_0^a \alpha_0^b
  \,.
\end{equation}
The measure for the fermion zero modes is the natural one induced by
the metric on $\Lie(\SG)$.

The superpotential $W$ is a holomorphic function of $\varphi_0$ and
gauge invariant, and as such invariant under $\SG_\C$.  The domain
$\gamma$ of the bosonic integration is also $\SG_\C$-invariant.  The
emergence of complex gauge symmetry is suggestive.  Sure enough, the
above integral may be interpreted as the path integral for a
zero-dimensional gauged sigma model with gauge group $\SG_\C$.  This
is a gauge theory described by a map $\varphi_0$ from a point to
$\gamma$, and its action is given by $-2\pi W/\eps$.  The
$\Omega$-deformation parameter $\eps$ plays the role of the Planck
constant, so the undeformed limit $\eps \to 0$ is the classical limit.

Since the integral \eqref{eq:LPI} is supposed to be a gauge fixed form
of the path integral for this bosonic theory, the fermionic piece
$S_0'$ in the exponent must be a gauge fixing action.  As we explained
already, the complex gauge symmetry can be gauge fixed by the
condition $f_{\GF,0} = \mm = 0$.  Denoting the ghosts for the real and
imaginary parts of $\SG_\C$ by $(b_0,c_0)$ and $(\beta_0,\alpha_0)$,
respectively, we can write the corresponding gauge fixing action as
\begin{equation}
  S_{\GF,0}
  =
  (f_{\GF,0}^a)^2
  + 2\iu (\ST_b \cdot f_{\GF,0}^a) b_0^a \ct_0^b
  + \mm_a^2
  + 2\iu \bigl((\iu \ST_b) \cdot \mm_a\bigr) \beta_0^a \alpha_0^b
  \,.
\end{equation}
Similarity between $S_0'$ and $S_{\GF,0}$ is obvious, but they do not
precisely match.  The last term in $S_{\GF,0}$ is
$-\iu\varphib_0 \{\ST_a, \ST_b\} \varphi_0 \beta_0^a \alpha_0^b$,
so up to a trivial rescaling of the ghosts, it differs from the last
term in $S_0'$ by a quantity which vanishes on $\mm^{-1}(0)$.
However, the effect of this discrepancy, if any, should be offset by
the one-loop determinant, as we can argue as follows.

The idea is to rescale the potential $\mm_a^2$ by a large factor via a
$Q_{V+\BRST}$-exact deformation (say, by rescaling the bilinear form
$\Tr$).  Then, the integral localizes to $\mm^{-1}(0)$ where the
discrepancy disappears.  Now we show that $\Delta_{\text{1-loop}}$ is
constant on $\mm^{-1}(0)$.%
\footnote{In general, the argument given below cannot be applied
  globally, and the one-loop determinant is not a constant but a flat
  section of a line bundle.  We will not address this issue here.}

First, we note that the intersection $\gamma \cap \mm^{-1}(0)$ is a
Lagrangian submanifold of $X$.  This is because
$\gamma \cap \mm^{-1}(0)$ is the union of $\SG$-orbits in
$\mm^{-1}(0)$ that make up the Lagrangian submanifold
$\Lag \subset \Mod$.  Having an isotropic image under the symplectic
reduction, $\gamma \cap \mm^{-1}(0)$ is itself isotropic.
Furthermore, it has dimension $\dim\SG + \dim\Lag = \dim X/2$.

Next, pick $\varphi_0 \in \gamma \cap \mm^{-1}(0)$ and choose an
orthonormal basis $(e_j)$, $j=1$, $\dotsc$, $\dim \SG$, of
$T_{\varphi_0}(\SG \cdot \varphi_0) \subset T_{\varphi_0}(\gamma \cap
\mm^{-1}(0))$.  We can extend it to an orthonormal basis $(e_k)$,
$k = 1$, $\dotsc$, $\dim\SG + \dim\Lag$, of
$T_{\varphi_0}(\gamma \cap \mm^{-1}(0))$.  As we saw earlier, the
vectors $Je_j$ are normal to $\mm^{-1}(0)$.  They are also tangent to
$\gamma$, so $(e_k, Je_j)$ is an orthonormal basis of
$T_{\varphi_0} \gamma$.  On the other hand, $(Je_k)$ is an orthonormal
basis of $N_{\varphi_0}(\gamma \cap \mm^{-1}(0))$.  Then,
$(e_k, Je_k)$ is an orthonormal basis of $T_{\varphi_0} X$, and
$((e_k - \iu Je_k)/\sqrt{2})$ is a unitary basis of
$T_{\varphi_0}^{1,0} X$.  We require the bases constructed here to
vary smoothly over $\gamma \cap \mm^{-1}(0)$.

In terms of this unitary basis, the boundary conditions are described
in a uniform manner, irrespective of the choice of the point
$\varphi_0 \in \gamma \cap \mm^{-1}(0)$.  For example, the condition
$(\iota_V\rho, \etab) \in T_{\varphi_0}\gamma \otimes \C$ says that
$\iota_V\rho^l = \etab^l$ for $l = \dim\SG + 1$, $\dotsc$,
$\dim\SG + \dim\Lag$.  Also, the quadratic terms in the nonzero modes,
from which the one-loop determinant is calculated, has a uniform
expression in a unitary basis.  Therefore, the one-loop determinant is
independent of $\varphi_0$.

\subsection{Localization on a plane}

We have just seen that when the spacetime is a disk of finite radius
and the boundary condition is given by a Lagrangian brane, the
quasi-topological sector of the $\Omega$-deformed B-twisted gauge
theory is equivalent to a zero-dimensional gauged sigma model with
complex gauge group whose target space is the support of the brane.

The case when the spacetime is a plane is similar but qualitatively
different.  It is similar in that the $\Omega$-deformed B-twisted
gauge theory in this case is still equivalent to a zero-dimensional
gauged sigma model with the same complex gauge group.  The target
space, however, is different due to the noncompactness of the
spacetime; it is no longer given by the brane itself.  Rather, it
consists of gradient flows generated by the
superpotential~\cite{Nekrasov:2018pqq}, as we now show.

\subsubsection{Path integral on a semi-infinite cylinder}

Let us deform the spacetime $D = \R^2$ into the shape of a cigar,
consisting of a semi-infinite cylinder capped with a hemisphere.  We
split the path integral on the cigar into two parts.  One is performed
on the hemisphere, and we already understand it well.  The other is on
the cylinder.  Our strategy is to impose some boundary condition at
infinity and see what state the latter path integral yields at the
other end of the cylinder.  Subsequently we feed this state into the
former path integral to deduce the result of the path integral on the
whole cigar.

Let $D_0$ and $D_\infty$ be the hemisphere and cylinder parts of $D$,
respectively.  As usual, we can deform the action by $Q_V$-exact terms
since this does not change the $Q_V$-cohomology class of the state at
the end.  Using this freedom we choose the metric to be such that
\begin{equation}
  g_{\theta\theta} = \frac{1}{|\eps|^2}
\end{equation}
on $D_\infty$ so that we have $\|V\| = 1$ on the cylinder.  Moreover,
we make $g_{rr}(r)$ decay sufficiently fast so that $D$ has a finite
area.

The action on $D$ is the sum of two $Q_V$-invariant integrals,
$S_{D_0}$ on $D_0$ and $S_{D_\infty}$ on~$D_\infty$.  It turns out
that $S_{D_\infty}$ is $Q_V$-exact: the part of the action that
depends on $W$ can be written as
\begin{equation}
  \delta_V \int_{D_\infty} \! \frac{\rmd\theta}{V^\theta}
  \rho \frac{\del W}{\del\varphi} \,.
\end{equation}
For the above choice of metric, the bosonic part of $S_{D_\infty}$ is
given by
\begin{multline}
  \int_{D_\infty} \! \rmd\rh \, \rmd\thetah \biggl(
  \Tr\biggl(
  -\frac12 F^{ij} F_{ij} - D^i\sigma^j D_i\sigma_j
  - \frac12 [\sigma^i, \sigma^j] [\sigma_i, \sigma_j]
  \biggr)
  \\
  +
  \frac12
  \biggl|\cDb_\rh\varphi
  - \frac{\eps}{|\eps|} \frac{\del\Wb}{\del\varphib}\biggr|^2
  + |D_\thetah\varphi|^2
  + |\sigma_\thetah \varphi|^2
  - \frac{1}{4} (\varphib \ST_a \varphi)^2
  \biggr)
  \,.
\end{multline}
Note that there are no boundary terms in this expression.

Let us ``squash'' the cigar in the longitudinal direction in such a
way that $g_{rr}$ is rescaled on $D_\infty$ as
\begin{equation}
  g_{rr} \to t^{-2} g_{rr} \,.
\end{equation}
At the same time, we also rescale $W$ as
\begin{equation}
  W \to tW \,.
\end{equation}
Both of these deformations are $Q_V$-exact.  In the limit
$t \to \infty$, the path integral on $D_\infty$ localizes to the locus
where
\begin{equation}
  F
  = D_r\sigma
  = D_\theta\sigma_r
  = [\sigma_r,\sigma_\theta]
  = \cDb_\rh\varphi
    - \frac{\eps}{|\eps|} \frac{\del\Wb}{\del\varphib}
  = 0 \,.
\end{equation}

For the path integral to be nonvanishing, the boundary condition at
$r = \infty$ must be compatible with the localization equations.
Then, for the vector multiplet, we should take the Neumann condition
as we did in the hemisphere case.  We can choose the gauge $A_r = 0$
on $D_\infty$, and in this gauge and with this boundary condition on
the vector multiplet, the above equations reduce to
\begin{equation}
  \CA_r
  = \del_r\CA_\theta
  = \del_\rh\varphi
    - \frac{\eps}{|\eps|} \frac{\del\Wb}{\del\varphib}
  = 0 \,.
\end{equation}

If $D_\infty$ were compact, with $r$ varying over a finite interval,
we would be able to shrink it to a very short cylinder so that the
localization equation for~$\varphi$ would imply that $\varphi$ does
not vary in the longitudinal direction.  In the case at hand, however,
$r$ is not bounded above and takes values in $[r_0,\infty)$ on
$D_\infty$ for some $r_0 > 0$.  If we introduce a new coordinate
\begin{equation}
  s = |\eps| \int_{r_0}^r \! \sqrt{g_{rr}} \, \rmd r \,,
\end{equation}
which ranges from $0$ to $\infty$, then in terms of this coordinate
the equation becomes
\begin{equation}
  \label{eq:grad-flow}
  \del_s\varphi - \frac{1}{\epsb} \frac{\del\Wb}{\del\varphib}
  = 0 \,.
\end{equation}
Therefore, $\varphi$ localizes to a solution of the gradient flow
equation generated by the function $\Re(W/\eps)$ on $X$ with respect
to the K\"ahler metric \eqref{eq:gX}.

We have found that the chiral multiplet scalar should approach a
gradient flow as $t \to \infty$.  As we will see, for the convergence
of path integral the flow must terminate at a fixed point.  So we pick
a submanifold $\gamma_\infty$ of the critical locus $\crit(W)$ of $W$
and demand
\begin{equation}
  \varphi \in \gamma_\infty
\end{equation}
at $r = \infty$.  The boundary condition at $r = \infty$ for the
chiral multiplet is the brane-type condition characterized by
$\gamma_\infty$.

Now we turn our attention to the state produced by the path integral
at the other boundary of $D_\infty$.  The wavefunction of this state
is sharply peaked on the localization locus.  In the limit
$t \to \infty$, the effect of including this wavefunction in the path
integral on~$D_0$ is to impose a boundary condition that forces the
bosonic fields to lie on the localization locus.  For the vector
multiplet this is the Neumann condition.

For the chiral multiplet, the boundary condition is a brane-type
condition whose support $\gamma_0$ consists of all points $p \in X$
such that there exists a gradient flow
$\varphi_s\colon \R_{\geq 0} \to X$ with $\varphi_0 = p$ and
$\varphi_\infty \in \gamma_\infty$, namely the union of all gradient
flow trajectories terminating on $\gamma_\infty$.  When
$\gamma_\infty$ is a nondegenerate critical point, $\gamma_0$ is known
as a Lefschetz thimble.

\subsubsection{Gradient flow trajectories as Lagrangian branes}

We have reduced the path integral on a plane to the path integral on
the hemisphere $D_0$ with a particular brane boundary condition.  For
the path integral on $D_0$ to be sensible, $\Re(W/\eps)$ had better be
bounded above on the brane support $\gamma_0$ so that the boundary
term does not diverge.  Furthermore, for the path integral to be
nonvanishing, $\gamma_0$ should be a Lagrangian brane, that is, there
should be a Lagrangian submanifold $\Lag_0$ of~$\Mod$ such that
$\gamma_0 = \pi^{-1}(\Lag_0)$.

These requirements are satisfied if we choose the brane support
$\gamma_\infty$ at $r = \infty$ appropriately~\cite{Hori:2000ck,
  Witten:2010zr}.  Since $W$ is invariant under $\SG_\C$ and so is the
gradient flow equation~\eqref{eq:grad-flow}, gradient flows in $X$
define gradient flows in $\Mod$ which are generated by $\Re(W/\eps)$ as
a function on~$\Mod$.  We pick a compact Lagrangian submanifold
$\Lag_\infty$ of $\crit(W) \subset \Mod$ and set
\begin{equation}
  \gamma_\infty = \pi^{-1}(\Lag_\infty) \,.
\end{equation}
Then we have $\gamma_0 = \pi^{-1}(\Lag_0)$, with $\Lag_0$ being the
union of all gradient flow trajectories in~$\Mod$ that terminate on
$\Lag_\infty$.

First of all, $\Re(W/\eps)$ is bounded above on $\gamma_0$ because it
is nondecreasing along a gradient flow:
\begin{equation}
  \del_s \Re\biggl(\frac{W}{\eps}\biggr)
  = \frac{1}{|\eps|^2}
    \frac{\del W}{\del\varphi} \frac{\del\Wb}{\del\varphib}
  \geq 0\,.
\end{equation}
As such, it attains the maximum value along each gradient flow when it
reaches $\gamma_\infty$, but this value is locally constant on
$\crit(W)$.

We can show that $\Lag_0$ is a middle-dimensional submanifold of
$\Mod$ as follows.  By the holomorphic Morse--Bott lemma, in a
neighborhood of any point $p \in \crit(W) \subset \Mod$ we can find
local holomorphic coordinates $(z^i)_{i=1}^{\dim_\C \Mod}$ such that
\begin{equation}
  W = W(p) + \sum_{i=1}^n (z^i)^2
\end{equation}
for some $n$.  The Hessian of $\Re(W/\eps)$ at $p$ has $n$ positive,
$n$ negative and $(\dim_\R\Mod - 2n)$ zero eigenvalues.  Hence, the
union of gradient flows trajectories terminating at $p$ is a
submanifold of dimension $n$.  Since~$\Lag_0$ is the union of such
submanifolds as $p$ varies over the $(\dim_\R\Mod/2 - n)$-dimensional
submanifold~$\Lag_\infty$, it has dimension $\dim_\R\Mod/2$.

To show that $\Lag_0$ is isotropic, we use the fact that gradient flows
are Hamiltonian flows generated by $\Im(W/\eps)$.  Indeed, if
$v = \del_s\varphi \del_\varphi + \del_s\varphib \del_\varphib$ is a
vector field generating a gradient flow, we have
\begin{equation}
  \label{eq:omega-W}
  \iota_v \omega_X
  =
  \frac{\iu}{2}
  \biggl(\frac{1}{\epsb}
  \frac{\del\Wb}{\del\varphib} \rmd\varphib
  - \rmd\varphi \frac{1}{\eps}
    \frac{\del W}{\del\varphi}\biggr)
  =
  \rmd\Im\biggl(\frac{W}{\eps}\biggr)
  \,.
\end{equation}
It follows that $\omega_X$ is preserved along the flows:
$\CL_v \omega_X = (\rmd \iota_v + \iota_v \rmd) \omega_X = 0$.  On the
other hand, any differential form on $\Lag_0$ is mapped to a
differential form on $\Lag_\infty$ upon pullback to $\Lag_\infty$ by
gradient flows.  Since $\omega_X$ vanishes on $\Lag_\infty$ by
construction, the invariance of $\omega_X$ under gradient flows
implies that $\omega_X$ vanishes when restricted to $\Lag_0$.

Combining what we have just found and the localization of the path
integral on the hemisphere, we arrive at the main result of this
section: The quasi-topological sector of a B-twisted gauge theory with
gauge group $\SG$, subjected to an $\Omega$-deformation on $\R^2$, is
equivalent to a zero-dimensional gauged sigma model with gauge
symmetry $\SG_\C$ whose action is $-2\pi W/\eps$ and target space is a
Lagrangian brane $\gamma_0 = \pi^{-1}(\Lag_0)$.  The Lagrangian
submanifold $\Lag_0$ of the K\"ahler quotient $\Mod$ consists of the
gradient flow trajectories generated by $\Re(W/\eps)$, terminating on
a chosen compact Lagrangian submanifold $\Lag_\infty$ of
$\crit(W) \subset \Mod$.

\section{Four-dimensional Chern--Simons theory from six dimensions}
\label{sec:4dCS-from-6d}

Let us apply the result obtained in the previous section to the
six-dimensional topological--holomorphic theory on
$D \times \Sigma \times C$, viewing it as a B-twisted gauge theory
on~$D$.

The chiral multiplet scalars of the theory form a partial
$G_\C$-connection
\begin{equation}
  \CA = \CA_m \rmd x^m + A_\zb \rmd\zb
\end{equation}
on $\Sigma \times C$.  The target space $X$ is therefore the space of
such connections, with $(\CA_m, A_\zb)$ providing holomorphic
coordinates.  The gauge group $\SG$ is the group of maps from
$\Sigma \times C$ to $G$, which is the group of gauge transformations
that are constant on $D$.  Looking at the chiral multiplet action, we
see that $X$ is endowed with the $\SG$-invariant K\"ahler metric
\begin{equation}
  g_X
  =
  -\frac{1}{2e^2} \int_{\Sigma \times C}
  \! \sqrt{g_\Sigma} \, \rmd^2x \, \rmd^2 z
  \Tr\bigl(
  \delta\CA^m \otimes \delta\CAb_m
  + \delta\CAb^m \otimes \delta\CA_m
  + \delta A_\zb \otimes \delta A_z
  + \delta A_z \otimes \delta A_\zb
  \bigr)
  \,,
\end{equation}
where $\sqrt{g_\Sigma} \, \rmd^2x$ is the volume form of $\Sigma$.
The superpotential is given by the integral \eqref{eq:W-6d}.  This is
not a fully gauge invariant expression; we will address this point
later.

Now we take $D = \R^2$ and turn on an $\Omega$-deformation using the
rotation symmetry.  Then, by localization the path integral reduces to
the integral
\begin{equation}
  \label{eq:PI-4dCS}
  \int \cD\CA
  \exp\biggl(\frac{\iu}{\pi\hbar}
  \int_{\Sigma \times C} \! \rmd z \wedge \ChS(\CA)\biggr)
  \,,
\end{equation}
with
\begin{equation}
  \hbar = -\frac{\epsilon e^2}{2\pi^2} \,.
\end{equation}
This is precisely the path integral for four-dimensional Chern--Simons
theory with gauge group $G_\C~$\cite{Costello:2013zra,
  Costello:2013sla, Costello:2017dso, Costello:2018gyb}.

Thus we conclude: the $\Omega$-deformed topological--holomorphic
theory on $\R^2 \times \Sigma \times C$ is equivalent to
four-dimensional Chern--Simons theory on $\Sigma \times C$.

We still have to identify the integration domain for the localized
path integral~\eqref{eq:PI-4dCS}.  With application to integrable
lattice models in mind, let us do so in the case when $\Sigma$ is a
flat torus $T^2$ and $C$ is an elliptic curve $E = \C/(\Z + \tau\Z)$.
Moreover, we take $G$ to be either $\U(N)$ or a connected and simply
connected compact Lie group.  We parametrize~$\Sigma$ with periodic
Cartesian coordinates $(x,y)$, and let $\CC_x$ and~$\CC_y$ denote the
homology cycles represented by loops in the $x$- and $y$-directions.
The one-cycles along $E$ are denoted by $\CC_a$ and $\CC_b$, with the
former corresponding to a path from $z = 0$ to~$1$ and the latter a
path from $z = 0$ to $\tau$.

The main task is to understand the critical locus of $W$ in the
K\"ahler quotient $\Mod$ of~$X$.  The K\"ahler form of $X$ is
\begin{equation}
  \omega_X
  =
  -\frac{\iu}{2 e^2} \int_{\Sigma \times C}
  \! \sqrt{g_\Sigma} \, \rmd^2x \, \rmd^2 z
  \Tr\bigl(
  \delta\CA^m \wedge \delta\CAb_m
  + \delta A_\zb \wedge \delta A_z
  \bigr)
  \,.
\end{equation}
A simple computation shows that the moment map $\upmu$ for the
$\SG$-action is given by the formula
\begin{equation}
  \langle\mm, \veps\rangle
  =
  -\frac{1}{e^2}
  \int_{\Sigma \times C}
  \! \sqrt{g_\Sigma} \, \rmd^2x \, \rmd^2 z
  \Tr\biggl(\veps
  \biggl(D^m\phi_m - \frac{\iu}{2} F_{z\zb}\biggr)\biggr) \,,
\end{equation}
where $\veps \in \Lie(\SG)$ is the parameter of gauge transformation.
Hence, the zero locus of $\mm$ is described by the condition
\begin{equation}
  \label{eq:stab}
  D^m\phi_m - \frac{\iu}{2} F_{z\zb} = 0 \,,
\end{equation}
and $\Mod$ is the quotient by the $\SG$-action of the space of partial
$G_\C$-connections satisfying this condition.  The critical locus of
$W$ is where the equations of motion hold:
\begin{equation}
  \label{eq:EOM}
  \CF_{mn} = \CF_{m\zb} = 0 \,.
\end{equation}

Imposing the conditions \eqref{eq:stab} and \eqref{eq:EOM} is the same
as requiring the vanishing of the integral
\begin{equation}
  \label{eq:int-locus}
  -\int_{\Sigma \times C} \! \sqrt{g_\Sigma} \, \rmd^2x \, \rmd^2 z
  \Tr\biggl(\frac12 \CFb^{mn} \CF_{mn}
  + \CFb^m{}_z \CF_{m\zb}
  + \biggl(D^m\phi_m - \frac{\iu}{2} F_{z\zb}\biggr)^2\biggr)
  \,.
\end{equation}
By integration by parts we can rewrite this integral as
\begin{multline}
  \label{eq:int-locus-bulk}
  -\int_{\Sigma \times C} \! \sqrt{g_\Sigma} \, \rmd^2x \, \rmd^2 z
  \Tr\biggl(
  \frac12 F^{mn} F_{mn}
  +  F^m{}_z F_{m\zb}
  - \frac14 F_{z\zb}^2
  \\
  + D^m\phi^n D_m\phi_n
  + D_z\phi^m D_\zb\phi_m
  + \frac12 [\phi^m, \phi^n][\phi_m, \phi_n]
  \biggr)
  \,.
\end{multline}
The integrand is again a sum of nonnegative terms and must vanish
separately.  On $\mm^{-1}(0) \cap \crit(W)$, therefore, $A$ is a flat
connection and $\phi$ satisfies
\begin{equation}
  \label{eq:Dphi=0}
  D_m \phi_n = D_z\phi_m = [\phi_m,\phi_n]
  = 0 \,.
\end{equation}

A flat connection $A$ on a principal $G$-bundle $P \to T^2 \times E$
is characterized, up to gauge transformation, by the holonomies around
the one-cycles in the base,
\begin{equation}
  P\exp\biggl(\int_{\CC_\bullet} A\biggr) \,,
  \quad
  \text{$\bullet = x$, $y$, $a$, $b$} \,.
\end{equation}
Since the fundamental group of $T^2 \times E$ is abelian, the
holonomies form a commuting quadruple of elements of $G$.

For $G = \U(N)$, the elements of the quadruple can be diagonalized
simultaneously.  Things are a little more complicated if $G$ is not
unitary.  In this case, these elements can be pairwise conjugated to
lie in a given maximal torus $\T$ of $G$, but in general it is not
possible to put all of them into $\T$.  Still, the moduli space of
commuting quadruples has a component in which all four elements belong
to the same maximal torus, and if the holonomies are generic they fall
in this component.  We will restrict our attention to this generic
situation.

The equations $D_m\phi_n = D_z\phi_m = 0$ imply that $\phi_m$ are left
invariant by the holonomies.  Under the genericity assumption, this
condition requires $\phi_m$ to be valued in the Lie algebra~$\tf$
of~$\T$, and the remaining equation $[\phi_m,\phi_n] = 0$ is
satisfied.  If we choose a gauge such that $A$ is represented by a
constant $\tf$-valued one-form, the same equations imply that $\phi$
is also a constant $\tf$-valued one-form.

The constant $\tf_\C$-valued one-form $\CA$ define local holomorphic
coordinates on $\crit(W) \subset \Mod$.  A better set of local
holomorphic coordinates is given by
\begin{equation}
  \uptau_x
  = \frac{1}{2\pi\iu} \int_{\CC_x} \CA
  \,,
  \qquad
  \uptau_y
  = \frac{1}{2\pi\iu} \int_{\CC_y} \CA
\end{equation}
and
\begin{equation}
  \label{eq:uplambda}
  \uplambda
  = -\frac{1}{2\pi\iu} \biggl(\tau \int_{\CC_a} A - \int_{\CC_b} A\biggr)
  = \frac{1}{4\pi} \int_E A_\zb \, \rmd^2z
 \,.
\end{equation}
These quantities are invariant under topologically trivial
$\T_\C$-valued gauge transformations.

Globally, topologically nontrivial gauge transformations induce the
identifications
\begin{equation}
  \uptau_x \sim \uptau_x + u \,,
  \qquad
  \uptau_y \sim \uptau_y + u \,,
  \qquad
  \uplambda \sim \uplambda + u \sim \uplambda + \tau u \,,
\end{equation}
with $2\pi\iu u \in \tf$ being an element of the kernel of the
exponential map $\exp\colon \gf \to G$.
Lastly, we must identify values related by the action of the Weyl
group $W(G)$ of $G$, which is part of the gauge symmetry.  Altogether,
the relevant part of $\crit(W) \subset \Mod$ is isomorphic to
\begin{equation}
  \label{eq:moduli}
  \bigl((\C^\times)^r \times (\C^\times)^r \times E^r\bigr)/W(G) \,,
\end{equation}
where $r$ is the rank of $G$.

As discussed in the previous section, the integration domain for the
localized path integral is the union of the gradient flow trajectories
terminating on a chosen compact Lagrangian submanifold $\Lag_\infty$
of $\crit(W) \subset \Mod$.  An obvious Lagrangian submanifold of the
moduli space~\eqref{eq:moduli} is the product of closed curves in each
factor.  For example, for the torus part
$(\C^\times)^r \times (\C^\times)^r$, we can set $\Im\uptau_x$ and
$\Im\uptau_y$ to constant elements of $\tf$.

It will prove useful to interpret the last factor of the moduli
space~\eqref{eq:moduli} in the language of holomorphic vector bundles.
At each point in $\Sigma$, the gauge bundle $\PB$ restricts to a
principal $G$-bundle over $E$.  Pick a unitary representation of $G$
and consider the vector bundle associated to this representation.  The
$G$-action extends to a $G_\C$-action, making it a $G_\C$-bundle.
Since the integrability condition $\delb_A^2 = 0$ is trivially
satisfied for a dimensional reason, given a connection $A$ there
always exists a holomorphic structure on this bundle in which the
Dolbeault operator $\delb$ coincides with $\delb_A$.  On
$\mm^{-1}(0) \cap \crit(W)$ where $A$ is flat, the bundle has degree
$0$ (that is, topologically trivial) and is semistable.  Conversely, a
semistable holomorphic vector bundle of degree $0$ arises in this way
from a flat unitary connection, according to the Narasimhan--Seshadri
theorem~\cite{MR0184252}.

The relation between the coordinates on $\Mod$ and the holomorphic
structure is as follows.  The associated vector bundle in question is
a quotient of a flat bundle over the universal cover $\C$ of $E$. Let
us take a gauge in which $A_\zb$ is constant and valued in $\tf_\C$.
Then, choosing a basis $(s_i(0))_{i=1}^n$ consisting of eigenvectors of
$A_\zb$ in the fiber at $z = 0$, we can define holomorphic sections
\begin{equation}
  \label{eq:hol-sect}
  s_i(z)
  =
  \exp\bigl((z - \zb) A_\zb\bigr) s_i(0)
  =
  \exp\biggl(-2\pi\iu \uplambda_i \frac{\Im z}{\Im\tau}\biggr) s_i(0)
  \,,
\end{equation}
where $\uplambda_i$ is the eigenvalue of $\uplambda$ associated with
$s_i(0)$.  These sections provide a basis for a local holomorphic
frame.  They obey the monodromy relations
\begin{equation}
  \label{eq:monodromy}
  s_i(z + 1) = s_i(z) \,, \qquad
  s_i(z + \tau) = \exp(-2\pi\iu\uplambda_i) s_i(z) \,,
\end{equation}
which determine the corresponding holomorphic transition functions.
Thus, the parameter~$\uplambda$ of the flat connection specifies
the holomorphic structure via monodromy of holomorphic sections.

Now that we have understood the integration domain, let us come back
to the more fundamental question: how do we make sense of the
superpotential in the first place when it lacks gauge invariance?
Fortunately, no problem arises if $G$ is connected, which we assume.

The point is that given a homotopy $\CAt\colon [0,1] \to X$ between
two connections $\CA_0$ and~$\CA_1$, we can define the difference of
$W$ evaluated for $\CA_0$ and $\CA_1$ in a gauge invariant manner:
\begin{equation}
  W(\CA_1) - W(\CA_0)
  =
  \frac{\iu}{e^2} \int_{[0,1] \times \Sigma \times C}
  \! \rmd z \wedge \Tr\bigl(\CFt \wedge \CFt\bigr) \,.
\end{equation}
Here $\CFt$ is the curvature of $\CAt$, regarded as a connection on
$[0,1] \times \Sigma \times C$.  By assumption, for any two gauge
equivalent connections $\CA_0$ and $\CA_1$ satisfying the equations of
motion~\eqref{eq:EOM}, there is a path $\gt\colon [0,1] \to G$ such
that $\gt(0)$ is the identity element and $\CA_1$ is the gauge
transform of $\CA_0$ by $\gt(1)$.  For the homotopy~$\CAt$ generated
by the action of $\gt$ on $\CA_0$, the right-hand side of the above
formula vanishes since the components of~$\rmd z \wedge \CFt$ along
$\Sigma \times C$ are zero throughout the interval~$[0,1]$.  Hence,
$W$ can be made gauge invariant for connections in $\crit(W)$.  Also
by the same formula, the value of $W(\CA_1)$ for a connection $\CA_1$
equipped with a homotopy to a connection $\CA_0$ in $\crit(W)$ is
determined from $W(\CA_0)$.  We only have to deal with such
connections because $\CA$ must approach a point on $\gamma_\infty$
as~$r \to \infty$, and $\gamma_\infty$ is a submanifold of $\crit(W)$.

If we choose $\gamma_\infty$ inside a connected component of
$\crit(W)$, the definition of $W$ on the relevant part of $X$ boils
down to a choice of a single constant as the value of $W$ in that
component.  This constant may be thought of as an overall
normalization factor for the path integral.

\section{Integrable lattice models from four-dimensional Chern--Simons
  theory}
\label{sec:ILM-4dCS}

Now that we have understood the six-dimensional origin of
four-dimensional Chern--Simons theory, let us focus on this theory
itself and explore its physical properties.  In this section we
explain how integrable lattice models and related mathematical
structures arise from nonlocal observables of the theory.  Throughout
this section we take $C = E$, except for the argument in
section~\ref{sec:LO-ILM} which works for all choices $C = \C$,
$\C^\times$ and $E$.  Also, we take $\Sigma = T^2$ whenever the
topology of $\Sigma$ matters.

\subsection{Line operators and integrable lattice models}
\label{sec:LO-ILM}

As in the ordinary Chern--Simons theory, the basic observables in
four-dimensional Chern--Simons theory are Wilson lines.  Recall that
in the six-dimensional topological--holomorphic theory there are
$Q$-invariant Wilson lines constructed from the partial
$G_\C$-connection $\CA$, which lie in the four-manifold
$M = D \times \Sigma$ and are supported at points on $C$.  For
$D = \R^2$ or a disk, these Wilson lines remain as good observables
even after the $\Omega$-deformation is turned on (that is, they are
$Q_V$-invariant) if they are supported on closed curves in $\Sigma$
and placed at the origin of $D$.  They descend to Wilson lines in
four-dimensional Chern--Simons theory.

\begin{figure}
  \centering
  \subfloat[\label{fig:lattice}]{
    \begin{tikzpicture}[scale=1.2]
      \fill[ws] (0,0) rectangle (3, 2);

      \begin{scope}[shift={(0.5,0)}]
        \draw[r->] (0,0) -- (0,2);
        \draw[r->] (1,0) -- (1,2);
        \draw[r->] (2,0) -- (2,2);
      \end{scope}

      \begin{scope}[shift={(0,0.5)}]
        \draw[r->] (0,0) -- (3,0);
        \draw[r->] (0,1) -- (3,1);
      \end{scope}

      \draw[frame] (0,0) rectangle (3,2);
    \end{tikzpicture}
  }
  \qquad
  \subfloat[\label{fig:pieces}]{
    \begin{tikzpicture}[scale=1.2]
      \fill[ws] (0,0) rectangle (3,2);

      \begin{scope}[shift={(0.75,0)}]
        \draw[r->] (0,0) -- (0,2);
        \draw[r->] (1,0) -- (1,2);
        \draw[r->] (2,0) -- (2,2);
      \end{scope}
      
      \begin{scope}[shift={(0,0.75)}]
        \draw[r->] (0,0) -- (3,0);
        \draw[r->] (0,1) -- (3,1);
      \end{scope}

      \begin{scope}[shift={(0,0.25)}]
        \draw[boundary, double] (0,0) -- (3,0);
        \draw[boundary, double] (0,1) -- (3,1);
      \end{scope}
      
      \begin{scope}[shift={(0.25,0)}]
        \draw[boundary, double] (0,0) -- (0,2);
        \draw[boundary, double] (1,0) -- (1,2);
        \draw[boundary, double] (2,0) -- (2,2);
      \end{scope}

      \draw[frame] (0,0) rectangle (3,2);

      \begin{scope}[shift={(0,0.25)}]
        \draw[thick, white] (-0.01,0) -- (3.01,0);
        \draw[thick, white] (-0.01,1) -- (3.01,1);
      \end{scope}
      
      \begin{scope}[shift={(0.25,0)}]
        \draw[thick, white] (0,-0.01) -- (0,2.01);
        \draw[thick, white] (1,-0.01) -- (1,2.01);
        \draw[thick, white] (2,-0.01) -- (2,2.01);
      \end{scope}
    \end{tikzpicture}
  }
  \qquad
  \subfloat[\label{fig:R-matrix}]{
    \begin{tikzpicture}[scale=0.75]
      \fill[ws] (0,0) rectangle (2, 2);

      \draw[r->] (1,0) -- (1,2);
      \draw[r->] (0,1) -- (2,1);
      
      \draw[boundary] (0,0) rectangle (2,2);
      
      \node[shift={(-0.15,0.15)}] at (0,2) {$a$};
      \node[shift={(-0.15,-0.15)}] at (0,0) {$b$};
      \node[shift={(0.15,-0.15)}] at (2,0) {$c$};
      \node[shift={(0.15,0.15)}] at (2,2) {$d$};

      \node[shift={(-0.25,0)}] at (0,1) {$\alpha$};
      \node[shift={(0,-0.25)}] at (1,0) {$\beta$};
      \node[shift={(0.25,0)}] at (2,1) {$$};
      \node[shift={(0,0.25)}] at (1,2) {$$};
    \end{tikzpicture}
  }
  \caption{(a) A lattice formed by line operators on $T^2$. (b)
    Decomposition of the lattice into square pieces.  (c) A single
    square piece with boundary conditions specified on the corners.}
  \label{fig:}
\end{figure}
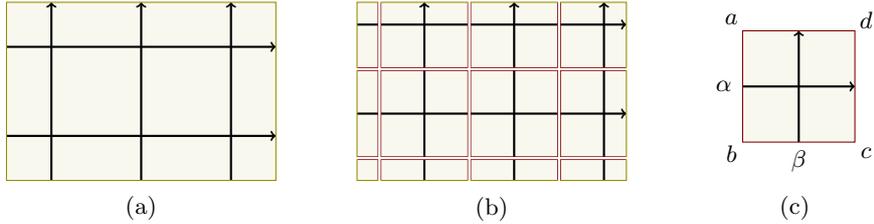

In the present setup, these Wilson lines wind around various
one-cycles of $\Sigma = T^2$.  More generally, suppose that there are
$m + n$ line operators $L_\alpha$, $\alpha = 1$, $\dotsc$, $m + n$,
the first $m$ of which are supported on the horizontal lines located
at $(y,z) = (y_\alpha, z_\alpha)$, while the last $n$ are supported on
the vertical lines at $(x,z) = (x_\alpha, z_\alpha)$.  These line
operators form an $m \times n$ square lattice on $T^2$.  The case with
$(m,n) = (2,3)$ is illustrated in Figure~\ref{fig:lattice}.

We are interested in the correlation function
\begin{equation}
  \biggvev{\prod_{\alpha=1}^{m+n} L_\alpha} \,.
\end{equation}
In order to compute this quantity, we break $T^2$ up into square
pieces, each containing precisely two intersecting segments of line
operators~\cite{Yagi:2015lha, Yagi:2016oum}.  See
Figure~\ref{fig:pieces} for an example of this decomposition.

Take a single such piece, containing line operators $L_\alpha$ and
$L_\beta$.  On the corners we pick boundary conditions,%
\footnote{To handle surfaces with corners in the framework of
  open-closed topological field theory, one may imagine cutting out
  the corners and replacing them with branes on which open strings
  have ends.  For each corner we are choosing a boundary condition
  that specifies the type of the brane sitting
  there~\cite{Yagi:2016oum}.}
which we label $a$, $b$, $c$ and $d$, as in Figure~\ref{fig:R-matrix}.
This determines Hilbert spaces assigned to the sides of the square.
Let $\V_{ab,\alpha}$ be the Hilbert space of states on an interval
with boundary conditions $a$ on the left end and $b$ on the right end,
intersected by $L_\alpha$ in the middle.  The path integral on the
square piece produces a linear map
\begin{equation}
  \label{eq:R-matrix}
  \RM_{\alpha\beta}
  \biggl(
  \begin{array}{cc}
    a & d \\
    b & c
  \end{array}
  \biggr)
  \colon
  \V_{ab,\alpha} \otimes \V_{bc,\beta}
  \to
   \V_{ad,\beta} \otimes \V_{dc,\alpha}
  \,.
\end{equation}
We call this operator an \emph{R-matrix}.

After computing the path integral on each square piece, we can glue
the pieces back together by composing the resulting R-matrices in an
appropriate way.  Finally, we sum over the boundary conditions
specified on the corners so that the fields are allowed to have all
possible behaviors at those points.%
\footnote{Here we are assuming that the vacuum state of the Hilbert
  space for a closed string (which is mapped to the identity operator
  under the state--operator correspondence) can be expanded in boundary
  states describing branes~\cite{Yagi:2016oum}.}

This procedure for computing the correlation function of line
operators may be thought of as defining the partition function of a
lattice model in statistical mechanics.  In this lattice model, state
variables (or ``spins'') are placed on the faces and edges of the
lattice of line operators.  The boundary conditions on the corners are
identified with the face variables, whereas basis vectors of the
Hilbert spaces on the sides of the squares are the edge variables.
The matrix elements of an R-matrix encode the local Boltzmann weights
for various configurations of states around a vertex of the lattice.
The partition function of the lattice model is the product of the
local Boltzmann weights, summed over all allowed state configurations.
This is precisely what we have to calculate to reconstruct the path
integral on the whole torus from those on the square pieces.

The crucial property that makes this interpretation useful is that the
theory is topological on $T^2$.%
\footnote{In reality, as we will see later, the topological invariance
  is broken due to a framing anomaly~\cite{Costello:2017dso,
    Costello:2018gyb}.  For the purpose of this discussion it suffices
  to consider the situation where the lines making up the lattice are
  straight and therefore the framing anomaly plays no role.}
This property ensures that the state space and local Boltzmann weights
of the lattice model are independent of the locations of the lines or
how we cut $T^2$ into pieces; only topology matters.

So far we have only used the structure of a two-dimensional
topological field theory to establish that a collection of line
operators gives rise to a lattice model.  Actually, our theory has
more than just this structure.  It is really four-dimensional, and
this fact has a profound implication~\cite{Costello:2013zra,
  Costello:2013sla}.

The two-dimensional topological invariance guarantees that the
partition function of the lattice model remains unchanged when one of
the lines, say a horizontal one, is moved up and down.  This is true
as long as it does not pass another horizontal line, at which point
the topology of the lattice changes.  In general, one excepts the
partition function of a quantum field theory to behave badly at a
singular configuration where two line operators sit on top of each
other.  In the present case, however, the line operators are
generically located at different points on $C$, and the partition
function should be perfectly smooth even when two lines coincide on
$T^2$ since they are separated on $C$.  The topological invariance
on~$T^2$ then implies that the partition function is left intact
when the positions of two lines are interchanged.

Another important point is that each line in the lattice carries a
continuous complex parameter, namely its coordinate on $C$.  In the
context of lattice models, this parameter is called the \emph{spectral
  parameter} of the line.  Hence, $\RM_{\alpha\beta}$ depends on the
spectral parameters~$z_\alpha$ and~$z_\beta$, and by translation
invariance it is a function of the difference $z_\alpha - z_\beta$.
Since the theory is holomorphic on $C$, it should satisfy
\begin{equation}
  [D_\zb, \RM_{\alpha\beta}] = 0
\end{equation}
so that gauge invariant quantities constructed from the R-matrices are
holomorphic in the spectral parameters.

These two properties -- the commutativity of any two parallel lines
and the existence of a spectral parameter assigned to each line -- are
what make a lattice model \emph{integrable}.  Let us quickly explain
why.

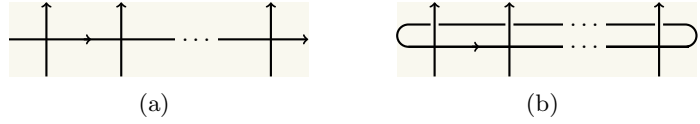
\begin{figure}
  \centering
  \subfloat[\label{fig:MM}]{
    \begin{tikzpicture}[scale=1, baseline=(x.base)]
      \node (x) at (0,0) {\vphantom{x}};
      
      \fill[ws] (0,-0.5) rectangle (4,0.5);
      
      \draw[r->] (0,0) -- (1.1,0);
      \draw[r->] (1,0) -- (4,0);
      \node[ws, inner sep=1pt] at (2.5,0) {$\,\dots$};
      
      \begin{scope}[shift={(0.5,-0.5)}]
        \draw[r->] (0,0) -- (0,1);
        \draw[r->] (1,0) -- (1,1);
        \draw[r->] (3,0) -- (3,1);
      \end{scope}
    \end{tikzpicture}
  }
  \qquad
  \subfloat[\label{fig:TM}]{
    \begin{tikzpicture}[scale=1, baseline=(x.base)]
      \node (x) at (0,0) {\vphantom{x}};
      
      \fill[ws] (0,-0.5) rectangle (4,0.5);
      
      \begin{scope}[shift={(0,-0.1)}]
        \draw[r->] (0.15,0) -- (1.1,0);
        \draw[r-] (1,0) -- (3.85,0) arc (-90:90:0.15)
        -- (0.15,0.3) arc (90:270:0.15) -- (3.9,0);

        \node[ws, inner sep=1pt] at (2.5,0) {$\,\dots$};
        \node[ws, inner sep=1pt] at (2.5,0.3) {$\,\dots$};
      \end{scope}

      \begin{scope}[shift={(0.5,-0.5)}]
        \draw[line width=3pt, olive!5] (0,0.65) -- ++(90:0.1);
        \draw[line width=3pt, olive!5] (1,0.65) -- ++(90:0.1);
        \draw[line width=3pt, olive!5] (3,0.65) -- ++(90:0.1);
        \draw[r->] (0,0) -- (0,1);
        \draw[r->] (1,0) -- (1,1);
        \draw[r->] (3,0) -- (3,1);
      \end{scope}
    \end{tikzpicture}
  }
  \caption{(a) A monodromy matrix and (b) a transfer matrix.}
  \label{fig:MM-TM}
\end{figure}

Formally, we can reformulate the above lattice model in such a way
that it no longer carries state variables on the faces: we simply
introduce big Hilbert spaces
\begin{equation}
  \V_\alpha = \bigoplus_{a,b} \V_{ab,\alpha}
\end{equation}
and extend the R-matrix \eqref{eq:R-matrix} to a linear map
\begin{equation}
  \RM_{\alpha\beta}(z_\alpha - z_\beta)
  \colon \V_\alpha \otimes \V_\beta \to \V_\beta \otimes \V_\alpha \,,
\end{equation}
setting the excess matrix elements to zero.  With this reformulation,
we can introduce the row-to-row \emph{monodromy matrices}
\begin{equation}
  T_\alpha(z_\alpha; z_{m+1}, \dotsc, z_{m+n})
  =
  \RM_{\alpha, m+n}(z_\alpha - z_{m+n})
  \circ_{\V_\alpha} \dotsb \circ_{\V_\alpha}
  \RM_{\alpha, m+1}(z_\alpha - z_{m+1})
\end{equation}
and \emph{transfer matrices}
\begin{equation}
  t_\alpha(z_\alpha; z_{m+1}, \dotsc, z_{m+n})
  =
  \Tr_{\V_\alpha}
  T_\alpha(z_\alpha; z_{m+1}, \dotsc, z_{m+n})
   \,,
\end{equation}
where the compositions and trace are taken in the space $\V_\alpha$
assigned to the horizontal edges in the $\alpha$th row.  These are
endomorphisms of
$\V_\alpha \otimes \V_{m+1} \otimes \dotsb \otimes \V_{m+n}$ and
$\V_{m+1} \otimes \dotsb \otimes \V_{m+n}$, respectively.
Graphically, a monodromy matrix is a horizontal line traversing
segments of vertical lines, and a transfer matrix is obtained when the
horizontal line makes a loop and comes back to the starting point; see
Figure~\ref{fig:MM-TM}.

Using the transfer matrices we can express the partition function as a
trace:
\begin{equation}
  \biggvev{\prod_{\alpha=1}^{m+n} L_\alpha}
  =
  \Tr_{\V_{m+1} \otimes \dotsb \otimes \V_{m+n}}
  (t_m \circ \dotsb \circ t_1) \,.
\end{equation}
If we think of the vertical direction as a time direction, we may
regard the transfer matrices $t_1$, $\dotsc$, $t_m$ as a sequence of
discrete ``time evolution operators'' acting on the ``total Hilbert
space'' $\V_{m+1} \otimes \dotsb \otimes \V_{m+n}$ of the lattice
model.

The commutativity of horizontal lines means that
transfer matrices commute:
\begin{equation}
  [t_\alpha(z_\alpha), t_\beta(z_\beta)] = 0 \,.
\end{equation}
(Here we have suppressed the dependence of the transfer matrices on
the spectral parameters assigned to the vertical lines.)  If we expand
$t_\alpha(z_\alpha)$ in the powers of $z_\alpha$, the expansion
coefficients are themselves operators on the total Hilbert space.  In
this way we obtain an infinite number of ``conserved charges'' which
commute with the time evolution operator $t_\beta(z_\beta)$.  Further
expanding $t_\beta(z_\beta)$ in $z_\beta$, we learn that these
conserved charges mutually commute.  In this sense the lattice model
is said to be integrable.

To recapitulate, the correlation function of a lattice of line
operators in four-dimensional Chern--Simons theory is the
partition function of an integrable lattice model defined on the same
lattice.  The integrability is a consequence of the topological
invariance on $T^2$ and the existence of the extra dimensions $C$.

\begin{figure}
  \centering
  \subfloat[\label{fig:unitarity}]{
    \vphantom{
      \begin{tikzpicture}[scale=0.75, baseline=(x.base)]
      \node (x) at (30:2) {\vphantom{x}};
      \fill[ws] (0,-0.5) rectangle ({3*sqrt(3)/2},2.5);
      \draw[r->] (-30:1) node[below] {$\gamma$} -- ++(0,3);
    \end{tikzpicture}
  }%
    \def\pathA{(0,0) to[out=0, in=180] (1,0.5) to[out=0, in=180] (2,0)}
    \def\pathB{(2,0.5) to[in=0,out=180] (1,0) to[in=0,out=180] (0,0.5)}
    \begin{tikzpicture}[scale=1.2, baseline=(x.base)]
      \node (x) at (1,0.25) {\vphantom{x}};
      
      \fill[ws] (0,-0.5) rectangle (2,1);
      
      \draw[r->] node[left] {$\beta$} \pathA;
      \draw[r<-] \pathB node[left] {$\alpha$};
      
      \node at (6/6,0.75) {$a$};
      \node at (1/6,0.25) {$b$};
      \node at (6/6,-0.25) {$c$};
      \node at (11/6,0.25) {$d$};
      \node at (6/6,0.25) {$e$};
    \end{tikzpicture}
    \ =
    \begin{tikzpicture}[scale=1.2, baseline=(x.base)]
      \node (x) at (1,0.25) {\vphantom{x}};
      
      \fill[ws] (0,-0.5) rectangle (2,1);
      
      \draw[r->] (0,0) node[left] {$\beta$} -- (2,0);
      \draw[r->] (0,0.5) node[left] {$\alpha$} -- (2,0.5);
      
      \node at (6/6,0.75) {$a$};
      \node at (1/6,0.25) {$b$};
      \node at (6/6,-0.25) {$c$};
      \node at (11/6,0.25) {$d$};
    \end{tikzpicture}
  }
  \qquad
  \subfloat[\label{fig:YBE}]{
    \begin{tikzpicture}[scale=0.75, baseline=(x.base)]
      \node (x) at (30:2) {\vphantom{x}};
      
      \fill[ws] (0,-0.5) rectangle ({3*sqrt(3)/2},2.5);
      
      \draw[r->] (0,0) node[left] {$\beta$} -- ++(30:3);
      \draw[r->] (0,2) node[left] {$\alpha$} -- ++(-30:3);
      \draw[r->] (-30:1) node[below] {$\gamma$} -- ++(0,3);
      
      \node (O) at ({sqrt(3)*4/6},1) {};
      \node at ($(O) + (120:{sqrt(3)*4/6})$) {$a$};
      \node at ($(O) + (180:{sqrt(3)*3/6})$) {$b$};
      \node at ($(O) + (240:{sqrt(3)*4/6})$) {$c$};
      \node at ($(O) + (300:{sqrt(3)*3/6})$) {$d$};
      \node at ($(O) + (360:{sqrt(3)*4/6})$) {$e$};
      \node at ($(O) + (60:{sqrt(3)*3/6})$) {$f$};
      \node at ($(O)$) {$g$};
    \end{tikzpicture}
    \ =
    \begin{tikzpicture}[scale=0.75, baseline=(x.base)]
      \node (x) at (30:1) {\vphantom{x}};
      
      \fill[ws] (0,-1) rectangle ({3*sqrt(3)/2},2);
      
      \draw[r->] (0,0) node[left] {$\beta$} -- ++(30:3);
      \draw[r->] (0,1) node[left] {$\alpha$} -- ++(-30:3);
      \draw[r->] (-30:2) node[below] {$\gamma$} -- ++(0,3);
      
      \node (O) at ({sqrt(3)*5/6},0.5) {};
      \node at ($(O) + (120:{sqrt(3)*3/6})$) {$a$};
      \node at ($(O) + (180:{sqrt(3)*4/6})$) {$b$};
      \node at ($(O) + (240:{sqrt(3)*3/6})$) {$c$};
      \node at ($(O) + (300:{sqrt(3)*4/6})$) {$d$};
      \node at ($(O) + (360:{sqrt(3)*3/6})$) {$e$};
      \node at ($(O) + (60:{sqrt(3)*4/6})$) {$f$};
      \node at ($(O)$) {$g$}; 
    \end{tikzpicture}
  }
  \caption{(a) The unitarity relation and (b) the Yang--Baxter equation.}
  \label{fig:unitarity-YBE}
\end{figure}
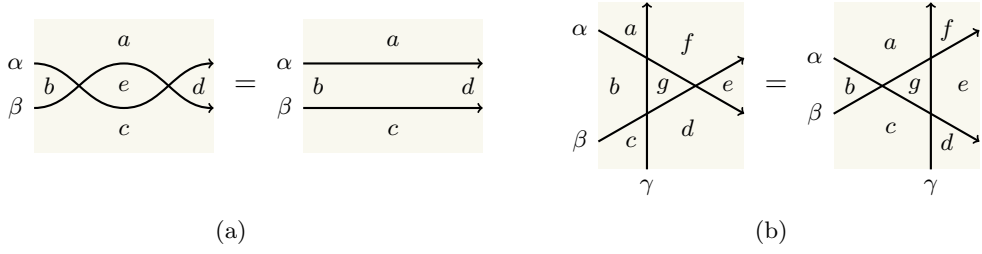

In fact, we can make a stronger statement.  A similar argument as
above leads to the conclusion that the R-matrices satisfy the
\emph{unitarity relation}
\begin{equation}
  \label{eq:unitarity}
  \sum_e
  \RM_{\beta\alpha}
  \biggl(
  \begin{array}{cc}
    a & d \\
    e & c
  \end{array}
  \biggm|
  z_\beta - z_\alpha
  \biggr)
  \RM_{\alpha\beta}
  \biggl(
  \begin{array}{cc}
    a & e \\
    b & c
  \end{array}
  \biggm|
  z_\alpha - z_\beta
  \biggr)
  \\
  =
  \delta_{bd} \id_{\V_{ab,\alpha} \otimes \V_{bc,\beta}}
\end{equation}
and the \emph{Yang--Baxter equation}
\begin{multline}
  \label{eq:YBE}
  \sum_g
  \RM_{\alpha\beta}
  \biggl(
  \begin{array}{cc}
    f & e \\
    g & d
  \end{array}
  \biggm|
  z_\alpha - z_\beta
  \biggr)
  \RM_{\alpha\gamma}
  \biggl(
  \begin{array}{cc}
    a & f \\
    b & g
  \end{array}
  \biggm|
  z_\alpha - z_\gamma
  \biggr)
  \RM_{\beta\gamma}
  \biggl(
  \begin{array}{cc}
    b & g \\
    c & d
  \end{array}
  \biggm|
  z_\beta - z_\gamma
  \biggr)
  \\
  =
  \sum_g
  \RM_{\beta\gamma}
  \biggl(
  \begin{array}{cc}
    a & f \\
    g & e
  \end{array}
  \biggm|
  z_\beta - z_\gamma
  \biggr)
  \RM_{\alpha\gamma}
  \biggl(
  \begin{array}{cc}
    g & e \\
    c & d
  \end{array}
  \biggm|
  z_\alpha - z_\gamma
  \biggr)
  \RM_{\alpha\beta}
  \biggl(
  \begin{array}{cc}
    a & g \\
    b & c
  \end{array}
  \biggm|
  z_\alpha - z_\beta
  \biggr)
  \,.
\end{multline}
The latter is an equality between two linear maps from
$\V_{ab,\alpha} \otimes \V_{bc,\beta} \otimes \V_{cd,\gamma}$ to
$\V_{af,\gamma} \otimes \V_{fe,\beta} \otimes \V_{ed,\alpha}$, and
each R-matrix is implicitly tensored with an identity operator.  These
relations imply the commutativity of transfer matrices, hence
integrability.  Their graphical representations are shown in
Figure~\ref{fig:unitarity-YBE}.

\subsection{Wilson lines  and dynamical R-matrices}
\label{sec:dR}

What kinds of R-matrices do we get if $L_\alpha$ are Wilson lines
\begin{equation}
  W_\alpha
  =
  \Tr_{V_\alpha} P\exp\biggl(\oint \CA\biggr)
\end{equation}
in representations $G \to \GL(V_\alpha)$?  To answer this question, we
recall how we defined the path integral for our theory.  The
computation is done in two steps.  First, we fix a $\tf_\C$-valued
gauge field $\CA^\infty$ that represents a point in the Lagrangian
submanifold $\Lag_\infty$ of the moduli space~\eqref{eq:moduli}, and
integrate over the gradient flow trajectories generated by the real
part of the action.  The result is a function on $\Lag_\infty$.
Subsequently, we integrate this function over~$\Lag_\infty$.

The first step can be well approximated by perturbation theory around
the background~$\CA^\infty$.  In perturbation theory, the
contributions to the correlation function come from the exchange of
gluons between Wilson lines.  (There are also vacuum and self-energy
diagrams which should be taken care of by renormalization.)  The
fluctuations from $\CA^\infty$ that we integrate over are massive by
construction.  So if we take advantage of the topological invariance
of the theory and rescale the metric on $T^2$ by a very large factor,
the contributions from gluons traveling a finite distance in $T^2$ are
suppressed.  This argument might fail if the coupling constant
increases as we take the large volume limit, but this does not happen
as our theory is actually infrared free.  Thus, quantum effects get
localized in the vicinity of the crossings of Wilson lines.
Accordingly, the correlation function factorizes into the product of
local contributions associated to the vertices of the lattice.  These
local contributions are the R-matrices of the lattice model.

While quantum effects are important only for interactions between
nearby Wilson lines, classical effects are not confined to short
distances.  A Wilson line may be thought of as a heavy, electrically
charged particle moving along a curve.  The state of this particle is
labeled by a weight $\upomega \in \tf_\C^*$ of its representation.
When two such particles encounter, they exchange gluons and their
states may change.  Hence, a state of the system under consideration
is specified by a set of weights assigned to the edges of the lattice.
Each of these edges sources an electromagnetic field, which does
affect charged objects at distant places.

As an example, consider a Wilson line in the state $\upomega$ along a
horizontal line $K$ at $y = z = 0$.  The part of the Wilson line felt
by faraway objects is
\begin{equation}
  \exp\biggl(\int_K \upomega(\CA^\infty)\biggr)
  =
  \exp\biggl(\int_{K \times I \times E}
  \Tr(\upomega \CA^\infty)
  \rmd\theta(y) \delta^2(z,\zb) \, \rmd^2z\biggr) \,,
\end{equation}
where $I$ is an interval in the $y$-direction such that
$K \times I \times E$ contains the Wilson line and the objects under
consideration, $\theta(y)$ is a step function such that
$\del_y\theta(y) = \delta(y)$, and we have identified $\tf_\C$ and
$\tf_\C^*$ via the bilinear form $\Tr$.  The presence of this factor
in the path integral has the same effect on those objects as shifting
$A^\infty_\zb$ by $-\pi\hbar\upomega \theta(y) \delta^2(z,\zb)$ over
$K \times I \times E$.  We must take this shift into account when
computing the R-matrices.

The above analysis shows that the R-matrices depend on the effective
background gauge field which differs from $\CA^\infty$ by a shift due
to the combined effect of all Wilson lines present in the system.  By
gauge symmetry, the R-matrices are functions of the parameter
\begin{equation}
  \uplambda \in \tf_\C^*
\end{equation}
for the effective background, defined by formula~\eqref{eq:uplambda}.
Its value jumps by $\hbar\upomega$ across a segment of Wilson line
carrying the state $\upomega$.  Drawing a Wilson line with a dashed
line, we can express this jump rule graphically as follows:
\begin{equation}
  \label{eq:lambda-change}
  \begin{tikzpicture}[scale=0.6, baseline=(x.base)]
    \node (x) at (1,1) {\vphantom{x}};

    \fill[ws] (0,0.3) rectangle (3,2.2);

    \draw[dr->] (0,1.25) node[left] {$\upomega$} -- (3,1.25);

    \node at (1.5,1.75) {$\uplambda$};
    \node at (1.5,0.75) {$\uplambda - \hbar\upomega$};
  \end{tikzpicture}
  \ .
\end{equation}
In lattice models, this parameter $\uplambda$ is called the
\emph{dynamical parameter}.  An R-matrix that has a dynamical
parameter is known as a \emph{dynamical R-matrix}.  The appearance of
dynamical R-matrices from Wilson lines was argued
in~\cite{Costello:2017dso} based on considerations in an effective
two-dimensional abelian gauge theory.

When one refers to an R-matrix depending on a dynamical parameter,
there is a potential confusion as to which point one is evaluating the
dynamical parameter at because its value varies from place to place.
We define the R-matrix
\begin{equation}
  \RM_{\alpha\beta}(z_\alpha - z_\beta, \uplambda)
  \colon V_\alpha \otimes V_\beta \to V_\beta \otimes V_\alpha
  \,,
\end{equation}
arising from the crossing of two Wilson lines $W_\alpha$ and
$W_\beta$, with respect to the dynamical parameter on the top-left
face:
\begin{equation}
  \label{eq:RF-g}
  \RM_{\alpha\beta}(z_\alpha - z_\beta, \uplambda)
  =
  \begin{tikzpicture}[scale=0.6, baseline=(x.base)]
    \node (x) at (1,1) {\vphantom{x}};

    \fill[ws] (0,0) rectangle (2,2);
    \draw[dr->] (0,1) node[left] {$\alpha$} --  (2,1);
    \draw[dr->] (1,0) node[below] {$\beta$} -- (1,2);

    \node at (0.5,1.5) {$\uplambda$};
  \end{tikzpicture}
  \ .
\end{equation}
The dynamical parameters on the other three faces are determined once
states are chosen on the edges.  Consistency at the bottom-right face
requires that the R-matrix has zero weight, that is,
$\RM_{\alpha\beta}$ commutes with the action of $\tf_\C$.

A priori, the R-matrix~\eqref{eq:RF-g} also depends on the components
of $\CA^\infty$ along $T^2$.  However, by a $\T_\C$-gauge
transformation we can make $\CA^\infty_x$ and $\CA^\infty_y$ vanish
everywhere except in the neighborhood of a single $x$-coordinate and a
single $y$-coordinate, respectively.  In this gauge the sole effect of
these components is to twist the periodic boundary conditions with the
gauge transformations by the corresponding holonomies.  Hence, we
conclude that the transfer matrices and the partition function are
given by
\begin{equation}
  t_\alpha
  =
  \Tr_{V_\alpha}\bigl(
  \exp(2\pi\iu\uptau_x)
  \RM _{\alpha, m+n}
  \circ_{V_\alpha} \dotsb \circ_{V_\alpha}
  \RM_{\alpha, m+1}
  \bigr)
\end{equation}
and
\begin{equation}
  \biggvev{\prod_{\alpha=1}^{m+n} W_\alpha}
  =
  \Tr_{V_{m+1} \otimes \dotsb \otimes V_{m+n}}
  \bigl(\exp(2\pi\iu\uptau_y) t_m \circ \dotsb \circ t_1\bigr) \,,
\end{equation}
where $\RM_{\alpha\beta}$ now refers to the R-matrix in the background
with $\CA^\infty_x = \CA^\infty_y = 0$.  The zero-weight property of
$\RM_{\alpha\beta}$ implies that the transfer matrix has zero weight.
In turn, this ensures that the partition function is independent of
the choice of the row in which the twist $\exp(2\pi\iu\uptau_y)$ is
inserted, as it must be by gauge invariance.  By symmetry the same can
be said about the choice of the column for $\exp(2\pi\iu\uptau_x)$.

Keeping track of how the dynamical parameter changes in the graphical
representation of the Yang--Baxter equation, we find that the
R-matrices arising from the crossings of Wilson lines obey
\begin{multline}
  \label{eq:dYBE}
  \RM_{\alpha\beta}(z_\alpha - z_\beta, \uplambda - \hbar h_\gamma)
  \RM_{\alpha\gamma}(z_\alpha - z_\gamma, \uplambda) 
  \RM_{\beta\gamma}(z_\beta - z_\gamma, \uplambda - \hbar h_\alpha) 
  \\
  =
  \RM_{\beta\gamma}(z_\beta - z_\gamma, \uplambda) 
  \RM_{\alpha\gamma}(z_\alpha - z_\gamma, \uplambda - \hbar h_\beta) 
  \RM_{\alpha\beta}(z_\alpha - z_\beta, \uplambda)
  \,.
\end{multline}
Here the notation $h_\alpha$ means that it is to be replaced with
$\upomega$ when the R-matrices act on a state with weight $\upomega$
in $V_\alpha$.  The Yang--Baxter equation of this form is known as the
\emph{dynamical Yang--Baxter equation}.

Four-dimensional Chern--Simons theory thus produces a dynamical
R-matrix $\RM_{\alpha\beta}$, specified by a choice of the gauge group
$G$ and a pair of representations $(V_\alpha, V_\beta)$ of $G$.  This
R-matrix has zero weight and satisfies the unitarity relation
\begin{equation}
  \label{eq:d-unitarity}
  \RM_{\beta\alpha}(z_\beta - z_\alpha, \uplambda)
  \RM_{\alpha\beta}(z_\alpha - z_\beta, \uplambda) 
  = \id_{V_\alpha \otimes V_\beta} \,.
\end{equation}

Furthermore, by perturbation theory we can compute the R-matrix order
by order in $\hbar$.  At each order (except the zeroth),
$\RM_{\alpha\beta}(z_\alpha - z_\beta, \uplambda)$ diverges at
$z_\alpha - z_\beta = 0$, which is the point corresponding to the
situation where $W_\alpha$ and $W_\beta$ intersect in the
four-dimensional spacetime.  At the first order, the divergence comes
from a diagram in which a single gluon travels between the two Wilson
lines in a neighborhood of the intersection, without going around
one-cycles of $E$.  Hence, if we gauge away $A^\infty_\zb$ in this
neighborhood, the singular behavior of the R-matrix is independent of
the dynamical parameter to first order in $\hbar$.

\subsection{\texorpdfstring{Dynamical R-matrices for $G = \U(N)$ and
    $\SU(N)$}{Dynamical R-matrices for G = U(N) and SU(N)}}

For $G = \U(N)$ and $(V_\alpha, V_\beta) = (\C^N,\C^N)$, with $\Tr$
taken to be the trace in the vector representation $\C^N$, Etingof and
Varchenko \cite{MR1645196} showed that a dynamical R-matrix with the
properties described above is unique to all orders in perturbation
theory, up to certain simple transformations and perturbative
corrections to $\tau$.  It is Felder's R-matrix for the elliptic
quantum group for $\slf_N$~~\cite{Felder:1994pb, Felder:1994be}, which
first appeared as the Boltzmann weight for an integrable lattice model
discovered by Jimbo, Miwa and Okado~\cite{MR908997, Jimbo:1987mu}.
For $N = 2$, the Jimbo--Miwa--Okado model reduces to the eight-vertex
solid-on-solid model~\cite{Baxter:1972wf}.

To state the result of \cite{MR1645196} more precisely, we need a
little preparation.

First of all, let us introduce some notations.  For $G = \U(N)$, the
complexified Cartan subalgebra $\tf_\C$ is the space of complex
diagonal matrices.  The standard basis for $\tf_\C$ consists of the
matrices $E_{ii}$, $i=1$, $\dotsc$, $N$, which have $1$ in the $(i,i)$
entry and~$0$ elsewhere.  The trace $\Tr$ identifies $E_{ii}$ with its
dual $E_{ii}^*$, so we can write the dynamical parameter as
\begin{equation}
  \uplambda = \sum_{i=1}^N \uplambda_i E_{ii}^* \,,
\end{equation}
using an $N$-tuple of complex numbers
$(\uplambda_1, \dotsc, \uplambda_N)$.  The standard basis vector $e_i$
of $\C^N$ has weight $\upomega_i = E_{ii}^*$.  The matrix elements of
an endomorphism $R$ of $\C^N \otimes \C^N$ are defined by
$R(e_i \otimes e_j) = \sum_{k,l=1}^N e_k \otimes e_l R_{ij}^{kl}$.

We also need Jacobi's first theta function
$\theta_1(z) = \theta_1(z|\tau)$.  In terms of the theta function with
characteristics
\begin{equation}
  \theta\biggl[
  \begin{array}{c}
    a \\ b
  \end{array}
  \biggr]
  (z|\tau)
  =
  \sum_{n=-\infty}^\infty
  e^{\pi\iu (n+a)^2 \tau + 2\pi\iu(n+a)(z+b)}
  \,,
\end{equation}
this is given by
\begin{equation}
  \theta_1(z|\tau)
  =
  -
  \theta\biggl[
  \begin{array}{c}
    1/2 \\ 1/2
  \end{array}
  \biggr]
  (z|\tau)
  \,.
\end{equation}
It is an odd function:
\begin{equation}
  \theta_1(-z) = -\theta_1(z) \,.
\end{equation}
From the identities
\begin{equation}
  \begin{aligned}
    \theta\biggl[
    \begin{array}{c}
      a \\ b
    \end{array}
    \biggr]
    (z+1|\tau)
    &=
    e^{2\pi\iu a}
    \theta\biggl[
    \begin{array}{c}
      a \\ b
    \end{array}
    \biggr]
    (z|\tau)
    \,,
    \\
    \theta\biggl[
    \begin{array}{c}
      a \\ b
    \end{array}
    \biggr]
    (z + c\tau|\tau)
    &=
    e^{-\pi\iu c^2\tau - 2\pi\iu c(z+b)}
    \theta\biggl[
    \begin{array}{c}
      a + c \\ b
    \end{array}
    \biggr]
    (z|\tau)
    \,,
  \end{aligned}
\end{equation}
it follows that $\theta_1$ has the following quasi-periodicity
property:
\begin{equation}
  \theta_1(z+1) = -\theta_1(z)
  \,, \qquad
  \theta_1(z+\tau) = -e^{-\pi\iu\tau - 2\pi\iu z} \theta_1(z)
  \,.
\end{equation}

We can now define Felder's R-matrix $\RF$.  This is an
$\End(\C^N \otimes \C^N)$-valued meromorphic function on
$\C \times \tf_\C^*$ such that
\begin{equation}
  \RM^{\mathrm{F}}(z,\uplambda) = P\RF(z,\uplambda)
\end{equation}
satisfies the dynamical Yang--Baxter equation \eqref{eq:dYBE} and the
unitarity relation~\eqref{eq:d-unitarity}.  Here
$P \in \End(\C^N \otimes \C^N)$ is the swap isomorphism:
$P(v \otimes w) = w \otimes v$.  The matrix elements
$\RF(z,\uplambda)_{ij}^{kl}$ vanishes unless $\{i,j\} = \{k,l\}$.  The
nonzero matrix elements are
\begin{equation}
  \RF(z, \uplambda)_{ii}^{ii}
  =
  1
  \,,
  \quad
  \RF(z, \uplambda)_{ij}^{ij}
  =
  \frac{\theta_1(z) \theta_1(\uplambda_{ij} + \hbar)}
  {\theta_1(z - \hbar) \theta_1(\uplambda_{ij})}
  \,,
  \quad
  \RF(z, \uplambda)_{ij}^{ji}
  =
  \frac{\theta_1(\hbar) \theta_1(z - \uplambda_{ij})}
  {\theta_1(z - \hbar) \theta_1(\uplambda_{ij})}
  \,,
\end{equation}
where $i \neq j$ and $\uplambda_{ij} = \uplambda_i - \uplambda_j$.

Finally, let $\RM^{\U(N)}$ be the R-matrix for the crossing of two
Wilson lines in the vector representation of $\U(N)$, and
$R^{\U(N)} = P\RM^{\U(N)}$.  As our aim is to relate $R^{\U(N)}$
and~$\RF$, we must identify $R^{\U(N)}$ with an
$\End(\C^N \otimes \C^N)$-valued meromorphic function
on~$\C \times \tf_\C^*$.  We do this by choosing a trivialization for
the rank-$N$ holomorphic vector bundle $V_\uplambda \to E$
corresponding to a flat gauge field characterized by the dynamical
parameter $\uplambda$.

Let us treat the dynamical parameter on the left side of a Wilson line
as the background gauge field experienced by the charged particle; for
instance, the Wilson line in diagram~\eqref{eq:lambda-change} is a
charged particle moving in the background $\uplambda$, which itself
sources a gauge field and shifts the background to
$\uplambda - \hbar\upomega$ on the right side.  If the spectral
parameter of this line is~$z$, a state of the charged particle is a
point in the fiber $V_\uplambda|_z$.  We identify the local
holomorphic frame $(s_i)_{i=1}^N$ of $V_\uplambda$ defined by formula
\eqref{eq:hol-sect} and the standard frame $(e_i)_{i=1}^N$ of the
trivial bundle $\C \times \C^N$.  With respect to this trivialization,
$R^{\U(N)}$ is an $\End(\C^N \otimes \C^N)$-valued function, and its
matrix elements are meromorphic functions of the spectral parameter
since $D_\zb s_i = 0$ and $[D_\zb, R^{\U(N)}] = 0$.  The matrix
element $R^{\U(N)}(z_1 - z_2,\uplambda)^{kl}_{ij}$ describes the
process in which the state $s_i(z_1) \otimes s_j(z_2)$ in
$V_\uplambda|_{z_1} \otimes V_{\uplambda - \hbar\upomega_i}|_{z_2}$
evolves into the state $s_k(z_1) \otimes s_l(z_2)$ in
$V_{\uplambda - \hbar\upomega_l}|_{z_1} \otimes V_{\uplambda}|_{z_2}$.

By choosing this gauge, we have set $A^\infty_\zb = 0$ and let the
monodromies of $s_i$ encode the dynamical parameter.  For a generic
value of $\uplambda$, all we can do now is to rescale $s_i$ by
separate factors, so the residual gauge symmetry (apart from the Weyl
group action) is given by $\T_C$-valued gauge transformations that are
constant on $E$.  Since we have also gauged away $\CA^\infty_x$ and
$\CA^\infty_y$, these gauge transformations must be constant on
$\Sigma$ as well.

According to a theorem of Etingof and Varchenko~\cite{MR1645196},
$R^{\U(N)}$, regarded as an $\End(\C^N \otimes \C^N)$-valued function
as above, is related to $\RF$ by a sequence of transformations.  Some
of these transformations can be understood as $\T_\C$-valued gauge
transformations which are meromorphic and possibly multivalued on $E$.
On a dynamical R-matrix
$R\colon \C \times \tf^*_\C \to \End(\C^N \otimes \C^N)$, the gauge
transformation $R \mapsto g \cdot R$ by a $\T_\C$-valued meromorphic
function $g$ on $\C \times \tf_\C^*$ acts by
\begin{equation}
  \label{eq:gauge-transf}
  g \cdot R(z_1 - z_2, \uplambda)
  =
  \bigl(g(z_1, \uplambda - \hbar h_2)
  \otimes g(z_2,\uplambda)\bigr)
  R(z_1 - z_2, \uplambda)
  \bigl(g(z_1, \uplambda)^{-1}
  \otimes g(z_2,\uplambda - \hbar h_1)^{-1}\bigr) \,.
\end{equation}
Under gauge transformations a unitary R-matrix is mapped to a unitary
R-matrix.

One of the transformations relevant for the theorem is the gauge
transformation by a multivalued function of the form
\begin{equation}
  g(z,\uplambda)^i_j
  =
  \delta^i_j
  e^{-z(\psi(\uplambda) - \psi(\uplambda - \hbar\upomega_i))} \,,
\end{equation}
with $\psi(\lambda)$ being a meromorphic function on $\tf_\C^*$.

Another transformation involves a closed meromorphic multiplicative
two-form $\varphi$ on~$\tf_\C^*$, which is a set $\{\varphi_{ij}\}$ of
meromorphic functions on $\tf_\C^*$ such that
$\varphi_{ij} = \varphi_{ji}^{-1}$ and
\begin{equation}
  \frac{
    \varphi_{ij}(\uplambda)
    \varphi_{jk}(\uplambda)
    \varphi_{ki}(\uplambda)
  }{
    \varphi_{ij}(\uplambda - \hbar\upomega_k)
    \varphi_{jk}(\uplambda - \hbar\upomega_i)
    \varphi_{ki}(\uplambda - \hbar\upomega_j)
  }
  =
  1 \,.
\end{equation}
Its action $R \mapsto \varphi \cdot R$ is given by
\begin{equation}
  \varphi \cdot R(z,\uplambda)_{ij}^{ij}
  =
  \varphi_{ij}(\uplambda) R(z,\uplambda)_{ij}^{ij} \,,
\end{equation}
with the other matrix elements unchanged.  This transformation is also
a gauge transformation, at least locally on $\tf_\C^*$.  Indeed,
locally we can write $\varphi$ as an exact form \cite{MR1957065}; in
other words, there exist meromorphic functions $\{\xi_i\}$ such that
\begin{equation}
  \varphi_{ij}(\uplambda)
  =
  \frac{\xi_i(\uplambda) \xi_j(\uplambda - \hbar\upomega_i)}
       {\xi_i(\uplambda - \hbar\upomega_j) \xi_j(\uplambda)} \,.
\end{equation}
Thus, the action of $\varphi$ is locally the gauge transformation with
$g(z,\lambda)^i_j = \delta^i_j \xi_i(\lambda)^{-1}$.

The other relevant transformations are the maps
\begin{equation}
  \label{eq:Weyl}
  R(z,\uplambda)
  \mapsto
  \sigma \otimes \sigma
  R(z,\sigma^{-1} \cdot \uplambda)
  (\sigma \otimes \sigma)^{-1} \,,
\end{equation}
with $\sigma$ being an element of the symmetric group $S_N$, acting on
$\tf_\C^*$ and $\C^N$ in the obvious ways;
\begin{equation}
  R(z,\uplambda)
  \mapsto
  f(z) R(z, \uplambda) \,,
\end{equation}
with $f$ a meromorphic function on $\C$ such that $f(z) f(-z) = 1$;
and
\begin{equation}
  \label{eq:c+mu}
  R(z,\uplambda)
  \mapsto
  R(bz, c\uplambda + \upmu) \,,
\end{equation}
with $b$, $c \in \C^\times$ and $\upmu \in \tf_\C^*$.

The map \eqref{eq:Weyl} is simply the action of the Weyl group, under
which our R-matrix should be invariant.  For $c \neq 1$, the
map~\eqref{eq:c+mu} changes the amount by which the dynamical
parameter jumps across Wilson lines.  So we have $c = 1$.

To constrain the remaining freedom, we look at the quasi-periodicity
of $R^{\U(N)}$.  In the gauge we are using, the holomorphic sections
$s_i$ obey the monodromy relations~\eqref{eq:monodromy}.  In view of
these relations, the matrix elements of $R^{\U(N)}$ have the
quasi-periodicity property
\begin{equation}
  \begin{aligned}
  R^{\U(N)}(z + 1, \uplambda)_{ij}^{kl}
  &= R^{\U(N)}(z, \uplambda)_{ij}^{kl} \,,
  \\
  R^{\U(N)}(z + \tau, \uplambda)_{ij}^{kl}
  &=
  e^{2\pi\iu (\uplambda_{ki} - \hbar (\upomega_l)_k)}
  R^{\U(N)}(z, \uplambda)_{ij}^{kl}
  =
  e^{-2\pi\iu (\uplambda_{lj} + \hbar(\upomega_i)_j)}
  R^{\U(N)}(z, \uplambda)_{ij}^{kl} \,.
  \end{aligned}
\end{equation}
Here $\upomega_i = \sum_{j=1}^N (\upomega_i)_j E_{jj}^*$, or
$(\upomega_i)_j = \delta_{ij}$.  For this to be the case, $R^{\U(N)}$
should take the form%
\footnote{For $R^{\U(N)}(z)_{ii}^{ii}$ to have the correct
  quasi-periodicity,
  $\psi(\uplambda) - 2\psi(\uplambda - \hbar\upomega_i) +
  \psi(\uplambda - 2\hbar\upomega_i)$ must be independent of
  $\uplambda$ for all $i$.  A function $\psi(\uplambda)$ that has this
  property and is invariant under the Weyl group action is a multiple
  of the trace $\sum_{i=1}^N \uplambda_i$, but the corresponding gauge
  transformation acts trivially on the R-matrix.  Then, the
  quasi-periodicity of $R^{\U(N)}(z)_{ii}^{ii}$ fixes that of $f$, and
  the quasi-periodicity of the other components determines the value
  of $b$ and tells that $\upmu_{ij}$ are integers.  Shifting
  $\uplambda_{ij}$ by integers does not affect $\RF$.}
\begin{equation}
  \label{eq:R-UN} 
  R^{\U(N)}(z,\uplambda)
  =
  f(z) \varphi \cdot \RF(-z,\uplambda) \,,
\end{equation}
with $f$ satisfying the quasi-periodicity relations
\begin{equation}
  f(z + 1) = f(z) \,,
  \qquad
  f(z + \tau) = e^{-2\pi\iu\hbar} f(z)
  \,.
\end{equation}
At this point there is nothing that constrains $\varphi$.

Let us turn to the case when the gauge group is $\SU(N)$.  In this
case $\tf_\C$ is the space of complex traceless diagonal matrices, so
the dynamical parameter for $\SU(N)$, which we call $\uplambdab$,
obeys the constraint
\begin{equation}
  \sum_{i=1}^N \uplambdab_i = 0 \,.
\end{equation}
The weight $\upomegab_i = \sum_{j=1}^N (\upomegab_i)_j E_{jj}^*$ of
$e_i$ is given by $(\upomegab_i)_j = \delta_{ij} - 1/N$.  We refer to
the background field configuration specified by a dynamical parameter
$\uplambdab$ as an \emph{$(N,0)$ background}, for a reason that will
become clear later.

To identify the R-matrix $R^{(N,0)}$ for the vector representation of
$\SU(N)$, we consider four-dimensional Chern--Simons theory for
$G = \U(N)$ and split the gauge field into the overall $\U(1)$ part
and the $\SU(N)$ part:
\begin{equation}
  \CA = \CA^{\U(1)} + \CA^{\SU(N)} \,.
\end{equation}
Correspondingly, the dynamical parameter splits as
\begin{equation}
  \uplambda = \uplambda_0 I^* + \uplambdab \,,
  \qquad
  \uplambda_0 = \frac{1}{N} \sum_{i=1}^N \uplambda_i \,,
\end{equation}
where $I^*$ is the dual of the identity matrix.  Since $\CA^{\U(1)}$
and $\CA^{\SU(N)}$ are decoupled in the theory, the total R-matrix is
the product of the R-matrices $R^{\U(1)}$ for the $\U(1)$ part and
$R^{(N,0)}$ for the $\SU(N)$ part:
\begin{equation}
  R^{\U(N)}(z,\uplambda)
  =
  R^{\U(1)}(z, \uplambda_0)
  R^{(N,0)}(z, \uplambdab) \,.
\end{equation}
The $\U(1)$ part is a scalar function of $z$ and $\uplambda_0$, while
the $\SU(N)$ part depends on $z$ and $\uplambdab$.

We know $R^{\U(N)}(z,\uplambda)_{ii}^{ii} = f(z)$ and therefore
$R^{\U(1)}$ is independent of~$\uplambda_0$.  The
formula~\eqref{eq:R-UN} for $R^{\U(N)}$ then implies that
$\varphi \cdot \RF$ is a function of $\uplambdab$ and not of
$\uplambda_0$.  As $\RF$ is independent of $\uplambda_0$, so is
$\varphi$.  Thus, we can write
\begin{equation}
  \label{eq:R(N,0)}
  R^{(N,0)}(z,\uplambdab)
  =
  f^{(N,0)}(z) \xi^{-1} \cdot \RF(-z,\uplambdab) \,,
\end{equation}
where we have expressed the action of $\varphi$ as the gauge
transformation by a diagonal matrix
$\xi^{-1} = \diag(\xi_1^{-1}, \dotsc, \xi_N^{-1})$ of meromorphic
functions of $\uplambdab$.  By considering the monodromies of
$R^{(N,0)}$ as in the $\U(N)$ case, we deduce
\begin{equation}
  \label{eq:fb-period}
  f^{(N,0)}(z + 1) = f^{(N,0)}(z) \,,
  \qquad
  f^{(N,0)}(z + \tau)
  = e^{-2\pi\iu\hbar(N-1)/N} f^{(N,0)}(z)
  \,.
\end{equation}
The unitarity relation requires
\begin{equation}
  \label{eq:unitarity-f(N,0)}
  f^{(N,0)}(z) f^{(N,0)}(-z) = 1 \,.
\end{equation}

In sections~\ref{sec:framing} and \ref{sec:junctions}, we will obtain
more conditions on $f^{(N,0)}$ from considerations on framing anomaly
and junctions of Wilson lines.

\subsection{Surface operators and nondynamical R-matrices}

Just as an electrically charged particle moving in spacetime creates a
Wilson line, the worldline of a magnetically charged particle is also
a line operator.  This operator is called an \emph{'t Hooft line
  operator} if the particle is a magnetic monopole.  More generally, a
dyon, which carries both electric and magnetic charges, creates a
Wilson--'t Hooft line operator~\cite{Kapustin:2005py}.

Suppose that in addition to Wilson lines, we have 't Hooft lines lying
in $\Sigma$ and supported at points on $E$ in four-dimensional
Chern--Simons theory.  The inclusion of 't Hooft lines in the path
integral means that the gauge field has a prescribed behavior such
that as the distance from any of these lines tends to zero, the gauge
field approaches the corresponding monopole configuration.

For monopoles to originate from $Q_V$-invariant configurations in six
dimensions and have a classical interpretation as a particle, their
field configurations, away from the points at which they are located,
should be solutions of the semistability condition~\eqref{eq:stab} and the
equations of motion~\eqref{eq:EOM}.  Although we analyzed these
equations in section \ref{sec:4dCS-from-6d}, there we assumed that all
fields were nonsingular, which may not be the case in the presence of
monopoles.  We have to reexamine the analysis to incorporate possible
singularities.

Let $U$ be the union of small disks in $E$, each centered at the
location of a monopole where some fields may become singular.
Performing integration by parts on the integral~\eqref{eq:int-locus}
as before, but this time taking $C = E \setminus U$, we find that this
integral equals the bulk integral~\eqref{eq:int-locus-bulk} plus the
boundary term
\begin{equation}
  \label{eq:bint-delC}
  -\int_{\Sigma \times \del C} \! \sqrt{g_\Sigma} \, \rmd^2x
  \Tr\bigl(2 \phi^m (F_{mz} \rmd z + F_{m\zb} \rmd\zb)\bigr)
  \,.
\end{equation}
For solutions of the equations of motion, this term equals
\begin{equation}
  \label{eq:bint-delC-2}
  \iu
  \int_{\Sigma \times \del C} \! \sqrt{g_\Sigma} \, \rmd^2x
  (\rmd z \del_z - \rmd\zb \del_\zb) \Tr(\phi^m \phi_m)
  =
  -\int_{\Sigma \times \del U} \! \sqrt{g_\Sigma} \, \rmd^2x
  \, \rmd\theta \, r\del_r
  \Tr(\phi^m \phi_m) \,,
\end{equation}
where $(r,\theta)$ are the polar coordinates around the monopoles
(defined by $2\zb = r e^{\iu\theta}$; recall the definition
\eqref{eq:z} of $z$).

We know that $\phi$ is constant in the absence of monopoles, so
$-\Tr(\phi^m \phi_m)$ should decay to a constant as $r$ increases.
Then the boundary term is nonpositive.  There are two possibilities:
either the boundary term remains nonzero as we send the radii of the
disks to zero, or it vanishes in this limit.

In the former case, the previous argument based on the positivity of
the terms in the integrand fails.  As a result, the characterization
of semistable solutions of the equations of motion is altered, a
complication we want to avoid.  We will not pursue this possibility in
this paper.

Therefore we consider the latter possibility.  The previous argument
then goes through, and the semistable solutions are still parametrized
by the same data as in the case with no monopoles, as long as we stay
away from singularities.  In particular, the curvature of the gauge
field vanishes everywhere except at the points on $E$ where the 't
Hooft lines are placed.  Such tightly confined magnetic fluxes are
familiar: they are Dirac strings attached to the monopoles.

\begin{figure}
  \centering
  \subfloat[]{
    \begin{tikzpicture}[align at bottom]
      \draw[-, very thick, color=midnightblue!50] (0,0) -- (2,0);
      \fill[midnightblue!80] (0,0) circle[radius=3pt];
 
      \draw[->] (1.5,0.07)
      arc [start angle=20, end angle=340, x radius=4pt, y radius=6pt];

      \node at (1.5,0.4) {$\theta$};

      \node (x) at (0,-0.2) {};
   \end{tikzpicture}
  }
  \qquad
  \subfloat[]{
    \begin{tikzpicture}[align at bottom]
      \begin{scope}
        \clip (0,0) to[out=120, in=-70] (0,2) --
        (1.5,2) -- (1.5,0) -- (0,0);
        
        \fill[dshaded] (-1,0) rectangle (3,2);
      \end{scope}
      
      \draw[z->] (0,0) to[out=120, in=-70] (0,2);
      \fill[midnightblue!80] (0,0) circle[radius=3pt];
    \end{tikzpicture}
  }
  \qquad
  \subfloat[\label{fig:monopole-antimonopole}]{
    \begin{tikzpicture}[align at bottom]
      \begin{scope}
        \fill[dshaded] (0,0) rectangle (1.5,2);
      \end{scope}
      
      \draw[z->] (0,0) -- (0,2);
      \draw[z->] (1.5,2) -- (1.5,0);
      
      \fill[midnightblue!80] (0,0) circle[radius=3pt];
      \fill[midnightblue!80] (1.5,2) circle[radius=3pt];
    \end{tikzpicture}
  }
  \caption{(a) A Dirac string emanating from a monopole.  (b) The
    motion of the monopole creates an 't Hooft operator bounding a
    surface operator.  (c) A surface operator formed by a Dirac string
    stretched between a monopole--antimonopole pair.  Here the
    antimonopole is represented as a monopole moving in the reverse
    direction.}
  \label{fig:monople}
\end{figure}
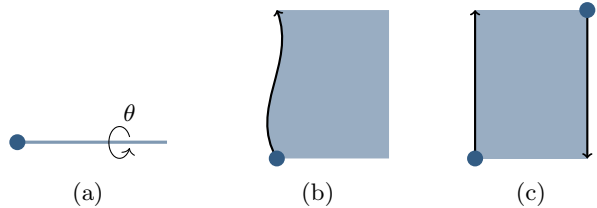

As the monopoles move, their Dirac strings sweep out surfaces.  Hence,
these 't Hooft lines are really the boundaries of \emph{surface
  operators}.  Since the spacetime is compact in the present setup,
every Dirac string emanating from a monopole must be eventually
absorbed by other monopoles.  For example, a Dirac string may be
suspended between a pair of monopoles with opposite charges.  The
introduction of 't Hooft lines thus divides $\Sigma$ into distinct
regions supporting various surface operators.  See
Figure~\ref{fig:monople} for illustrations.

The signature of a confined magnetic flux is the Aharonov--Bohm
effect, the phase shift in the wavefunction as an electrically charged
particle travels around the flux.  Near the location of a Dirac string
in $E$, the gauge field behaves as
\begin{equation}
  A = \iu \alpha \, \rmd\theta + \dotsb \,,
\end{equation} 
where $\iu \alpha \in \tf$ and the ellipsis refers to terms less
singular than $1/r$ as $r \to 0$.  The gauge transformation by
$g = \exp(\iu u \theta)$, with $2\pi\iu u \in \ker \exp|_\tf$, shifts
$\alpha$ by $u$, so the singular behavior of the gauge field is
characterized by the holonomy $\exp(2\pi\iu\alpha)$ around the
singularity.  Surface operators that induce nontrivial monodromies in
fields are often called Gukov--Witten surface
operators~\cite{Gukov:2006jk}.

In the familiar story of monopoles, one requires this monodromy to be
the identity so that the Dirac sting is unobservable, and this leads
to the quantization of monopole charges.  Here, the quantization
condition needs not be satisfied.  If the monodromy is nontrivial, the
Dirac string is physical, hence so is the surface operator it creates.
In that case the 't Hooft line is not a genuine line operator as it
cannot exist by itself without having to bound a physical surface
operator.

To better understand these surface operators, consider first a
situation in which none of them are present in the system.  Part of
the data specifying a semistable solution of the equations of motion is a
flat $G$-bundle over $E$.  Such a bundle is characterized by the
holonomies $a$, $b$ of the gauge field around the one-cycles $\CC_a$,
$\CC_b$ of $E$.  They satisfy the relation
\begin{equation}
  aba^{-1} b^{-1} = e \,,
\end{equation}
where $e$ is the identity element of $G$.

\begin{figure}
  \centering
  \begin{tikzpicture}[scale=1.2]
    \fill[ws] (0,0) rectangle (3, 2);
    
    \fill[dshaded] (0,0) rectangle (0.25,2);
    \fill[dshaded] (0.75,0) rectangle (1.25,2);
    \fill[dshaded] (1.75,0) rectangle (2.25,2);
    \fill[dshaded] (2.75,0) rectangle (3,2);

    \begin{scope}[shift={(0.5,0)}]
      \draw[dr->] (0,0) -- (0,2);
      \draw[dr->] (1,0) -- (1,2);
      \draw[dr->] (2,0) -- (2,2);
    \end{scope}
    
    \begin{scope}[shift={(0,0.5)}]
      \draw[dr->] (0,0) -- (3,0);
      \draw[dr->] (0,1) -- (3,1);
    \end{scope}
      
    \begin{scope}[shift={(0.25,0)}]
      \draw[z->] (0,2) -- (0,0);
      \draw[z->] (0.5,0) -- (0.5,2);
      \draw[z->] (1,2) -- (1,0);
      \draw[z->] (1.5,0) -- (1.5,2);
      \draw[z->] (2,2) -- (2,0);
      \draw[z->] (2.5,0) -- (2.5,2);
    \end{scope}
    
    \draw[frame] (0,0) rectangle (3,2);
  \end{tikzpicture}
  \caption{A lattice of Wilson lines in the presence of surface
    operators.}
  \label{fig:lattice-so}
\end{figure}
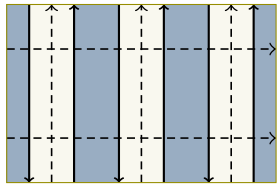

Now suppose that we put surface operators at a point $p \in E$,
covering some region of~$\Sigma$ whose boundaries extend in the
$y$-direction, as in Figure~\ref{fig:lattice-so}.  As a result of the
introduction of the surface operators, the holonomies are modified in
this region, where instead of the above relation they obey
\begin{equation}
  \label{eq:ababalpha}
  a b a^{-1} b^{-1} = \exp(2\pi\iu\alpha) \,.
\end{equation}

While a pair $(a,b)$ satisfying this modified relation corresponds to
a flat $G$-bundle over $E \setminus \{p\}$, this bundle cannot be
extended to a flat $G$-bundle over all of $E$.  The right-hand side
becomes the identity only if we project the equation to the quotient
of $G$ by a normal subgroup $N$ containing $\exp(2\pi\iu\alpha)$.
This means that $(a,b)$ still describes a flat bundle over~$E$ only if
the structure group can be reduced to $G/N$.  The surface operators
thus modify the gauge bundle in a rather drastic way.

This modification of the gauge bundle is of a special
kind~\cite{Kapustin:2006pk}.  The surface operators map a solution of
the equations of motion to another solution.  In particular, we have
\begin{equation}
  \CF_{x\zb} = F_{x\zb} - \iu D_\zb\phi_x = 0
\end{equation}
throughout $\Sigma$, provided that we are away from $p \in E$.  In a
gauge in which $A_x = 0$, this equation reads
\begin{equation}
  \del_x A_\zb = \iu D_\zb\phi_x \,.
\end{equation}
This shows that along the $x$-direction $A_\zb$ varies by gauge
transformations, and the holomorphic structure defined by $A_\zb$
remains unchanged.  Therefore, the holomorphic vector bundles
associated to a unitary representation of $G$, before and after the
modification, are isomorphic on $E \setminus \{p\}$.

If the normal subgroup $N$ acts trivially in the chosen
representation, the modified bundle can be extended to $E$ as a
$(G/N)_\C$-bundle.  In this situation the surface operators modify a
holomorphic $(G/N)_\C$-bundle over $E$ to another holomorphic
$(G/N)_\C$-bundle over~$E$, which is isomorphic to the original one on
$E \setminus \{p\}$.  Such a modification of a holomorphic vector
bundle over a Riemann surface is known as a \emph{Hecke modification}.

In relation to integrable lattice models, the case of particular
interest is when $G = \SU(N)$ and
\begin{equation}
  \label{eq:m-charge}
  \alpha
  =
  \diag\biggl(1 - \frac{1}{N},
  -\frac{1}{N}, \dotsc, -\frac{1}{N}\biggr) \,,
\end{equation}
or any choice of $\alpha$ related to this one by a permutation of the
diagonal entries.%
\footnote{In the rest of this section we only consider 't Hooft lines
  and surface operators whose charges have $\pm 1 - 1/N$ in the first
  entry.  The distinction between surface operators with charges
  related by permutations is meaningful in a generic background.}
In this case the monodromies are represented by
$N \times N$ matrices $A$, $B$ with determinant $1$, satisfying the
relation
\begin{equation}
  \label{eq:ABA-1B-1}
  ABA^{-1} B^{-1} = e^{-2\pi\iu/N} I \,.
\end{equation}
The right-hand side of this relation is a generator of the center
$\Z_N$ of $\SU(N)$.  Thus, the pair $(A, B)$ defines a flat vector
bundle over $E$ with structure group $\PSU(N) = \SU(N)/\Z_N$.

The reason this surface operator is interesting is that
relation~\eqref{eq:ABA-1B-1} determines $(A,B)$ uniquely up to gauge
transformation: we can take them to be the matrices defined by
\begin{equation}
  A e_k = e^{\pi\iu(N - 1)/N} e^{-2\pi\iu k/N} e_k
  \,,
  \qquad
  B e_k = e_{k+1}
  \,.
\end{equation}
In other words, the flat $\PSU(N)$-bundle over $E$ has no moduli.
Consequently, the R-matrix arising from the crossing of a pair of
Wilson lines in this background has no dynamical parameter, and
satisfies the ordinary Yang--Baxter equation
\begin{equation}
  \label{eq:ndYBE}
  \RM_{\alpha\beta}(z_\alpha - z_\beta)
  \RM_{\alpha\gamma}(z_\alpha - z_\gamma) 
  \RM_{\beta\gamma}(z_\beta - z_\gamma) 
  =
  \RM_{\beta\gamma}(z_\beta - z_\gamma) 
  \RM_{\alpha\gamma}(z_\alpha - z_\gamma) 
  \RM_{\alpha\beta}(z_\alpha - z_\beta)
  \,.
\end{equation}
With respect to holomorphic frames on the holomorphic vector bundles
associated to the representations of the Wilson lines, the R-matrix is
represented by a matrix of meromorphic function on $E$.

Although the form of the R-matrix generally depends on the location
$p \in E$ of the surface operator in which the Wilson lines are
placed, for a suitable choice of holomorphic frames this dependence
disappears.  (We implicitly assumed that such a choice was made when
we wrote down the Yang--Baxter equation above.)  This is because if we
change the location of the surface operator from $p$ to $p'$, the
associated bundles over $E \setminus \{p\}$ change to new bundles over
$E \setminus \{p'\}$, but the two sets of bundles are isomorphic on
$E \setminus \{p,p'\}$ since they are both isomorphic there to the set
of bundles we originally had before the introduction of the surface
operator.  It follows that there exists a choice of holomorphic frames
on the relevant bundles with respect to which the form of the R-matrix
remains unchanged under the shift in the location, at all point in
$E \setminus \{p, p'\}$, hence on the whole~$E$.

Let us call the field configuration for this surface operator the
\emph{$(N,1)$ background}, and let $R^{(N,1)}$ denote the R-matrix for
the crossing of two Wilson lines in the vector representation in this
background:
\begin{equation}
  \label{eq:R(N,1)}
  R^{(N,1)}(z_1 - z_2)
  =
  \begin{tikzpicture}[scale=0.6, baseline=(x.base)]
    \node (x) at (1,1) {\vphantom{x}};

    \fill[dshaded] (0,0) rectangle (2,2);
    \draw[dr->] (0,1) node[left] {$z_1$} --  (2,1);
    \draw[dr->] (1,0) node[below] {$z_2$} -- (1,2);
  \end{tikzpicture}
  \ .
\end{equation}
The associated bundle over $E \setminus \{p\}$ has
holomorphic sections
\begin{equation}
  \st_i(z) = P\exp\biggl(-\int_0^z A\biggr) \st_i(0) \,.
\end{equation}
(We have to be a little careful about the choice of the counter for
the integral in the exponent because of the singularity of $A$.)  With
respect to the holomorphic frame $(\st_i)_{i=1}^N$, we have
$A_\zb = 0$ identically and the dependence on the location of the
surface operator disappears.  In this frame $R^{(N,1)}$ is an
$\End(\C^N \otimes \C^N)$-valued meromorphic function with the
quasi-periodicity property
\begin{equation}
  \begin{aligned}
    R^{(N,1)}(z + 1)
    &= A_1 R^{(N,1)}(z) A_1^{-1}
    = A_2^{-1} R^{(N,1)}(z) A_2
    \,,
    \\
    R^{(N,1)}(z + \tau)
    &=
    B_1 R^{(N,1)}(z) B_1^{-1}
    = B_2^{-1} R^{(N,1)}(z) B_2
    \,.
  \end{aligned}
\end{equation}
We have introduced the notation $X_1 = X \otimes I$ and
$X_2 = I \otimes X$ for $X \in \End(\C^N)$.

There is a well-known R-matrix which almost has the same
quasi-periodicity.  The Baxter--Belavin R-matrix
$\RB$~\cite{Baxter:1971cr, Baxter:1972hz, Belavin:1981ix} is a unitary
solution of the Yang--Baxter equation~\eqref{eq:ndYBE} satisfying the
relations
\begin{equation}
  \begin{aligned}
    \RB(z + 1)
    &= A_1 \RB(z) A_1^{-1}
    = A_2^{-1} \RB(z) A_2
    \,,
    \\
    \RB(z + \tau)
    &=
    e^{2\pi\iu\hbar (N-1)/N}
    B_1 \RB(z) B_1^{-1}
    =
    e^{2\pi\iu\hbar (N-1)/N}
    B_2^{-1} \RB(z) B_2
    \,.
  \end{aligned}
\end{equation}
It is an $\End(\C^N \otimes \C^N)$-valued meromorphic function whose
matrix elements are given by~\cite{Richey:1986vt}
\begin{equation}
  \RB(z)_{ij}^{kl}  
  =
  \delta_{i+j, k+l}
  \dfrac{\theta_1(\hbar)}{\theta_1(z + \hbar)}
  \dfrac{\theta^{(k-l)}(z + \hbar)}{\theta^{(k - i)}(\hbar)
    \theta^{(i-l)}(z)} 
  \dfrac{\prod_{m=0}^{N-1} \theta^{(m)}(z)}{\prod_{n=1}^{N-1} \theta^{(n)}(0)}
  \,,
\end{equation}
where the indices are understood modulo $N$ and
\begin{equation}
  \theta^{(j)}(z|\tau,N)
  =
  \theta\biggl[
  \begin{array}{c}
    1/2 - j/N \\ 1/2
  \end{array}
  \biggr]
  (z|N\tau)
  \,.
\end{equation}
For $N = 2$, these matrix elements reduce to the local Boltzmann
weights for the eight-vertex model~\cite{Baxter:1971cr,
  Baxter:1972hz}.

Comparing the quasi-periodicity of $R^{(N,1)}$ and $\RB$, it is fairly
natural to identify these two R-matrices:
\begin{equation}
  \label{eq:R(N,1)-RB}
  R^{(N,1)}(z) = f^{(N,1)}(z) \RB(z) \,.
\end{equation}
Here $f^{(N,1)}$ is a function that accounts for the slight
discrepancy in the quasi-periodicity, and satisfies
$f^{(N,1)}(z) f^{(N,1)}(-z) = 1$ so that the unitarity is preserved.
In fact, as explained in~\cite{Costello:2018gyb}, a theorem proved by
Belavin and Drifeld~\cite{MR674005} on the classification of the
solutions of the classical Yang--Baxter equation ensures that
$R^{(N,1)}$ must be of this form to all orders in $\hbar$, up to
reparametrizations of $\hbar$.

We can also consider the \emph{$(N,-1)$ background} created by the
surface operator with the opposite charge,
\begin{equation}
  \alpha
  =
  \diag\biggl(-1 + \frac{1}{N},
  \frac{1}{N}, \dotsc, \frac{1}{N}\biggr)
  \,,
\end{equation}
and identify the R-matrix $R^{(N,-1)}$ that arises from the crossing
of Wilson lines in this background.  Graphically we distinguish the
$(N,-1)$ background from the $(N,1)$ background by using a different
color:
\begin{equation}
  \label{eq:R(N,-1)}
  R^{(N,-1)}(z_1 - z_2)
  =
  \begin{tikzpicture}[scale=0.6, baseline=(x.base)]
    \node (x) at (1,1) {\vphantom{x}};

    \fill[lshaded] (0,0) rectangle (2,2);
    \draw[dr->] (0,1) node[left] {$z_1$} --  (2,1);
    \draw[dr->] (1,0) node[below] {$z_2$} -- (1,2);
  \end{tikzpicture}
  \ .
\end{equation}

Since $A^{-1} BA B^{-1} = e^{2\pi\iu/N} I$, in an appropriate gauge
this R-matrix should obey the quasi-periodicity relations
\begin{equation}
  \begin{aligned}
    R^{(N,-1)}(z + 1)
    &= A_1^{-1} R^{(N,-1)}(z) A_1
    = A_2 R^{(N,-1)}(z) A_2^{-1}
    \,,
    \\
    R^{(N,-1)}(z + \tau)
    &=
    B_1 R^{(N,-1)}(z) B_1^{-1}
    = B_2^{-1} R^{(N,-1)}(z) B_2
    \,.
  \end{aligned}
\end{equation}
Noting that $A^T = A$ and $B^T = B^{-1}$, we see that
$R^{(N,-1)}$ can be written as
\begin{equation}
  R^{(N,-1)}(z) = f^{(N,-1)}(z) \RB(z)^T \,,
\end{equation}
where $\RB(z)^T$ is the transpose of $\RB(z)$.

We have already encountered a function that has the right properties
to be $f^{(N,1)}$ or $f^{(N,-1)}$: the function $f^{(N,0)}$ which
enters the definition \eqref{eq:R(N,0)} of the dynamical R-matrix
$R^{(N,0)}$.  We will argue in section~\ref{sec:intertwining} that the
three functions are actually equal.

The relation between modification of bundles and that of R-matrices
discussed here had been previously considered in~\cite{Levin:2008em,
  Levin:2012ih}.  In particular, the R-matrices in the presence of
more general surface operators were studied in~\cite{Levin:2012ih}.
In general, the R-matrices depend on $l$~moduli, with
$0 \leq l \leq N-1$.

\subsection{Intertwining operators and vertex--face correspondences}
\label{sec:intertwining}

Once we have new line operators, we can construct new R-matrices.
Especially interesting are the R-matrices that correspond to a Wilson
line crossing an 't Hooft line and moving into a surface operator.
The Yang--Baxter equations involving two Wilson lines and one 't~Hooft
line, such as the ones illustrated in Figure~\ref{fig:VFC}, show that
these R-matrices intertwine the dynamical R-matrix and nondynamical
ones.  This kind of relation between dynamical and nondynamical
R-matrices is known as a \textit{vertex--face
  correspondence}~\cite{Baxter:1972wf, MR908997, Jimbo:1987mu}, for
the two R-matrices may be regarded as the Boltzmann weights for
lattice models of ``face type'' and ``vertex type,'' respectively.

\begin{figure}
  \centering
  \subfloat[\label{fig:VFC+}]{
  \begin{tikzpicture}[scale=0.65, baseline=(x.base)]
    \node (x) at (30:2) {\vphantom{x}};

    \fill[ws] (-30:1) rectangle ++({-sqrt(3)/2},3);
    \fill[dshaded] (-30:1) rectangle ++({sqrt(3)},3);

    \draw[dr->] (0,0) node[left] {$z_2$} -- ++(30:3);
    \draw[dr->] (0,2) node[left] {$z_1$} -- ++(-30:3);
    \draw[z->] (-30:1) node[below] {$w$} -- ++(0,3);

    \node (O) at ({sqrt(3)*4/6},1) {};
    \node at ($(O) + (120:{sqrt(3)*9/12})$) {$\uplambdab$};
  \end{tikzpicture}
  \ =
  \begin{tikzpicture}[scale=0.65, baseline=(x.base)]
    \node (x) at (30:1) {\vphantom{x}};

    \fill[ws] (-30:2) rectangle ++({-sqrt(3)},3);
    \fill[dshaded] (-30:2) rectangle ++({sqrt(3)/2},3);

    \draw[dr->] (0,0) node[left] {$z_2$} -- ++(30:3);
    \draw[dr->] (0,1) node[left] {$z_1$} -- ++(-30:3);
    \draw[z->] (-30:2) node[below] {$w$} -- ++(0,3);

    \node (O) at ({sqrt(3)*5/6},0.5) {};
    \node at ($(O) + (120:{sqrt(3)*7/12})$) {$\uplambdab$};
  \end{tikzpicture}
  }
  \qquad
  \subfloat[\label{fig:VFC-}]{
  \begin{tikzpicture}[scale=0.65, baseline=(x.base)]
    \node (x) at (30:2) {\vphantom{x}};

    \fill[dshaded] (-30:1) rectangle ++({-sqrt(3)/2},3);
    \fill[ws] (-30:1) rectangle ++({sqrt(3)},3);

    \draw[dr->] (0,0) node[left] {$z_2$} -- ++(30:3);
    \draw[dr->] (0,2) node[left] {$z_1$} -- ++(-30:3);
    \draw[z<-] (-30:1) node[below] {$w$} -- ++(0,3);

    \node (O) at ({sqrt(3)*4/6},1) {};
    \node at ($(O) + (60:{sqrt(3)*7/12})$) {$\uplambdab$};
  \end{tikzpicture}
  \ =
  \begin{tikzpicture}[scale=0.65, baseline=(x.base)]
    \node (x) at (30:1) {\vphantom{x}};

    \fill[dshaded] (-30:2) rectangle ++({-sqrt(3)},3);
    \fill[ws] (-30:2) rectangle ++({sqrt(3)/2},3);

    \draw[dr->] (0,0) node[left] {$z_2$} -- ++(30:3);
    \draw[dr->] (0,1) node[left] {$z_1$} -- ++(-30:3);
    \draw[z<-] (-30:2) node[below] {$w$} -- ++(0,3);

    \node (O) at ({sqrt(3)*5/6},0.5) {};
    \node at ($(O) + (60:{sqrt(3)*9/12})$) {$\uplambdab$};
  \end{tikzpicture}
  }
  \caption{Vertex--face correspondences between $R^{(N,0)}$ and
    $R^{(N,1)}$.}
  \label{fig:VFC}
\end{figure}

Let $S$ be the intertwining operator between $R^{(N,0)}$ and
$R^{(N,1)}$ that arises from the the crossing of a Wilson line in the
vector representation and an 't Hooft line of charge
$\diag(1-1/N, -1/N, \dotsc, -1/N)$:
\begin{equation}
  S(z - w, \uplambdab)
  =
  \begin{tikzpicture}[scale=0.6, baseline=(x.base)]
    \node (x) at (1,1) {\vphantom{x}};

    \fill[ws] (0,0) rectangle (1,2);
    \fill[dshaded] (1,0) rectangle (2,2);

    \draw[dr->] (0,1) node[left]{$z$} -- (2,1);
    \draw[z->] (1,0) node[below] {$w$} -- (1,2);

    \node at (0.5,1.5) {$\uplambdab$};
  \end{tikzpicture}
  \ .
\end{equation}
By translation invariance $S$ is a function of the difference of the
spectral parameters of the two lines, which we have written as $z$ and
$w$ here; unlike the location of the bulk of the surface operator,
that of the 't Hooft line is a physical parameter.  It also depends on
the value of the dynamical parameter $\uplambdab$ in the region
adjacent to the 't Hooft line.

With respect to the local holomorphic frames we have been using for
the relevant bundles, $S$ is an $\End(\C^N)$-valued meromorphic
function and satisfies the quasi-periodicity relations
\begin{equation}
  S(z + 1, \uplambdab) = A S(z, \uplambdab)
  \,,
  \qquad
  S(z + \tau, \uplambdab) = B S(z, \uplambdab) \exp(-2\pi\iu \uplambdab)
  \,.
\end{equation}
In perturbation theory, we expect $S(z,\uplambdab)$ to have poles at
$z = 0$ where the Wilson and 't~Hooft lines intersect in the
four-dimensional spacetime.

A matrix $\Phi(z,\uplambdab)$ that has the right quasi-periodicity and
pole structure is given by~\cite{Levin:2008em}
\begin{equation}
  \Phi(z,\uplambdab)^j_i
  =
  \frac{\theta^{(j)}(z + N \uplambdab_i)}
  {\theta_1(z)^{1/N}}
  \,.
\end{equation}
In fact, $\Phi$ is an intertwining operator relating $\RF$ and
$\RB$~\cite{MR908997}:
\begin{multline}
  \label{eq:RBPhiPhi}
  \RB(z_1 - z_2)
  \Phi_1(z_1,\uplambdab)
  \Phi_2(z_2, \uplambdab - \hbar h_1)
  \\
  =
  \Phi_2(z_2,\uplambdab)
  \Phi_1(z_1, \uplambdab - \hbar h_2)
  \Theta^{-1} \cdot
  \RF(z_2 - z_1, \uplambdab)
  \,.
\end{multline}
Here $\Theta = \diag(\Theta_1, \dotsc, \Theta_N)$ is the diagonal
matrix of meromorphic functions with
\begin{equation}
  \Theta_i(\lambda)
  =
  \prod_{j (\neq i)} \theta_1(\lambda_{ij}) \,,
\end{equation}
acting on $\RF$ by the gauge transformation \eqref{eq:gauge-transf}.

Given the expression~\eqref{eq:R(N,0)} for $R^{(N,0)}$, the above
consideration suggests that we have
\begin{equation}
  S(z,\uplambdab)
  =
  \Phi(z + d,\uplambdab)
  (\Theta^{-1} \xi)(\uplambdab)
  \gb(z,\uplambdab)
  \,,
\end{equation}
where $d \in \C$ and $\gb$ is a diagonal matrix of meromorphic
functions on $E \times \tf^*$ that acts trivially on $R^{(N,0)}$.  Let
us further assume that the two functions $f^{(N,0)}$ and $f^{(N,1)}$ in
formulas~\eqref{eq:R(N,0)} and~\eqref{eq:R(N,1)-RB} are equal:
\begin{equation}
  f^{(N,0)} = f^{(N,1)} \,.
\end{equation}
Then, with this form of $S$, the following vertex--face correspondence
holds:
\begin{multline}
  \label{eq:RSS}
  R^{(N,1)}(z_1 - z_2)
  S_1(z_1 - w,\uplambdab)
  S_2(z_2 - w, \uplambdab - \hbar h_1)
  \\
  =
  S_2(z_2 - w,\uplambdab)
  S_1(z_1 - w, \uplambdab - \hbar h_2)
  R^{(N,0)}(z_1 - z_2, \uplambdab)
  \,.
\end{multline}
The two sides of this relation are represented by the diagrams in
Figure~\ref{fig:VFC+}.

A Wilson line coming out of the surface operator produces another
intertwining operator:
\begin{equation}
  S'(z - w, \uplambdab)
  =
  \begin{tikzpicture}[scale=0.6, baseline=(x.base)]
    \node (x) at (1,1) {\vphantom{x}};

    \fill[dshaded] (0,0) rectangle (1,2);
    \fill[ws] (1,0) rectangle (2,2);

    \draw[dr->] (0,1) node[left] {$z$} -- (2,1);;
    \draw[z<-] (1,0) node[below] {$w$} -- (1,2);

    \node at (1.5,1.5) {$\uplambdab$};
  \end{tikzpicture}
  \ .
\end{equation}
It satisfies the relation
\begin{multline}
  \label{eq:RS'S'}
  R^{(N,0)}(z_1 - z_2, \uplambdab)
  S'_2(z_2 - w, \uplambdab - \hbar h_1)
  S'_1(z_1 - w, \uplambdab)
  \\
  =
  S'_1(z_1 - w, \uplambdab - \hbar h_2)
  S'_2(z_2 - w, \uplambdab)
  R^{(N,1)}(z_1 - z_2)
  \,,
\end{multline}
which is the vertex--face correspondence in Figure~\ref{fig:VFC-}.  This
relation suggests that $S'$ is essentially the inverse of $S$.  Hence,
we propose that it can be written as
\begin{equation}
  \label{eq:S'}
  S'(z,\uplambdab)
  =
  \chi(z,\uplambdab)
  S(z + \delta,\uplambdab)^{-1}
  \,,
\end{equation}
where $\delta \in \C$ and $\chi$ is a diagonal matrix of meromorphic
functions that acts trivially on~$R^{(N,0)}$.  We will determine
$\delta$ and $\chi$ in sections~\ref{sec:framing}
and~\ref{sec:junctions}.

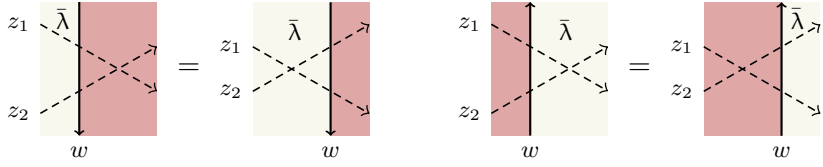
\begin{figure}
  \centering
  \subfloat[]{
  \begin{tikzpicture}[scale=0.65, baseline=(x.base)]
    \node (x) at (30:2) {\vphantom{x}};

    \fill[ws] (-30:1) rectangle ++({-sqrt(3)/2},3);
    \fill[lshaded] (-30:1) rectangle ++({sqrt(3)},3);

    \draw[dr->] (0,0) node[left] {$z_2$} -- ++(30:3);
    \draw[dr->] (0,2) node[left] {$z_1$} -- ++(-30:3);
    \draw[z<-] (-30:1) node[below] {$w$} -- ++(0,3);

    \node (O) at ({sqrt(3)*4/6},1) {};
    \node at ($(O) + (120:{sqrt(3)*9/12})$) {$\uplambdab$};
  \end{tikzpicture}
  \ =
  \begin{tikzpicture}[scale=0.65, baseline=(x.base)]
    \node (x) at (30:1) {\vphantom{x}};

    \fill[ws] (-30:2) rectangle ++({-sqrt(3)},3);
    \fill[lshaded] (-30:2) rectangle ++({sqrt(3)/2},3);

    \draw[dr->] (0,0) node[left] {$z_2$} -- ++(30:3);
    \draw[dr->] (0,1) node[left] {$z_1$} -- ++(-30:3);
    \draw[z<-] (-30:2) node[below] {$w$} -- ++(0,3);

    \node (O) at ({sqrt(3)*5/6},0.5) {};
    \node at ($(O) + (120:{sqrt(3)*7/12})$) {$\uplambdab$};
  \end{tikzpicture}
  }
  \qquad
  \subfloat[]{
  \begin{tikzpicture}[scale=0.65, baseline=(x.base)]
    \node (x) at (30:2) {\vphantom{x}};

    \fill[lshaded] (-30:1) rectangle ++({-sqrt(3)/2},3);
    \fill[ws] (-30:1) rectangle ++({sqrt(3)},3);

    \draw[dr->] (0,0) node[left] {$z_2$} -- ++(30:3);
    \draw[dr->] (0,2) node[left] {$z_1$} -- ++(-30:3);
    \draw[z->] (-30:1) node[below] {$w$} -- ++(0,3);

    \node (O) at ({sqrt(3)*4/6},1) {};
    \node at ($(O) + (60:{sqrt(3)*7/12})$) {$\uplambdab$};
  \end{tikzpicture}
  \ =
  \begin{tikzpicture}[scale=0.65, baseline=(x.base)]
    \node (x) at (30:1) {\vphantom{x}};

    \fill[lshaded] (-30:2) rectangle ++({-sqrt(3)},3);
    \fill[ws] (-30:2) rectangle ++({sqrt(3)/2},3);

    \draw[dr->] (0,0) node[left] {$z_2$} -- ++(30:3);
    \draw[dr->] (0,1) node[left] {$z_1$} -- ++(-30:3);
    \draw[z->] (-30:2) node[below] {$w$} -- ++(0,3);

    \node (O) at ({sqrt(3)*5/6},0.5) {};
    \node at ($(O) + (60:{sqrt(3)*9/12})$) {$\uplambdab$};
  \end{tikzpicture}
  }
  \caption{Vertex--face correspondences between $R^{(N,0)}$ and
    $R^{(N,-1)}$.}
  \label{fig:VFC-1}
\end{figure}

The intertwining operators involving the $(N,-1)$ background can be
identified in a similar manner.  Let us write
\begin{equation}
  \St(z - w, \uplambdab)
  =
  \begin{tikzpicture}[scale=0.6, baseline=(x.base)]
    \node (x) at (1,1) {\vphantom{x}};

    \fill[ws] (0,0) rectangle (1,2);
    \fill[lshaded] (1,0) rectangle (2,2);

    \draw[dr->] (0,1) node[left]{$z$} -- (2,1);
    \draw[z<-] (1,0) node[below] {$w$} -- (1,2);

    \node at (0.5,1.5) {$\uplambdab$};
  \end{tikzpicture}
  \ ,
  \qquad
  \St'(z - w, \uplambdab)
  =
  \begin{tikzpicture}[scale=0.6, baseline=(x.base)]
    \node (x) at (1,1) {\vphantom{x}};

    \fill[lshaded] (0,0) rectangle (1,2);
    \fill[ws] (1,0) rectangle (2,2);

    \draw[dr->] (0,1) node[left] {$z$} -- (2,1);;
    \draw[z->] (1,0) node[below] {$w$} -- (1,2);

    \node at (1.5,1.5) {$\uplambdab$};
  \end{tikzpicture}
\end{equation}
and define a matrix $\Phit(z,\uplambdab)$ by
\begin{equation}
  \Phit(z,\uplambdab)_i^j
  = \Phi(z,-\uplambdab + \hbar\upomegab_j)_j^i
  \,.
\end{equation}
If we assume
\begin{equation}
  f^{(N,0)} = f^{(N,-1)}
\end{equation}
and
\begin{equation}
  \St(z,\uplambdab)
  =
  \St'(z + \deltat,\uplambdab)^{-1} \chit(z,\uplambdab)
  \,,
  \qquad
  \St'(z,\uplambdab)
  =
  \gt(z,\uplambdab) \xi^{-1}(\uplambdab) \Phit(z+\dt,\uplambdab)
\end{equation}
for some diagonal matrices $\gt(z,\uplambdab)$ and
$\chit(z,\uplambdab)$ acting trivially on $R^{(N,0)}$, then using
the identities
\begin{equation}
  \RF(z,-\uplambdab)^T = \Theta^{-1} \cdot \RF(z,\uplambdab) \,,
  \qquad
  \RF\bigl(z,\uplambdab + \hbar(h_1+h_2)\bigr)
  = \RF(z,\uplambdab)
\end{equation}
we can verify that $\St$ and $\St'$ furnish the vertex--face
correspondences between $R^{(N,0)}$ and $R^{(N,-1)}$, shown in
Figure~\ref{fig:VFC-1}.

\subsection{L-operators}

Now consider a surface operator stretched between two antiparallel 't
Hooft lines, and a Wilson line traversing it.  This configuration
defines an \emph{L-operator} $L^{(N,0)}$:
\begin{equation}
  \label{eq:L(N,0)}
  L^{(N,0)}(z - w, z - w')
  =
  \begin{tikzpicture}[xscale=0.6, yscale=0.6, baseline=(x.base)]
    \node (x) at (0,1) {\vphantom{x}};

    \fill[ws] (0,0) rectangle (2,2);
    \fill[dshaded] (2/3,2) rectangle (4/3,0);

    \draw[dr->] (0,1) node[left] {$z$} -- (2,1);
    \draw[z->] (2/3,0) node[below] {$w$}-- (2/3,2);
    \draw[z->] (4/3,2)  -- (4/3,0) node[below] {$w'$};
  \end{tikzpicture}
  \ .
\end{equation}
The Wilson line shifts the dynamical parameters on the two sides of
the surface operator by amounts depending on the states on the left
and right edges.  Hence, we may think of $L^{(N,0)}$ as a matrix whose
entries are difference operators.

More precisely, we define the matrix element $L^{(N,0)}(z - w, z - w')^j_i$
to be a difference operator acting on a Weyl-invariant meromorphic
function $f$ on $\tf_\C^* \times \tf_\C^*$ as
\begin{equation}
  L^{(N,0)}(z - w, z - w')^j_i f(\uplambdab, \upmub)
  =
  S'(z - w, \upmub)^j_k S(z - w', \uplambdab)^k_i
  f(\uplambdab - \hbar\upomegab_i, \upmub - \hbar\upomegab_j)
  \ .
\end{equation}
Then, the vertex--face correspondences \eqref{eq:RSS} and
\eqref{eq:RS'S'} imply that $L^{(N,0)}$ satisfies the following
\emph{RLL relation} with $R^{(N,0)}$:
\begin{multline}
  \label{eq:RLL(N,0)}
  R^{(N,0)}(z_1 - z_2, \upmub)
  L^{(N,0)}_1(z_1 - w, z_1 - w')
  L^{(N,0)}_2(z_2 - w, z_2 - w')
  \\
  =
  \mathopen{:}
  L^{(N,0)}_2(z_2 - w, z_2 - w')
  L^{(N,0)}_1(z_1 - w, z_1 - w')
  R^{(N,0)}(z_1 - z_2, \uplambdab)
  \mathclose{:}
  \,.
\end{multline}
The normal ordering sign $\mathopen{:} \ \mathclose{:}$ means that the
matrix elements of $R^{(N,0)}$ should be placed in the leftmost
position so as not to be acted on by the L-operators.  This relation
is depicted in Figure~\ref{fig:RLL(N,0)}.

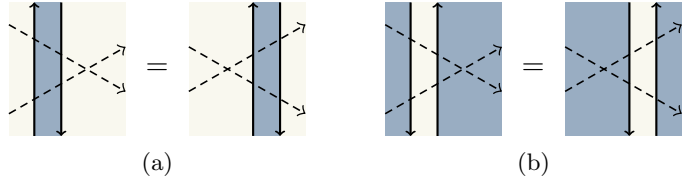
\begin{figure}
  \centering
  \subfloat[\label{fig:RLL(N,0)}]{
  \begin{tikzpicture}[scale=0.65, baseline=(x.base)]
    \node (x) at (30:2) {\vphantom{x}};

    \fill[ws] (-30:1) rectangle ++({-sqrt(3)/2},3);
    \fill[ws] (-30:1) rectangle ++({sqrt(3)},3);
    \fill[dshaded, shift={(-0.3,0)}] (-30:1) rectangle ++(0.6,3);

    \draw[dr->] (0,0) -- ++(30:3);
    \draw[dr->] (0,2) -- ++(-30:3);
    \draw[z->, shift={(-0.3,0)}] (-30:1) -- ++(0,3);
    \draw[z<-, shift={(0.3,0)}] (-30:1) -- ++(0,3);

    \node (O) at ({sqrt(3)*4/6},1) {};
  \end{tikzpicture}
  \ = \
  \begin{tikzpicture}[scale=0.65, baseline=(x.base)]
    \node (x) at (30:1) {\vphantom{x}};

    \fill[ws] (-30:2) rectangle ++({-sqrt(3)},3);
    \fill[ws] (-30:2) rectangle ++({sqrt(3)/2},3);
    \fill[dshaded, shift={(-0.3,0)}] (-30:2) rectangle ++(0.6,3);

    \draw[dr->] (0,0) -- ++(30:3);
    \draw[dr->] (0,1) -- ++(-30:3);
    \draw[z->, shift={(-0.3,0)}] (-30:2) -- ++(0,3);
    \draw[z<-, shift={(0.3,0)}] (-30:2) -- ++(0,3);

    \node (O) at ({sqrt(3)*5/6},0.5) {};
  \end{tikzpicture}
  }
  \qquad
  \subfloat[\label{fig:RLL(N,1)}]{
  \begin{tikzpicture}[scale=0.65, baseline=(x.base)]
    \node (x) at (30:2) {\vphantom{x}};

    \fill[dshaded] (-30:1) rectangle ++({-sqrt(3)/2},3);
    \fill[dshaded] (-30:1) rectangle ++({sqrt(3)},3);
    \fill[ws,shift={(-0.3,0)}] (-30:1) rectangle ++(0.6,3);

    \draw[dr->] (0,0) -- ++(30:3);
    \draw[dr->] (0,2) -- ++(-30:3);
    \draw[z<-, shift={(-0.3,0)}] (-30:1) -- ++(0,3);
    \draw[z->, shift={(0.3,0)}] (-30:1) -- ++(0,3);

    \node (O) at ({sqrt(3)*4/6},1) {};
  \end{tikzpicture}
  \ = \ 
  \begin{tikzpicture}[scale=0.65, baseline=(x.base)]
    \node (x) at (30:1) {\vphantom{x}};

    \fill[dshaded] (-30:2) rectangle ++({-sqrt(3)},3);
    \fill[dshaded] (-30:2) rectangle ++({sqrt(3)/2},3);
    \fill[ws, shift={(-0.3,0)}] (-30:2) rectangle ++(0.6,3);

    \draw[dr->] (0,0) -- ++(30:3);
    \draw[dr->] (0,1) -- ++(-30:3);
    \draw[z<-, shift={(-0.3,0)}] (-30:2) -- ++(0,3);
    \draw[z->, shift={(0.3,0)}] (-30:2) -- ++(0,3);

    \node (O) at ({sqrt(3)*5/6},0.5) {};
  \end{tikzpicture}
  }
  \caption{RLL relations for (a) $R^{(N,0)}$ and (b) $R^{(N,1)}$.}
  \label{fig:RLL}
\end{figure}

Interchanging the intertwining operators we get another L-operator:
\begin{equation}
  \label{eq:L(N,1)}
  L^{(N,1)}(z - w, z - w')
  =
  \begin{tikzpicture}[xscale=0.6, yscale=0.6, baseline=(x.base)]
    \node (x) at (0,1) {\vphantom{x}};

    \fill[dshaded] (0,0) rectangle (2,2);
    \fill[ws] (2/3,2) rectangle (4/3,0);

    \draw[dr->] (0,1) node[left] {$z$} -- (2,1);
    \draw[z<-] (2/3,0) node[below] {$w$} -- (2/3,2);
    \draw[z<-] (4/3,2) -- (4/3,0) node[below] {$w'$};
  \end{tikzpicture}
  \,.
\end{equation}
This is a matrix of difference operators acting on Weyl-invariant
meromorphic functions on $\tf_\C^*$ by
\begin{equation}
  \label{eq:L(N,1)-SS'}
  L^{(N,1)}(z - w, z - w')^j_i f(\uplambdab)
  =
  S(z - w, \uplambdab)^j_k S'(z - w', \uplambdab)^k_i
  f(\uplambdab - \hbar\upomegab_k)
  \,.
\end{equation}
It satisfies the RLL relation
\begin{multline}
  \label{eq:RLL(N,1)}
  R^{(N,1)}(z_1 - z_2)
  L^{(N,1)}_1(z_1 - w, z_1 - w')
  L^{(N,1)}_2(z_2 - w, z_2 - w')
  \\
  =
  L^{(N,1)}_2(z_2 - w, z_2 - w')
  L^{(N,1)}_1(z_1 - w, z_1 - w')
  R^{(N,1)}(z_1 - z_2)
  \,,
\end{multline}
which is the relation shown in Figure~\ref{fig:RLL(N,1)}.

A good way to think about the L-operators is that they are R-matrices
associated with the crossings of Wilson lines and ``thick'' line
operators, where the latter are composed of pairs of antiparallel 't
Hooft lines and carry infinite-dimensional representations.  For
example, $L^{(N,1)}$ is an R-matrix whose vertical line carries an
infinite-dimensional representation on the space of Weyl-invariant
meromorphic functions on $\tf_\C^*$.

Being constructed from the same intertwining operators, the two
L-operators $L^{(N,0)}$ and $L^{(N,1)}$ lead to the same transfer
matrix:
\begin{equation}
  \Tr_{\C^N}\Bigl(L^{(N,0)}_k \dotsm L^{(N,0)}_1\Bigr)
  =
  \Tr_{\C^N}\Bigl(L^{(N,1)}_k \dotsm L^{(N,1)}_1\Bigr)
  \,.
\end{equation}
This is a difference operator acting on the space of Weyl-invariant
meromorphic functions on $(\tf_\C^*)^{\otimes k}$.  By considering
Wilson lines in various representations, we get a number of such
difference operators which commute with each other.  For $k = 1$,
these difference operators are~\cite{MR1463830} the conserved charges
of the elliptic Ruijsenaars--Schneider model of type
$A_{N-1}$~\cite{Ruijsenaars:1986pp}.

The RLL relation \eqref{eq:RLL(N,1)} is, roughly speaking, the
defining relation for the elliptic quantum algebra
$\CA_{q,p}(\widehat{\slf}_N)$~\cite{Foda:1994xm, Fan:1997qd,
  Avan:1997cu} at level zero, with
$(q,p) = (e^{2\pi\iu\hbar}, e^{2\pi\iu\tau})$.  (It should be
supplemented with the relation that sets the quantum determinant of
the L-operator to~$1$.) The algebra $\CA_{q,p}(\widehat{\slf}_N)$ is
generated by the matrix elements of the L-operator, and is the
elliptic counterpart of the Yangian double $\cD Y_\hbar(\slf_N)$ and
the quantum affine algebra $U_q(\widehat{\slf}_N)$.  The coalgebra
structure making $\CA_{q,p}(\widehat{\slf}_N)$ a quantum group was
given in~\cite{Jimbo:1999zz}.

If it is further required that the dependence of the L-operator on the
spectral parameter takes a certain special form, the RLL relation
encodes the defining relations for the $\Z_N$ Sklyanin
algebra~\cite{Sklyanin:1982tf, MR783721}.  This is a two-parameter
deformation of the universal enveloping algebra $U(\slf_N)$ of
$\slf_N$, and reduces to the quantum group $U_q(\slf_N)$ in the limit
$\tau \to \iu\infty$~\cite{MR1025214, MR1138880, Hou:1991kx}.
Essentially, our L-operator $L^{(N,1)}$ gives an infinite-dimensional
representation of the $\Z_N$ Sklyanin algebra in terms of difference
operators~\cite{Sklyanin:1983ig, MR1247840, MR1230795}.  For $N = 2$,
this representation corresponds to a Verma module of $\slf_2$ whose
highest weight is determined by the difference $w - w'$ of the
spectral parameters of the two 't Hooft lines~\cite{Sklyanin:1982tf}.

In a similar way, the other L-operator $L^{(N,0)}$
provides~\cite{Yagi:2017hmj} an infinite-dimensional representation of
Felder's elliptic quantum group
$E_{q,p}(\slf_N)$~~\cite{Felder:1994pb, Felder:1994be}.  Alternative
formulations of (a central extension of) $E_{q,p}(\slf_N)$ are
discussed in~\cite{Jimbo:1999zz, MR1637805, Jimbo:1998bi,
  Konno:2016fmh}.

\subsection{Framing anomaly}
\label{sec:framing}

Up until now we have discussed four-dimensional Chern--Simons theory
on $\Sigma \times C$ assuming it is perfectly topological on $\Sigma$,
as suggested by the form of the action which makes no reference to a
metric on $\Sigma$.  As a matter of fact, this assumption is a little
too naive.  When it comes to actually performing the path integral,
one needs to introduce a metric on $\Sigma$ for gauge fixing and
regularization.  The introduction of metric can potentially spoil the
topological invariance.  This is indeed what happens, but in a
somewhat subtle manner.

A manifestation of this quantum anomaly is the fact that the equation
that seemingly represents the equivalence between two diagrams, shown
in Figure~\ref{fig:c-unitarity-(N,1)}, does not quite hold:
\begin{equation}
  R^{(N,1)}(z_1 - z_2)_{kj}^{ml}
  R^{(N,1)}(z_2 - z_1)_{li}^{nk}
  \neq
  \delta_i^m \delta_j^n
  \,.
\end{equation}
Instead, $R^{(N,1)}$ satisfies the \emph{crossing--unitarity relation}
\cite{Richey:1986vt}
\begin{equation}
  \label{eq:c-unitarity-(N,1)}
  R^{(N,1)}\biggl(z - \frac{1}{2} N\hbar\biggr)_{kj}^{ml}
  R^{(N,1)}\biggl(-z - \frac{1}{2} N\hbar\biggr)_{li}^{nk}
  =
  \delta_i^m \delta_j^n
  \,,
\end{equation}
provided that we have
\begin{equation}
  \label{eq:fb-fb}
  f^{(N,0)}(z) f^{(N,0)}(-z - N\hbar)
  =
  \frac{\theta_1(z + \hbar) \theta_1(z + (N-1)\hbar)}
       {\theta_1(z) \theta_1(z + N\hbar)}
  \,.
\end{equation}
Somehow the arguments of the R-matrices used in this relation have to
be shifted by $-N\hbar/2$ compared to the ordinary unitarity
relation.

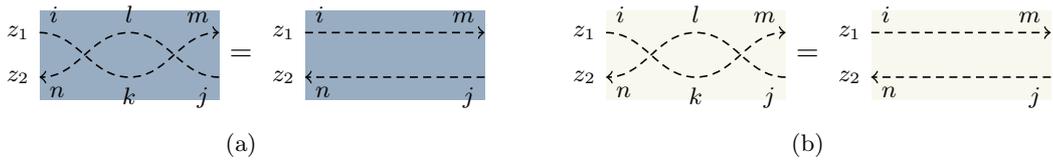
\begin{figure}
  \centering
  \def\pathA{(2,0) to[out=180, in=0] (1,0.5) to[out=180, in=0](0,0)}
  \def\pathB{(0,0.5) to[in=180, out=0] (1,0) to[in=180,out=0] (2,0.5)}
  \subfloat[\label{fig:c-unitarity-(N,1)}]{
    \begin{tikzpicture}[scale=1.2, baseline=(x.base)]
      \node (x) at (1,0.25) {\vphantom{x}};
    
      \fill[dshaded] (0,-0.25) rectangle (2,0.75);
      
      \draw[dr->] \pathA;
      \draw[dr->] \pathB;

      \node[left] at (0,0.5) {$z_1$};
      \node[left] at (0,0) {$z_2$};

      \node[above right] at (0,0.5) {$i$};
      \node[above left] at (2,0.5) {$m$};

      \node[below right] at (0,0) {$n$};
      \node[below left] at (2,0) {$j$};

      \node[below] at (1,0) {$k$};
      \node[above] at (1,0.5) {$l$};
    \end{tikzpicture}
    =
    \begin{tikzpicture}[scale=1.2, baseline=(x.base)]
      \node (x) at (1,0.25) {\vphantom{x}};
      
      \fill[dshaded] (0,-0.25) rectangle (2,0.75);
      
      \draw[dr->] (2,0) -- (0,0);
      \draw[dr->] (0,0.5) -- (2,0.5);

      \node[left] at (0,0.5) {$z_1$};
      \node[left] at (0,0) {$z_2$};

      \node[above right] at (0,0.5) {$i$};
      \node[above left] at (2,0.5) {$m$};

      \node[below right] at (0,0) {$n$};
      \node[below left] at (2,0) {$j$};
    \end{tikzpicture}
  }
  \qquad
  \subfloat[\label{fig:c-unitarity-(N,0)}]{
    \begin{tikzpicture}[scale=1.2, baseline=(x.base)]
      \node (x) at (1,0.25) {\vphantom{x}};
    
      \fill[ws] (0,-0.25) rectangle (2,0.75);
      
      \draw[dr->] \pathA;
      \draw[dr->] \pathB;

      \node[left] at (0,0.5) {$z_1$};
      \node[left] at (0,0) {$z_2$};

      \node[above right] at (0,0.5) {$i$};
      \node[above left] at (2,0.5) {$m$};

      \node[below right] at (0,0) {$n$};
      \node[below left] at (2,0) {$j$};

      \node[below] at (1,0) {$k$};
      \node[above] at (1,0.5) {$l$};
    \end{tikzpicture}
    =
    \begin{tikzpicture}[scale=1.2, baseline=(x.base)]
      \node (x) at (1,0.25) {\vphantom{x}};
      
      \fill[ws] (0,-0.25) rectangle (2,0.75);
      
      \draw[dr->] (2,0) -- (0,0);
      \draw[dr->] (0,0.5) -- (2,0.5);

      \node[left] at (0,0.5) {$z_1$};
      \node[left] at (0,0) {$z_2$};

      \node[above right] at (0,0.5) {$i$};
      \node[above left] at (2,0.5) {$m$};

      \node[below right] at (0,0) {$n$};
      \node[below left] at (2,0) {$j$};
    \end{tikzpicture}
  }
  \caption{Crossing--unitarity relations in (a) the $(N,1)$ background
    and (b) an $(N,0)$ background.}
  \label{fig:c-unitarity}
\end{figure}

This shift is due to \emph{framing anomaly}.  As an analysis carried
out in~\cite{Costello:2017dso} revealed, an anomaly breaks the gauge
invariance of a Wilson line when the line curves in the $(N,1)$
background.  For this anomaly to be canceled, the spectral parameter
must be shifted by $-\Delta\varphi N\hbar/2\pi$, where $\Delta\varphi$
is the angle by which the Wilson line bends.  Note that in order to
talk about the angle of a curve, one must endow $\Sigma$ with a
framing, that is, a choice of a trivialization of the tangent bundle.
The only closed surface that admits a framing is $T^2$, hence our
choice $\Sigma = T^2$.

\begin{figure}
  \centering
    \begin{tikzpicture}[scale=1, baseline=(x.base)]
      \node (x) at (0,0) {\vphantom{x}};

      \fill[dshaded] (0,-0.9) rectangle (3,0.9);
      
      \draw[dr->] (0,0.35) node[left, yshift=2pt] {$z_1$} -- (3,-0.65);
      \draw[dr->] (0,0) node[left] {$z_2$}
      -- (2,0) arc (-90:-60:0.25) -- ++(30:1.009);

      \draw ($(1.05,0)+(0:0.8)$) arc (0:-19.5:0.8)
      node[shift={(0.25,0.1)}] {$\varphi$};
    \end{tikzpicture}
    \
    =
    \begin{tikzpicture}[scale=1, baseline=(x.base)]
      \node (x) at (0,0.4) {\vphantom{x}};

      \fill[dshaded] (0,-0.5) rectangle (3,1.3);
      
      \draw[dr->] (0,1) node[left, yshift=2pt] {$z_1$} -- (3,0);
      \draw[dr->] (0,0) node[left] {$z_2$}
      -- (2,0) arc (-90:-60:0.25) -- ++(30:1.009)
      node[right, yshift=2pt] {$z_2 - \Delta\varphi N\hbar/2\pi$};
      \draw ($(2.49,0.245)+(30:0.35)$) arc (30:-30:0.35)
      node[shift={(0.7,0.2)}] {$\varphi + \Delta\varphi$};
    \end{tikzpicture}
    \caption{Translation of Wilson lines leads to the same operator.}
  \label{fig:R-angle}
\end{figure}
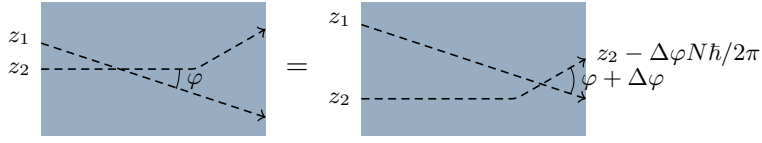

In turn, the framing anomaly implies, under the assumption that the
topological invariance on $\Sigma$ is otherwise unbroken, that the
R-matrix should really depend on the angle at which two Wilson lines
cross.  This is because as these lines curve, the R-matrix should
change by shifting the argument so as to compensate for the shift in
the spectral parameters. In Figure~\ref{fig:R-angle}, two diagrams are
shown in which a straight Wilson line intersects another Wilson line
which initially goes straight but at one point bends by angle
$\Delta\varphi$.  If the straight Wilson lines in the two diagrams are
parallel, these diagrams should represent the same operator.  From
this equality we deduce that the R-matrix $R^{(N,1)}_\varphi$ for
Wilson lines crossing at angle $\varphi$ satisfies the relation
\begin{equation}
  R^{(N,1)}_\varphi(z)
  =
  R^{(N,1)}_0\Bigl(z - \frac{\varphi}{2\pi} N\hbar\Bigr)
  \,.
\end{equation}

The unitarity relation \eqref{eq:d-unitarity}, as we formulated it,
does not involve any shift in the spectral parameter.  This is
possible only if the equation refers to the situation where the two
lines are almost parallel.  Therefore, the R-matrix $R^{(N,1)}$ that
appears in this equation corresponds to the case $\varphi = 0$:
\begin{equation}
  R^{(N,1)}
  =
  R^{(N,1)}_0
  \,.
\end{equation}
The crossing--unitarity relation \eqref{eq:c-unitarity-(N,1)}, on the
other hand, corresponds to the case when two lines are almost
antiparallel, which explains the shift by $-N\hbar/2$.

It turns out that the framing anomaly in an $(N,0)$ background is more
complicated.  To see why, consider the crossing--unitarity relation
shown in Figure~\ref{fig:c-unitarity-(N,0)}.  For $i \neq n$, the
left-hand side is nonvanishing only when $i = k = m$ and $j = l = n$.
If the sole effect of the framing anomaly were to shift the spectral
parameter just as in the $(N,1)$ background, then the left-hand side
in this case would be
\begin{equation}
  \label{eq:c-unitarity}
  R^{(N,0)}\biggl(z_1 - z_2 - \frac{1}{2} N\hbar,
  \uplambdab + \hbar\upomegab_j\biggr)_{ij}^{ij}
  R^{(N,0)}\biggl(z_2 - z_1 - \frac{1}{2} N\hbar,
  \uplambdab + \hbar\upomegab_j\biggr)_{ji}^{ji} \,.
\end{equation}
This equals
\begin{equation}
  \label{eq:c-unitarity-excess}
  \frac{\theta_1(\uplambdab_{ij}) \theta_1(\uplambdab_{ij} - 2\hbar)}
       {\theta_1(\uplambdab_{ij} - \hbar)^2}
\end{equation}
and not $1$ as required by the relation.

Apparently, the matrix elements of the R-matrix $R^{(N,0)}_\pi$ for
$\varphi = \pi$ differs from those of $R^{(N,0)} = R^{(N,0)}_0$ not
only by the shift in the spectral parameter, but also by some factors
which are ratios of theta functions containing~$\uplambdab$.  Let us
determine these factors.

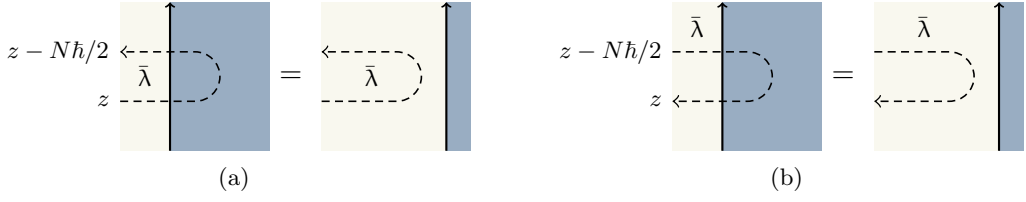
\begin{figure}
  \centering
  \subfloat[\label{fig:S'S}]{
    \begin{tikzpicture}[scale=1, baseline=(x.base)]
      \node (x) at (0,1) {\vphantom{x}};
      
      \fill[dshaded] (2/3,0) rectangle (2,2);
      \fill[ws] (0,0) rectangle (2/3,2);

      \draw[z->] (2/3,0) -- (2/3,2);
      
      \draw[dr->] (0,2/3) node[left] {$z$}
      -- (1,2/3) arc (-90:90:1/3)
      -- (0,4/3) node[left] {$z - N\hbar/2$};
      
      \node at (1/3,1) {$\uplambdab$};
    \end{tikzpicture}
    =
    \begin{tikzpicture}[scale=1, baseline=(x.base)]
      \node (x) at (0,1) {\vphantom{x}};
      
      \fill[dshaded] (5/3,0) rectangle (2,2);
      \fill[ws] (0,0) rectangle (5/3,2);

      \draw[z->] (5/3,0) -- (5/3,2);
      
      \draw[dr->] (0,2/3) -- (1,2/3) arc (-90:90:1/3) -- (0,4/3);
      
      \node at (2/3,1) {$\uplambdab$};
    \end{tikzpicture}
  }
  \qquad
  \subfloat[\label{fig:S'S-2}]{
    \begin{tikzpicture}[scale=1, baseline=(x.base)]
      \node (x) at (0,1) {\vphantom{x}};
      
      \fill[dshaded] (2/3,0) rectangle (2,2);
      \fill[ws] (0,0) rectangle (2/3,2);

      \draw[z->] (2/3,0) -- (2/3,2);
      
      \draw[dr->] (0,4/3) node[left] {$z - N\hbar/2$}
      -- (1,4/3) arc (90:-90:1/3)
      -- (0,2/3) node[left] {$z$};
      
      \node at (1/3,5/3) {$\uplambdab$};
    \end{tikzpicture}
    =
    \begin{tikzpicture}[scale=1, baseline=(x.base)]
      \node (x) at (0,1) {\vphantom{x}};
      
      \fill[dshaded] (5/3,0) rectangle (2,2);
      \fill[ws] (0,0) rectangle (5/3,2);

      \draw[z->] (5/3,0) -- (5/3,2);
      
      \draw[dr->] (0,4/3) -- (1,4/3) arc (90:-90:1/3) -- (0,2/3);
      
      \node at (2/3,5/3) {$\uplambdab$};
    \end{tikzpicture}
  }
  \caption{Unitarity relations involving a Wilson line and an 't Hooft
    line.}
  \label{fig:unitarity-WH}
\end{figure}

First, consider the equality between two diagrams shown in
Figure~\ref{fig:S'S}.  On the left-hand side, a Wilson line enters the
$(N,1)$ background and makes a left turn.  The spectral parameter gets
shifted by $-N\hbar/2$, and the line comes out to an $(N,0)$
background.  The right-hand side would be the identity operator if it
were placed in the $(N,1)$ background.  In the $(N,0)$ background,
however, the framing anomaly replaces it with a diagonal matrix
$\diag(\chi_1, \dotsc, \chi_N)$, which is a function of $\uplambdab$
but not of $z$ because of translation invariance.  So we get the
equality
\begin{equation}
  S'\biggl(z - \frac12 N\hbar, \uplambdab\biggr)^k_j
  S(z, \uplambdab)^j_i
  =
  \chi_k(\uplambdab) \delta^k_i
  \,.
\end{equation}
Comparing this equation with the expression \eqref{eq:S'} for $S'$, we
see
\begin{equation}
  \delta = \frac{1}{2} N\hbar \,,
  \qquad
  \chi(z,\uplambdab)^k_i = \chi_k(\uplambdab) \delta^k_i \,.
\end{equation}
It should be emphasized here that we have defined the intertwining
operators $S$, $S'$ using Wilson and 't~Hooft lines crossing at the
right angle.

Next, suppose that the Wilson line instead makes a right turn, as in
Figure~\ref{fig:S'S-2}.  Then, the right-hand side is replaced with
$\chi(\uplambdab)^{-1}$ because one can straighten out a line that
makes successive left and right turns, without altering the initial
and the final directions.  Thus we get another relation between $S$
and $S'$:
\begin{equation}
  S'(z, \uplambdab - \hbar\upomegab_i + \hbar\upomegab_k)^k_j
  S\biggl(z - \frac12 N\hbar, \uplambdab\biggr)^j_i
  =
  \chi_k(\uplambdab)^{-1} \delta^k_i
  \,.
\end{equation}
These two relations imply
\begin{equation}
  S^{-1}(z + N\hbar, \uplambdab + \hbar\upomegab_k)^k_j
  S(z, \uplambdab + \hbar\upomegab_i)^j_i
  =
  \chi_k(\uplambdab + \hbar\upomegab_k)^{-2}
  \delta^k_i
  \,.
\end{equation}

The left-hand side of this equation contains
\begin{equation}
  \Phi^{-1}(z + N\hbar + d, \uplambdab + \hbar\upomegab_k)^k_j  
  \Phi(z + d, \uplambdab + \hbar\upomegab_i)^j_i
  \,.
\end{equation}
According to the formula~\cite{MR1463830}
\begin{multline}
  \frac{\theta_1(z)^{1/N}}{\theta_1(z + N\hbar)^{1/N}}
  \Phi^{-1}(z + N\hbar,\upmub)^k_j \Phi(z,\uplambdab)^j_i
  \\
  =
  \frac{\theta_1(z + (N-1)\hbar + \uplambdab_i - \upmub_k - (N-1)/2)}
  {\theta_1(z + N\hbar - (N-1)/2)}
  \prod_{l (\neq k)}
  \frac{\theta_1(\uplambdab_i - \upmub_l - \hbar)}
       {\theta_1(\mub_{kl})}
  \,,
\end{multline}
this factor vanishes for $i \neq k$.  Setting $i = k$, we find
\begin{multline}
  \frac{\gb_k(z + N\hbar, \uplambdab + \hbar\upomegab_k)}
       {\gb_k(z, \uplambdab + \hbar\upomegab_k)}
  =
  \frac{\theta_1(z + N\hbar +d)^{1/N}}
       {\theta_1(z +d)^{1/N}}
  \frac{\theta_1(z + (N - 1)\hbar + d - (N-1)/2)}
       {\theta_1(z + N\hbar + d - (N-1)/2)}
  \\
  \times
  \chi_k(\uplambdab + \hbar\upomegab_k)^2
  \prod_{l (\neq k)}
  \frac{\theta_1(\uplambdab_{kl})}
       {\theta_1(\uplambdab_{kl} + \hbar)}
  \,.
\end{multline}
Since $z$ and $\uplambdab$ appear in separate factors on the
right-hand side, $\gb_k$ takes the form%
\footnote{If we write
  $\gb_k(z,\uplambdab) = h_k(z) \eta_k(z,\uplambdab)$, with $h_k$ as
  given below, then $\eta_k$ is a doubly periodic meromorphic function
  of $z$ satisfying
  $\eta_k(z, \uplambdab) = \eta_k(z + N\hbar, \uplambdab)$.  Assuming
  that any pair from $1$, $\tau$ and $N\hbar$ are linearly independent
  in $\C$, this implies that $\eta_k$ is independent of $z$ as there
  are no triply periodic meromorphic functions other than constants.}
\begin{equation}
  \gb_k(z,\uplambdab) = h_k(z) \eta_k(\uplambdab)
  \,.
\end{equation}
Then we have
\begin{align}
  \chi_k(\uplambdab)^2     
  &=
  C_k
  \prod_{l (\neq k)}
  \frac{\theta_1(\uplambdab_{kl})}
       {\theta_1(\uplambdab_{kl} - \hbar)}
  \,,
  \\
  \label{eq:hb-over-hb}
  \frac{h_k(z + N\hbar)}{h_k(z)}
  &=
  C_k
  \frac{\theta_1(z + N\hbar + d)^{1/N}}
       {\theta_1(z + d)^{1/N}}
  \frac{\theta_1(z + (N - 1)\hbar + d - (N-1)/2)}
       {\theta_1(z + N\hbar + d - (N-1)/2)}
\end{align}
for some constants $C_k$.

The requirement that $\gb$ acts trivially on $R^{(N,0)}$ translates to
the constraints
\begin{equation}
  \frac{h_i(z_1) h_j(z_2)}{h_j(z_1) h_i(z_2)}
  =
  \frac{\eta_i(\uplambda) \eta_j(\uplambda - \hbar\upomega_i)}
       {\eta_i(\uplambda - \hbar\upomega_j) \eta_j(\uplambda)}
  \,.
\end{equation}
This equation tells that the left-hand side cannot depend on $z_1$ or
$z_2$, so we have
\begin{equation}
  h_k(z) = c_k h(z)
\end{equation}
for some function $h$ and constants $c_k$.  Absorbing $c_k$ into
$\eta_k$, we can set
\begin{equation}
  h_k = h \,,
  \qquad
  C_k = C
\end{equation}
for some constant $C$.

We will see in section~\ref{sec:junctions} that $C = 1$.  Then we
can write
\begin{equation}
  \chi_k(\uplambdab)
  =
  \prod_{i < j}
  \frac{\theta_1(\uplambdab_{ij})^{1/2}}
  {\theta_1((\uplambdab - \hbar\upomegab_k)_{ij})^{1/2}} \,.
\end{equation}
This shows that in the definition \eqref{eq:L(N,1)-SS'} of the
difference operator $L^{(N,1)}$, what the factor $\chi$ contained in
$S'$ does is just to apply conjugation with the operator that acts on
a function $f(\uplambdab)$ by multiplication by
$\prod_{i < j} \theta_1(\uplambdab_{ij})^{1/2}$.  Therefore, it does
not affect the algebra generated by the L-operator.

Having determined $\chi$, we finally consider the same relation as in
Figure~\ref{fig:R-angle} but placed in an $(N,0)$ background.  Taking
$\varphi = 0$ and $\Delta\varphi = \pi$, we conclude
\begin{equation}
  \label{eq:R^(N,0)_pi}
  R^{(N,0)}_\pi(z,\uplambdab)_{ij}^{kl}
  =
  \frac{\chi_l(\uplambdab)}{\chi_j(\uplambdab - \hbar\upomegab_i)}
  R^{(N,0)}\biggl(z - \frac12 N\hbar, \uplambdab\biggr)_{ij}^{kl}
  \,.
\end{equation}
The prefactor on the right-hand side cancels the extra
factor~\eqref{eq:c-unitarity-excess} in the crossing--unitarity
relation, as it should.  The unitarity relation for $R^{(N,0)}_\pi$
also readily follows from this relation.

\subsection{Junctions of Wilson lines}
\label{sec:junctions}

Although the Yang--Baxter equations and various other relations put
strong constraints on the forms of the R-matrices and the intertwining
operators, we have not been able to fix some ambiguities.  While the
determination of the matrix $\xi$ is not so crucial as it drops out
from gauge invariant expressions, the function
$f^{(N,0)} = f^{(N,\pm1)}$ does affect physical quantities.  We can
determine this function by considering junctions of Wilson
lines~\cite{Costello:2018gyb}.

\begin{figure}
  \centering
  \subfloat[\label{fig:junction}]{
  \begin{tikzpicture}[align at bottom]
    \fill[ws] (-0.809,-0.95) rectangle (1,0.95);

    \node[draw, circle, fill=white, inner sep=0pt, minimum size=10pt]
    (eps) at (0,0) {$\veps$};

    \draw[dr->] (eps) -- (0:1);
    \draw[dr->] (eps) -- (72:1);
    \draw[dr->] (eps) -- (144:1);
    \draw[dr->] (eps) -- (216:1);
    \draw[dr->] (eps) -- (288:1);

    \draw (0:0.5) arc (0:72:0.5);
    \node[right] at (50:0.55) {$2\pi/N$};
  \end{tikzpicture}
  }
  \qquad
  \subfloat[\label{fig:junction-h}]{
  \begin{tikzpicture}[align at bottom]
    \fill[ws] (-0.8,-1.2) rectangle (1.5,1.2);

    \node at (-0.5,0) {$\uplambdab$};

    \node[draw, circle, fill=white, inner sep=0pt, minimum size=10pt]
    (eps) at (0,0) {$\veps$};

    \draw[dr->] (eps) -- (1.5,0)
    node [right] {$z - 2\hbar$};
    
    \draw[dr->] (eps)
    to [out=72, in=180, looseness=1] (0.5,0.4) -- (1.5,0.4)
    node [right] {$z - \hbar$};

    \draw[dr->] (eps)
    to[out=144, in=180, looseness=1.2] (0.05,0.8) -- (1.5,0.8)
    node [right] {$z$};
    
    \draw[dr->] (eps)
    to[out=216, in=180, looseness=1.2] (0.05,-0.8) -- (1.5,-0.8)
    node [right] {$z - 4\hbar$};

    \draw[dr->] (eps)
    to [out=288, in=180, looseness=1] (0.5,-0.4) -- (1.5,-0.4)
    node [right] {$z - 3\hbar$};
  \end{tikzpicture}
  }

  \caption{(a) A junction of Wilson lines for $N = 5$.  (b) The
    spectral parameters are shifted due to the framing anomaly when
    the Wilson lines are bent. }
\end{figure}
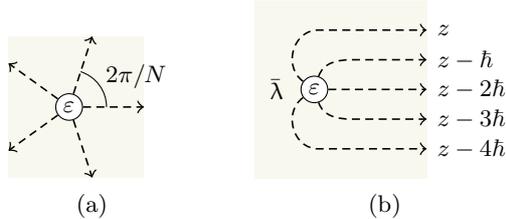

In gauge theory, one can join Wilson lines by contracting the ends of
the lines with an invariant tensor of the gauge group.  In the case of
$G = \SU(N)$, we use a completely antisymmetric tensor $\veps$ to
construct a junction of $N$ Wilson lines in the vector representation:
\begin{equation}
  \label{eq:junction}
  \veps^{i_1 \dotso i_N} (W_1)^{j_1}_{i_1} \dotsm (W_N)^{j_N}_{i_N} \,.
\end{equation}
An example for $N = 5$ is
shown in Figure~\ref{fig:junction}.

While in the path integral the junction is described by the constant
tensor $\veps$, it can receive quantum corrections in the effective
description we are using.  This is natural because states on the
Wilson lines participating in a junction live in holomorphic vector
bundles that are inequivalent due to the jumps of the spectral
parameter, and the notion of determinant has to be modified.

Now, take a junction and bend the Wilson lines so that they all extend
horizontally to the right, as in~Figure~\ref{fig:junction-h}.  At the
junction the lines have the same spectral parameter, but as they curve
their spectral parameters get shifted because of the framing anomaly.
It was found in~\cite{Costello:2017dso} that quantum mechanically a
configuration of Wilson lines suffers from an anomaly unless the lines
make equal angles at the junctions.  Therefore, in the region where
the lines are horizontal, the spectral parameters of adjacent lines
must differ by $\hbar$.  Let these parameters be $z$, $z - \hbar$,
$\dotsc$, $z - (N-1)\hbar$ from top to bottom.

As we have seen already, in addition to the shifts in the spectral
parameters, bending of Wilson lines in an $(N,0)$ background also
induces some factors of theta functions containing the dynamical
parameter.  We have determined these factors only in the case when the
lines make 180-degree turns, which can be useful only for $N = 2$.

Rather than trying to determine the quantum corrections to the
junction and the framing anomaly for general angles separately, let us
encapsulate both of these effects into a single tensor
$\veps_\hbar(\uplambdab)$.  This is the operator representing the
diagram in Figure~\ref{fig:junction-h}.  It is still totally
antisymmetric since the contributions to the path integral from terms
in the junction~\eqref{eq:junction} vanish if $i_m = i_n$ for some
$(m,n)$.

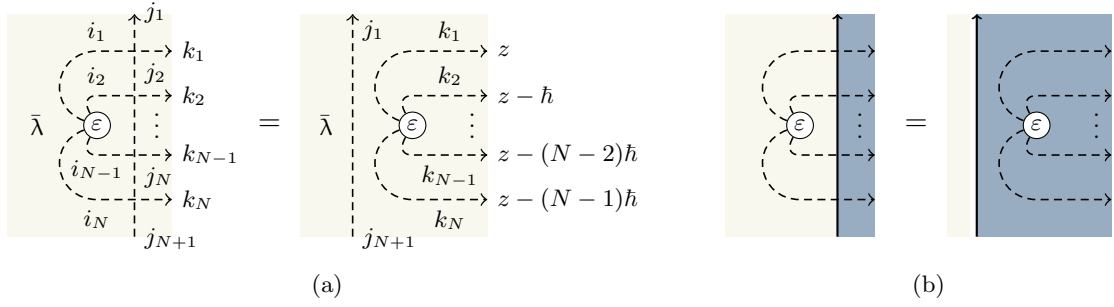
\begin{figure}
  \centering
  \subfloat[\label{fig:qdet}]{
  \begin{tikzpicture}[scale=1, baseline=(x.base)]
    \node (x) at (0,1) {\vphantom{x}};

    \fill[ws] (-1.2,-0.5) rectangle (1,2.5);

    \draw[dr->] (0.5,-0.5) -- (0.5,2.5);

    \node[above right] at (0.5, 2.25) {$j_1$};
    \node[right] at (0.5, 1.7) {$j_2$};
    \node[right] at (0.5, 0.3) {$j_N$};
    \node[below right] at (0.5, -0.25) {$j_{N+1}$};

    \node at (0.8,1.1) {$\vdots$};
    
    \node[right] (1) at (1,2) {};
    \node[right] (2) at (1,1.4) {};
    \node[right] (N-1) at (1,0.6) {};
    \node[right] (N) at (1,0) {};

    \node[draw, circle, fill=white, inner sep=0pt, minimum size=10pt]
    (eps) at (0,1) {$\veps$};

    \draw[dr->] (eps)
    to [out=160, in=180, looseness=1.5] (0,2)
    node[above] {$i_1$} -- (1) node[right=-4pt] {$k_1$};

    \draw[dr->] (eps)
    to [out=120, in=180, looseness=1.5] (0,1.4)
    node[above] {$i_2$} -- (2) node[right=-4pt] {$k_2$};

    \draw[dr->] (eps)
    to [out=240, in=180, looseness=1.5] (0,0.6)
    node[below=-2pt] {$i_{N-1}$} -- (N-1) node[right=-4pt] {$k_{N-1}$};

    \draw[dr->] (eps)
    to [out=200, in=180, looseness=1.5] (0,0)
    node[below] {$i_N$} -- (N) node[right=-4pt] {$k_N$};

    \node at (-0.8,1) {$\uplambdab$};
  \end{tikzpicture}
  = \
  \begin{tikzpicture}[baseline=(x.base)]
    \node (x) at (0,1) {\vphantom{x}};

    \fill[ws] (-1.5,-0.5) rectangle (1,2.5);

    \draw[dr->] (-0.8,-0.5) -- (-0.8,2.5);

    \node[above right] at (-0.8,2) {$j_1$};
    \node[below right] at (-0.8,-0.25) {$j_{N+1}$};

    \node at (0.8,1.1) {$\vdots$};
    
    \node[right] (1) at (1,2) {};
    \node[right] (2) at (1,1.4) {};
    \node[right] (N-1) at (1,0.6) {};
    \node[right] (N) at (1,0) {};

    \node[draw, circle, fill=white, inner sep=0pt, minimum size=10pt]
    (eps) at (0,1) {$\veps$};

    \draw[dr->] (eps)
    to [out=160, in=180, looseness=1.5] (0,2)
    -- node[above] {$k_1$} (1) node[right=-4pt] {$z$};

    \draw[dr->] (eps)
    to [out=120, in=180, looseness=1.5] (0,1.4)
    -- node[above] {$k_2$} (2) node[right=-4pt] {$z - \hbar$};

    \draw[dr->] (eps)
    to [out=240, in=180, looseness=1.5] (0,0.6)
    -- node[below] {$k_{N-1}$} (N-1) node[right=-4pt] {$z - (N-2)\hbar$};

    \draw[dr->] (eps)
    to [out=200, in=180, looseness=1.5] (0,0)
    -- node[below] {$k_N$} (N) node[right=-4pt] {$z - (N-1)\hbar$};

    \node at (-1.15,1) {$\uplambdab$};
  \end{tikzpicture}
  }
  \qquad
  \subfloat[\label{fig:qdet-so}]{
  \begin{tikzpicture}[scale=1, baseline=(x.base)]
    \node (x) at (0,1) {\vphantom{x}};

    \fill[ws] (-1,-0.5) rectangle (0.5,2.5);
    \fill[dshaded] (0.5,-0.5) rectangle (1,2.5);

    \draw[z->] (0.5,-0.5) -- (0.5,2.5);

    \node at (0.8,1.1) {$\vdots$};
    
    \node[right] (1) at (1,2) {};
    \node[right] (2) at (1,1.4) {};
    \node[right] (N-1) at (1,0.6) {};
    \node[right] (N) at (1,0) {};

    \node[draw, circle, fill=white, inner sep=0pt, minimum size=10pt]
    (eps) at (0,1) {$\veps$};

    \draw[dr->] (eps)
    to [out=160, in=180, looseness=1.5] (0,2)
    -- (1);

    \draw[dr->] (eps)
    to [out=120, in=180, looseness=1.5] (0,1.4)
    -- (2);

    \draw[dr->] (eps)
    to [out=240, in=180, looseness=1.5] (0,0.6)
    -- (N-1);

    \draw[dr->] (eps)
    to [out=200, in=180, looseness=1.5] (0,0)
     -- (N);

  \end{tikzpicture}
  = \
  \begin{tikzpicture}[baseline=(x.base)]
    \node (x) at (0,1) {\vphantom{x}};

    \fill[ws] (-1.2,-0.5) rectangle (-0.9,2.5);
    \fill[dshaded] (-0.8,-0.5) rectangle (1,2.5);

    \draw[z->] (-0.8,-0.5) -- (-0.8,2.5);

    \node at (0.8,1.1) {$\vdots$};
    
    \node[right] (1) at (1,2) {};
    \node[right] (2) at (1,1.4) {};
    \node[right] (N-1) at (1,0.6) {};
    \node[right] (N) at (1,0) {};

    \node[draw, circle, fill=white, inner sep=0pt, minimum size=10pt]
    (eps) at (0,1) {$\veps$};

    \draw[dr->] (eps)
    to [out=160, in=180, looseness=1.5] (0,2)
    -- (1);

    \draw[dr->] (eps)
    to [out=120, in=180, looseness=1.5] (0,1.4)
    -- (2);

    \draw[dr->] (eps)
    to [out=240, in=180, looseness=1.5] (0,0.6)
    -- (N-1);

    \draw[dr->] (eps)
    to [out=200, in=180, looseness=1.5] (0,0)
    -- (N);

  \end{tikzpicture}
  \vphantom{
    \begin{tikzpicture}[scale=1, baseline=(x.base)]
      \fill[ws] (-1.2,-0.5) rectangle (1,2.5);
      \node[below right] at (0.5, -0.25) {$j_{N+1}$};
    \end{tikzpicture}
  }%
  }
  \caption{(a) Quantum determinant relation for $R^{(N,0)}$ (b)
    Quantum determinant relation with a surface operator.}
\end{figure}

To this collection of Wilson lines let us introduce an additional
Wilson line, almost parallel to the horizontal lines.  The familiar
field theory argument then suggests that the relation shown in
Figure~\ref{fig:qdet} should hold.  (For the ease of visualization we
have drawn the additional Wilson line vertically.)  The left-hand side
of this relation, evaluated for
$(k_1, k_2, \dotsc, k_N) = (1,2,\dotsc,N)$, is the \emph{quantum
  determinant} of $R^{(N,0)}$.

To determine $\veps_\hbar(\uplambdab)$, we look at a similar relation,
in which the vertical Wilson line is replaced with an 't Hooft line;
see Figure~\ref{fig:qdet-so}.  The right-hand side of this relation
contains a junction in the $(N,1)$ background.  Since the $(N,1)$
background has no moduli, the antisymmetric tensor can only receive
quantum corrections that rescale it by an overall factor, which can be
absorbed by rescaling of the antisymmetric tensor used to define the
junction in the path integral.  Thus, we have
\begin{equation}
  \veps_\hbar(\uplambdab)^{i_1 \dotso i_N}
  \prod_{n=1}^N
  S\biggl(z - (n-1)\hbar,
  \uplambdab - \hbar\sum_{m=1}^{n-1} \upomegab_{i_m}\biggr)_{i_n}^{k_n}
  =
  \veps^{k_1 \dotso k_N}
  \,,
\end{equation}
or
\begin{multline}
  \label{eq:eps-eps}
  \veps_\hbar(\uplambdab)^{i_1 \dotso i_N}
  \prod_{n=1}^N h\bigl(z - (n-1)\hbar\bigr)
  \frac{\theta^{(k_n)}(z + N \uplambdab_{i_n} + d)}
  {\theta_1(z - (n-1)\hbar + d)^{1/N}}
  (\Theta^{-1} \xi \eta)
  \biggl(\uplambdab - \hbar\sum_{m=1}^{n-1} \upomegab_{i_m}\biggr)_{i_n}
  \\
  =
  \veps^{k_1 \dotso k_N}
  \,.
\end{multline}

A key to crack this equation is the following determinant
formula~\cite{MR1236389}:
\begin{equation}
  \det\bigl(\theta^{(j)}(z + N \uplambdab_{i})\bigr)_{i,j=1,\dotsc,N}
  =
  C_{N,\tau} \theta_1\bigl(z - (N-1)/2\bigr)
  \prod_{i < j} \theta_1(\uplambdab_{ij})
  \,.
\end{equation}
Here $C_{N,\tau}$ is a constant that depends only on $N$ and $\tau$.
From this formula it follows
\begin{multline}
  \frac{1}{\theta_1(z + d - (N-1)/2)}
  \prod_{n=1}^N
  \frac{\theta_1(z + d - (n-1)\hbar)^{1/N}}{h(z - (n-1)\hbar)}
  \\
  =
  \frac{1}{N!} \veps_{i_1 \dotso i_N}
  C_{N,\tau} \prod_{i < j} \theta_1(\uplambdab_{ij})
  \veps_\hbar(\uplambdab)^{i_1 \dotso i_N}
  \prod_{n=1}^N
  (\Theta^{-1} \xi \eta)
  \biggl(\uplambdab - \hbar\sum_{m=1}^{n-1} \upomegab_{i_m}\biggr)_{i_n} \,,
\end{multline}
and the two sides are equal to some constant $D$ which can depend on
$\hbar$.  Thus we get
\begin{equation}
  \label{eq:eps-eps-sigma}
  \veps_\hbar^{i_1 \dotso i_N}(\uplambdab)
  \prod_{n=1}^N
  (\Theta^{-1} \xi \eta)_{i_n}
  \biggl(\uplambdab - \hbar\sum_{m=1}^{n-1} \upomegab_{i_m}\biggr)
  =
  C_{N,\tau}^{-1} D
  \veps^{i_1 \dotso i_N}
  \prod_{i < j} \theta_1(\uplambdab_{ij})^{-1}
\end{equation}
and
\begin{equation}
  \frac{1}{\theta_1(z + d - (N-1)/2)}
  \prod_{n=1}^N
  \frac{\theta_1(z - (n-1)\hbar + d)^{1/N}}{h(z - (n-1)\hbar)}
  = D \,.
\end{equation}
The last equation is consistent with relation~\eqref{eq:hb-over-hb}
only if
\begin{equation}
  C_k = 1
\end{equation}
for all $k$.

Let us go back to the quantum determinant relation in
Figure~\ref{fig:qdet}.  For the calculation of the quantum determinant
of $R^{(N,0)}$, we can perform a gauge transformation to put the
R-matrix in a convenient form.  If we apply the gauge transformation
by $\Theta^{-1} \xi \eta$, the tensor used at the junction becomes
precisely the left-hand side of relation \eqref{eq:eps-eps-sigma}.
Moreover, since $\eta$ acts on $R^{(N,0)}$ trivially, we have
\begin{equation}
  \Theta^{-1} \xi \eta \cdot R^{(N,0)}(z, \uplambdab)_{ij}^{ij}
  =
  f^{(N,0)}(z)
  \Theta^{-1} \cdot
  \RF(-z, \uplambdab)_{ij}^{ij}
  =
  f^{(N,0)}(z)
  \frac{\theta_1(z) \theta_1(\uplambdab_{ij} - \hbar)}
  {\theta_1(z + \hbar) \theta_1(\uplambdab_{ij})}
\end{equation}
and therefore
\begin{equation}
  \Theta^{-1} \xi \eta \cdot R^{(N,0)}(\hbar,\uplambdab)_{ij}^{ij}
  = f^{(N,0)}(\hbar) \RF(-\hbar,\uplambdab)_{ji}^{ij}
  = \Theta^{-1} \xi \eta \cdot R^{(N,0)}(\hbar,\uplambdab)_{ji}^{ij} \,.
\end{equation}
From this we deduce that for generic values of $\uplambdab$, the
kernel of $\Theta^{-1} \xi \eta \cdot R^{(N,0)}(\hbar,\uplambdab)$ is
$\bigwedge^2 \C^N$.  In this gauge, the left-hand side of the quantum
determinant relation is antisymmetric under an exchange of final
states on adjacent horizontal lines, as we can see by making those
lines cross and using the Yang--Baxter equation.  Hence, it is
completely antisymmetric in the final states on all horizontal lines.

Making use of this antisymmetry we can arrange the final states so
that $k_1 = j_1$.  Then, the only contribution to the quantum
determinant comes from the case when $j_n = j_1 = k_1$ and $i_n = k_n$
for all $n$, and the quantum determinant relation reduces to the
equation
\begin{multline}
  C_{N,\tau}^{-1} D \veps^{k_1 \dotso k_N}
  \prod_{i < j} \theta_1(\uplambdab_{ij})^{-1}
  \prod_{n=1}^N
  \Theta^{-1} \xi \eta \cdot R^{(N,0)}\biggl(z - (n-1)\hbar,
  \uplambdab - \hbar\sum_{m=1}^{n-1} \upomegab_{k_m}\biggr)_{k_n j_1}^{k_n j_1}
  \\
  =
  C_{N,\tau}^{-1} D \veps^{k_1 \dotso k_N}
  \prod_{i < j}
  \theta_1\bigl((\uplambdab - \hbar\upomegab_{j_1})_{ij}\bigr)^{-1}
  \,.
\end{multline}
All constants and functions of $\uplambdab$ in the equation cancel
out, leaving
\begin{equation}
  \label{eq:prodfb}
  \frac{\theta_1(z - (N-1)\hbar)}{\theta_1(z)}
  \prod_{n=1}^N
  f^{(N,0)}\bigl(z - (n-1)\hbar\bigr)
  =
  1
  \,.
\end{equation}
This is consistent with the quasi-periodicity property
\eqref{eq:fb-period}, as well as with the unitarity
condition~\eqref{eq:unitarity-f(N,0)} and the crossing--unitarity
condition~\eqref{eq:fb-fb}.  The same equation is obtained if one sets
the quantum determinant of $R^{(N,1)}$ to $1$.

\section{String theory realization and dualities}
\label{sec:brane-construction}

In the final section we discuss a realization of four-dimensional
Chern--Simons theory and the associated integrable lattice models in
string theory.  The embedding into string theory allows us to invoke
its powerful dualities.  Using these dualities, we relate the field
theory setup considered in the previous sections to other setups which
have been extensively studied in relation to quantum integrable
systems.  The string theory realization thus provides a unified
perspective on a number of phenomena in which the same integrable
systems arise from apparently different theories.

\subsection{\texorpdfstring{Brane construction of the
    $\Omega$-deformed topological--holomorphic theory}{Brane
    construction of the Ω-deformed topological--holomorphic theory}}

Consider a stack of $N$ D5-branes in Type IIB superstring theory.  If
the spacetime is flat Minkowski space $\R^{9,1}$, the low energy
dynamics of the branes is described by six-dimensional $\CN = (1,1)$
super Yang--Mills theory with gauge group $\U(N)$.  Discarding the
decoupled degrees of freedom associated with the center-of-mass motion
of the D5-branes, we obtain the theory with gauge group $\SU(N)$.

If, instead, the spacetime is $T^*M \times C$ and the D5-branes wrap
the zero section of $T^*M$ and $C$, then the effective worldvolume
theory is topologically twisted along $M$~\cite{Bershadsky:1995qy}.
(Here, as before, $M$ is a four-manifold and $C$ is either $\C$,
$\C^\times$ or an elliptic curve $E$.)  In fact, it is the twisted
$\CN = (1,1)$ super Yang--Mills theory whose $Q$-invariant sector is
the topological--holomorphic theory on $M \times C$, constructed in
section~\ref{sec:6dTHT}.  The reason is that the four bosonic fields
parametrizing the positions of the branes in the fiber directions of
$T^*M$ are not scalars as in the untwisted theory.  Rather, at each
point on~$C$, they are components of a one-form on $M$.  Turning the
four scalar fields into a one-form on $M$ is precisely what the
topological twisting for the topological--holomorphic theory does.

Our goal is to understand how to introduce an $\Omega$-deformation to
this brane construction of the topological--holomorphic theory.  More
specifically, we take $M = \R^2 \times \Sigma$ and $C = E$, and wish
to turn on an $\Omega$-deformation in the worldvolume theory using the
rotation symmetry of $\R^2$.

To this end, suppose that we could realize the desired
$\Omega$-deformation, and subsequently dimensionally reduced the
$\Omega$-deformed theory on $E$.  Then, we would obtain an
$\Omega$-deformation of the GL-twisted $\CN = 4$ super Yang--Mills
theory on $\R^2 \times \Sigma$.  This $\Omega$-deformation is,
however, different from the one commonly considered in the study of
four-dimensional $\CN = 2$ supersymmetric gauge theories.

The standard $\Omega$-deformation~\cite{Nekrasov:2002qd,
  Nekrasov:2003rj} is compatible with the Donaldson--Witten twist
\cite{Witten:1988ze}.  Upon dimensional reduction on $\Sigma$, the
Donaldson--Witten twist descends to the A-twist of $\CN = (2,2)$
supersymmetric theories in two dimensions~\cite{Bershadsky:1995vm}.
On the other hand, as we have seen already, the
topological--holomorphic theory reduces to a B-twisted theory in two
dimensions, not an A-twisted one.

From the GL-twisted $\CN = 4$ super Yang--Mills theory we can obtain
either of these twists in two dimensions, depending on the choice of
the supercharge we use to define a topological theory.  In the
four-dimensional theory, the two types of twists are related by
\emph{S-duality}~\cite{Harvey:1995tg, Bershadsky:1995vm,
  Kapustin:2006pk}.  This means that the $\Omega$-deformation of the
topological--holomorphic theory descends to the S-dual of the standard
$\Omega$-deformation of the GL-twisted $\CN = 4$ super Yang--Mills
theory.

A nice thing about the standard $\Omega$-deformation of an $\CN = 2$
supersymmetric gauge theory is that it has a transparent geometric
construction.  First, we lift the theory to an $\CN = (1,0)$
supersymmetric gauge theory in six dimensions.  The lifted theory is
defined on the product $M \times E$.  Then, we twist this product so
that when we go around the one-cycles of $E$, we do not come back to
the point we started from, but arrive at a point that is shifted by
the action of an isometry of $M$.  Finally, we perform the dimensional
reduction of the lifted theory down to four dimensions.  The resulting
four-dimensional theory is deformed compared to the original one
because of the twisting of the product.

This procedure can be incorporated in our brane construction
straightforwardly~\cite{Hellerman:2011mv, Reffert:2011dp,
  Hellerman:2012zf, Orlando:2013yea}.  In our setup, the D5-branes are
supported on the product $\R^2 \times \Sigma \times E$ sitting in the
ten-dimensional spacetime $T^*\R^2 \times T^*\Sigma \times E$, where
$\R^2$ is the zero section of $T^*\R^2$ and, for the purpose of this
discussion, we can take $\Sigma$ to be the zero section of
$T^*\Sigma$.

If we apply T-duality on $E$, the D5-branes turn into D3-branes.  In
the limit where $E$ shrinks to a point, the low energy dynamics of
these D3-branes is described by the GL-twisted $\CN = 4$ super
Yang--Mills theory on $\R^2 \times \Sigma$.  To introduce the
$\Omega$-deformation, we modify the geometry before applying the
T-duality.  Viewing $\R^2 \times \Sigma \times E$ as a flat
$\R^2$-bundle over $\Sigma \times E$, we twist it so that the fiber is
rotated by some angles as it is transported along the one-cycles of
$E$.  For supersymmetry to be preserved, we must simultaneously rotate
the fiber of $T^*\R^2$ in the opposite direction.  Now, T-duality on
$E$ produces a D3-brane configuration realizing the GL-twisted
$\CN = 4$ super Yang--Mills theory, subjected to the standard
$\Omega$-deformation.

To obtain the brane setup for the $\Omega$-deformed
topological--holomorphic theory, all we have to do is to apply
S-duality to this D3-brane configuration, which leaves the D3-branes
intact but acts nontrivially on the background, and then T-duality on
the dual elliptic curve $E^\vee$ to turn the D3-branes back into
D5-branes.

Let us describe this construction more precisely, following the chain
of dualities step by step.  We use radial coordinates $(r,\vtheta)$
and $(\rho,\vphi)$ for the base and fiber of $T^*\R^2$, respectively,
and parametrize $E$ with real coordinates $(x^4,x^5)$ defined up to
the identification
\begin{equation}
  (x^4,x^5)
  \sim (x^4 + 2\pi R, x^5)
  \sim (x^4 + 2\pi R \tau_1, x^5 - 2\pi R \tau_2) \,,
\end{equation}
with $\tau_2 > 0$.  With respect to the complex coordinate
$z = (x^4 - \iu x^5)/2$, the modular parameter of $E$ is
$\tau = \tau_1 + \iu\tau_2$.

Our starting point is the D5-branes supported on a twisted product of
$T^*\R^2$ and $E$.  In terms of the periodic coordinates $y^1$, $y^2$
defined by
\begin{equation}
  x^4 = R(y^1 + \tau_1 y^2) \,,
  \qquad
  x^5 = -R\tau_2 y^2 \,,
\end{equation}
we can construct this space via the identification
\begin{equation}
  \label{eq:twisted-product}
  (\vtheta, \vphi, y^1, y^2)
  \sim (\vtheta + 2\pi\veps_1, \vphi - 2\pi\veps_1,  y^1 + 2\pi, y^2)
  \sim (\vtheta + 2\pi\veps_2, \vphi - 2\pi\veps_2,  y^1, y^2 + 2\pi) \,,
\end{equation}
with some parameters $\veps_1$, $\veps_2 \in \R$.  The spacetime
metric is given by
\begin{equation}
  g
  =
  \rmd r^2 + r^2 \rmd\vtheta^2
  + \rmd\rho^2 + \rho^2 \rmd\vphi^2
  + (\rmd x^4)^2 + (\rmd x^5)^2
  + g_{T^*\Sigma}
  \,,
\end{equation}
where $g_{T^*\Sigma}$ is a Ricci flat metric on $T^*\Sigma$.  
We take the dilaton to be a constant:
\begin{equation}
  \Phi = \Phi_0 \,.
\end{equation}
The other background fields, the Kalb--Ramond two-form field $B_2$ and
the RR $p$-form fields $C_p$, are all set to zero.

The first step in the chain of dualities is T-duality on $E$.  For
this step it is convenient to introduce angle variables
\begin{equation}
  \theta = \vtheta - \veps_1 y^1 - \veps_2 y^2 \,,
  \qquad
  \phi = \vphi + \veps_1 y^1 + \veps_2 y^2 \,,
\end{equation}
which disentangle the identification~\eqref{eq:twisted-product}:
\begin{equation}
  (\theta, \phi, y^1, y^2)
  \sim (\theta, \phi,  y^1 + 2\pi, y^2)
  \sim (\theta, \phi,  y^1, y^2 + 2\pi)
  \,.
\end{equation}
With these coordinates we can use the standard formulas for
T-duality~\cite{Giveon:1994fu, Johnson:2003gi}.

The action of T-duality on $g$ and $B_2$ can be expressed concisely in
terms of the tensor $g + B_2$.  We write it in the block matrix form
as
\begin{equation}
  g + B_2
  =
  \begin{pmatrix}
    K & N \\
    M & L
  \end{pmatrix}
  \,,
\end{equation}
where $K$ represents the block whose indices involve only $y^1$ and
$y^2$.  Under T-duality in the $y^1$- and $y^2$-directions, $g + B_2$
is transformed to $\gt + \Bt_2$, with the corresponding blocks given
by
\begin{equation}
  \Kt = K^{-1} \,,\qquad
  \Lt = L - MK^{-1} N \,, \qquad
  \Mt = MK^{-1} \,, \qquad
  \Nt = -K^{-1} N \,.
\end{equation}
The dilaton is shifted as
\begin{equation}
  \Phit = \Phi_0 - \frac12 \ln \det K \,.
\end{equation}

Since $B_2 = 0$ initially, $K$ and $L$ are symmetric while $M^T = N$.
Then, $\Kt$ and $\Lt$ are symmetric and $\Nt = -\Mt^T$.  The T-duality
thus turns the metric into a block diagonal form and induces a nonzero
B-field:
\begin{equation}
  \gt
  =
  \begin{pmatrix}
    \Kt & 0 \\
    0 & \Lt
  \end{pmatrix}
  \,,
  \qquad
  \Bt_2
  =
  \begin{pmatrix}
    0 & -\Mt^T \\
    \Mt & 0
  \end{pmatrix}
  \,.
\end{equation}
An explicit calculation shows
\begin{equation}
  \begin{aligned}
    \gt
    &=
    \rmd r^2
    + r^2 \rmd\theta^2
    + \rmd\rho^2
    + \rho^2 \rmd\phi^2
    - \frac{|\veps|^2}{\Delta^2} (r^2 \rmd\theta - \rho^2 \rmd\phi)^2
    + g_{T^*\Sigma}
    \\ &\qquad\qquad
    + \frac{4}{R^4 \tau_2^2 \Delta^2}
      \Bigl((r^2 + \rho^2) \bigl(\Im(\vepsb \, \rmd\zeta)\bigr)^2
      + \rmd\zeta \, \rmd\zetab\Bigr)
    \,,
    \\
    \Bt_2
    &=
    \frac{2}{R^2 \tau_2 \Delta^2}
    (r^2 \rmd\theta - \rho^2 \rmd\phi) \wedge \Re(\vepsb \, \rmd\zeta)
    \,,
    \\
    \Phit
    &=
    \Phi_0 - \ln(R^2 \tau_2 \Delta) \,,
  \end{aligned}
\end{equation}
where we have defined
\begin{equation}
  \veps = \frac{\tau\veps_1 - \veps_2}{R\tau_2} \,,
  \qquad
  \zeta = \frac{R}{2} (\tau y^1 - y^2) \,,
  \qquad
  \Delta^2 = 1 + |\veps|^2 (r^2 + \rho^2) \,.
\end{equation}
This is the NS fluxtrap background studied in~\cite{Hellerman:2011mv,
  Reffert:2011dp, Hellerman:2012zf, Orlando:2013yea}.

Next, we apply S-duality.  This step changes the metric and the
dilaton to
\begin{equation}
  \gh = e^{-\Phit} \gt \,,
  \qquad
  \Phih = -\Phit \,,
\end{equation}
and exchanges the B-field and the RR two-form:
\begin{equation}
  \Bh_2 = \Ct_2 \,,
  \qquad
  \Ch_2 = -\Bt_2 \,.
\end{equation}
This background is called the RR fluxtrap~\cite{Orlando:2013yea}.

Finally, we apply T-duality in the $y^1$- and $y^2$-directions again.
The resulting metric and dilaton are
\begin{equation}
  \begin{aligned}
    \check{g}
    &=
    R^2 \tau_2 \Delta e^{-\Phi_0}
    \biggl(
    \rmd r^2
    + r^2 \rmd\theta^2
    + \rmd\rho^2
    + \rho^2 \rmd\phi^2
    - \frac{|\veps|^2}{\Delta^2} (r^2 \rmd\theta - \rho^2 \rmd\phi)^2
    + g_{T^*\Sigma}
    \biggr)
    \\ &\qquad\qquad
    +
    \frac{4 e^{\Phi_0}}{ R^2 \tau_2 \Delta}
    \Bigl((r^2 + \rho^2) \bigl(\Im(\vepsb \, \rmd z)\bigr)^2
      + \rmd z \, \rmd\zb\Bigr)
    \,,
    \\
    \check\Phi
    &=
    \ln(R^2 \tau_2 \Delta) \,.
  \end{aligned}
\end{equation}
On the RR two-form this step acts as a $90$-degree rotation on the
$y^1$-$y^2$ plane, sending $\rmd\zeta$ to $\rmd z$:
\begin{equation}
  \check{C}_2
  =
  \frac{2}{R^2 \tau_2 \Delta^2}
  (r^2 \rmd\theta - \rho^2 \rmd\phi) \wedge \Re(\vepsb \, \rmd z)
  \,.
\end{equation}
Based on the argument we have given above, we claim that this is the
background in which a stack of D5-branes realizes the
$\Omega$-deformed topological--holomorphic theory.

In principle, we should be able to verify this claim by comparing the
Dirac--Born--Infeld (DBI) action for the worldvolume theory of the
D5-branes and the action for the $\Omega$-deformed
topological--holomorphic theory.  In practice, this is not as easy as
it may sound because the two actions only need to coincide up to
$Q$-exact terms and a nontrivial field redefinition.  Here we content
ourselves with confirming that the DBI action reproduces some
important terms.

The metric on the D5-brane worldvolume is
\begin{equation}
  \label{eq:g-D5}
  \check{g}_{\text{D5}}
  =
  R^2 \tau_2 \Delta_0 e^{-\Phi_0}
  \biggl(
  \rmd r^2
  + \frac{r^2}{\Delta_0^2} \rmd\theta^2
  + g_{\Sigma}
  \biggr)
  +
  \frac{4 e^{\Phi_0}}{ R^2 \tau_2 \Delta_0}
  \Bigl(r^2 \bigl(\Im(\vepsb \, \rmd z)\bigr)^2
  + \rmd z \, \rmd\zb\Bigr)
  \,,
\end{equation}
where $\Delta_0 = 1 + |\veps|^2 r^2$.  For this metric to reduce at
$\veps = 0$ to the one we used for the topological--holomorphic theory,
we must take
\begin{equation}
  e^{\Phi_0} = R^2 \tau_2  \,.
\end{equation}
Then, we have
\begin{equation}
  \sqrt{\check{g}_{\text{D5}}} \, \rmd^6 x
  = \sqrt{g_\Sigma} \, \rmd^6 x \,,
\end{equation}
where $\rmd^6 x = \rmd x^0 \wedge \dotsb \wedge \rmd x^5$, with
$x^0 + \iu x^1 = re^{\iu\theta}$ and $(x^2,x^3)$ being coordinates on
$\Sigma$.

The DBI action, expanded to quadratic order in derivatives, contains
the terms
\begin{equation}
  -\frac{(2\pi\alpha')^2}{2R^2 \tau_2} T_5
  \int_{\R^2 \times \Sigma \times E}
  \! \sqrt{g_\Sigma} \, \rmd^6 x
  \Tr\biggl(
  \frac{1}{\Delta_0^2} F^{rm} F_{rm}
  + F^{\theta m} F_{\theta m}
  + \frac{1}{2\Delta_0^2} F^{mn} F_{mn}\biggr) \,.
\end{equation}
Here $(2\pi \alpha')^{-1}$ is the string tension, $T_5$ is the
D5-brane tension, and indices are raised with respect to the metric
$\rmd r^2 + r^2 \rmd\theta^2 + g_\Sigma + (\rmd x^4)^2 + (\rmd
x^5)^2$.  We identify these terms with the kinetic terms
$|D_\rh\varphi|^2/(1 + \|V\|^2) + |D_\thetah\varphi|^2$ for
$\varphi = \CA_m$ and the potential term
$|\del W/\del\varphi|^2/(1 + \|V\|^2)$ for $\varphi = A_\zb$ in the
bosonic part \eqref{eq:bosonic-terms} (with $t = 1$) of the action for
the $\Omega$-deformed topological--holomorphic theory.  Thus we find
\begin{equation}
  \frac{1}{e^2}
  =
  \frac{(2\pi\alpha')^2}{2R^2 \tau_2} T_5
  \,,
  \qquad
  |\eps| = |\veps|
  \,.
\end{equation}

The RR two-form induces the Wess--Zumino term
\begin{equation}
  -\iu \frac{(2\pi\alpha')^2}{2} \mu_5
  \int_{\R^2 \times \Sigma \times E} \check C_2 \wedge \Tr(F \wedge F)
  \,,
\end{equation}
where $\mu_5$ is the D5-brane charge.  This term contains
\begin{equation}
  \biggl(\frac{(2\pi\alpha')^2}{2 R^2 \tau_2} e^2 \mu_5\biggr)
  \cdot
  2 \iu \Im
  \int_{\R^2}
  r \, \rmd r \wedge \rmd\theta
  \wedge
  \frac{\vepsb r}{\Delta_0^2}
  \del_r  \biggl(-\frac{\iu}{e^2} \int_{\Sigma \times E} \rmd z \wedge \ChS(A)\biggr)
 \,.
\end{equation}
We see it within the terms
$2\iu \Im(\Vb^\thetah \del_\rh W)/(1 + \|V\|^2)$.  Comparing the
coefficients of~$\del_r W$, we identify
\begin{equation}
  \eps = \veps \,.
\end{equation}
For the overall factor to be equal to $1$, we must have $T_5 = \mu_5$.
This is the BPS condition for D5-branes.

{\subsection{Wilson lines and surface operators}

Let us construct integrable lattice models in the above string theory
setup.  For $\Sigma = T^2$, the ten-dimensional spacetime is
\begin{equation}
  T^*\R^2 \times T^*\Sigma \times E
  \iso
  \R^2 \times T^2  \times E \times \R^2_{67} \times \R^2_{89} \,,
\end{equation}
where $\R^2_{67}$ and $\R^2_{89}$ are the fibers of $T^*\R^2$ and
$T^*\Sigma$, respectively.  The subscripts refer to the coordinates
for these spaces which are consistent with the ones used in
section~\ref{sec:6dTHT}.  We use coordinates $(x,y)$ for $T^2$ and a
complex coordinate $z$ on $E$.

Four-dimensional Chern--Simons theory for $G = \SU(N)$ is realized by
$N$ D5-branes $\text{D5}_i$, $i=1$, $\dotsc$, $N$, supported on
\begin{equation}
  \R^2 \times T^2 \times E \times \{0\} \times \{(\phi_x^i,\phi_y^i)\}
  \subset
  \R^2 \times T^2  \times E \times \R^2_{67} \times \R^2_{89} \,.
\end{equation}
Without loss of generality we may assume
\begin{equation}
  \phi_x^1 \leq \phi_x^2 \leq \dotsb \leq \phi_x^N \,.
\end{equation}
The coordinates $(\phi_x^i,\phi_y^i)$ of $\text{D5}_i$ in $\R^2_{89}$
determine the imaginary part of the background value of the complex
gauge field $\CA_x \rmd x + \CA_y \rmd y$.  Together with the real
part, given by the values of the gauge fields on $\text{D5}_i$ along
$T^2$, they specify the twisted periodic boundary conditions of the
lattice models.  In the absence of the $\Omega$-deformation, the
D5-branes would preserve half of the thirty-two supercharges of Type
IIB superstring theory.

The construction of integrable lattice models requires Wilson lines
and surface operators bounded by 't Hooft lines.  To be concrete, let
us consider a lattice similar to the one illustrated in
Figure~\ref{fig:lattice-so}.  It consists of $m$ horizontal and $n$
vertical Wilson lines in the vector representation of $\SU(N)$, as
well as $k$ vertical strips of surface operators.

In general, Wilson lines in the worldvolume theory of a stack of $N$
D-branes are created by fundamental strings ending on the D-branes.
The end of a semi-infinite open string behaves as a charged particle
with infinite mass.  There are $N$ choices for the D-brane on which
the string ends, and these correspond to the possible states of the
charged particle.  Thus, a single open string creates a Wilson line in
the vector representation.  For Wilson lines in other representations,
there are more elaborate constructions which involve multiple strings
and additional branes~\cite{Yamaguchi:2006tq, Gomis:2006sb,
  Gomis:2006im, Maruyoshi:2016caf}.

Adopting this construction, we see that the horizontal Wilson lines
are realized by fundamental strings $\text{F1}^\dashrarrow_\alpha$,
$\alpha = 1$, $\dotsc$, $m$, ending on one of the D5-branes at
$(y,z) = (y_\alpha,z_\alpha)$ and extending in the negative
$x^8$-direction.  If the $\alpha$th Wilson line is in the $i_\alpha$th
state, $\text{F1}^\dashrarrow_\alpha$ ends on $\text{D5}_{i_\alpha}$.  The
vertical Wilson lines are created by fundamental strings
$\text{F1}^\dashuarrow_\beta$, $\beta = 1$, $\dotsc$, $n$, ending on
$\text{D5}_{i_\beta}$ at $(x,z) = (x_\beta,z_\beta)$ and extending in
the negative $x^9$-direction.  To be compatible with the
$\Omega$-deformation, these strings must sit at the origins of $\R^2$
and $\R^2_{67}$.  In the undeformed situation,
$\text{F1}^\dashrarrow_\alpha$ would break half of the sixteen supercharges
preserved by the D5-branes, and $\text{F1}^\dashuarrow_\beta$ would
further break half of the surviving eight supercharges.

The brane realization for the 't Hooft lines can be identified from
the fact that 't~Hooft lines in $\CN = 4$ super Yang--Mills theory in
four dimensions are the S-duals of Wilson lines.  As such, in the
worldvolume theory of D3-branes these lines are created by D1-branes,
which are the S-duals of fundamental strings.  Since D3-branes are
what the D5-branes become if we compactify $\R^2$ to a torus and apply
T-duality along its one-cycles, 't Hooft lines in the worldvolume
theory of the D5-branes are created by the T-duals of those D1-branes,
namely D3-branes.

Therefore, the 't Hooft lines going upward in
Figure~\ref{fig:lattice-so} are created by semi-infinite D3-branes
$\text{D3}_\gamma$, $\gamma = 1$, $\dotsc$, $k$, coming from
$x^8 = -\infty$ and hitting $\text{D5}_{i_\gamma}$ at
$(x,z) = (x_\gamma^\uarrow, z^\uarrow_\gamma)$.  The
choices~$i_\gamma$ of the D5-branes that these D3-branes hit determine
the charges of the 't~Hooft lines: for $G = \U(N)$, the $\gamma$th
't~Hooft line has charge $\diag(0, \dotsc, 0, 1, 0, \dotsc, 0)$, with
$1$ in the $i_\gamma$th entry.  Throwing away the center-of-mass
degrees of freedom of the D5-branes makes the charge traceless,
replacing it with the fractional charge~\eqref{eq:m-charge} or its
permutation.  This brane realization of monopoles is the S-dual of the
one studied in~\cite{Cherkis:1997aa}.

If a D3-brane creating an 't Hooft line curves in $T^2$, it also has
to curve in $\R^2_{89}$ by the same angle to preserve supersymmetry.
In particular, an 't Hooft line going downward is created by a
D3-brane hitting one of the D5-branes from the positive
$x^8$-direction.  This observation suggests the following construction
for the strips of surface operators.

When $\text{D3}_\gamma$ comes from $x^8 = -\infty$ and hits
$\text{D5}_{i_\gamma}$, it makes a right turn to move along~$T^2$, and
the two branes form a bound state.  This D3--D5 bound state creates
the surface operator whose left boundary is the $\gamma$th upward 't
Hooft line.  While maintaining the bound state, $\text{D3}_\gamma$ can
gradually shift its position in $E$.  When $\text{D3}_\gamma$ reaches
$(x,z) = (x_\gamma^\darrow, z^\darrow_\gamma)$, it makes a left
turn and leaves $\text{D5}_{i_\gamma}$.  Then $\text{D3}_\gamma$ goes
off to $x^8 = +\infty$, yielding the downward 't~Hooft line on the
right boundary of the surface operator,

The D3-branes break half of the four supercharges preserved by the
other branes.  We refer to the semi-infinite parts of
$\text{D3}_\gamma$ responsible for the upward and downward 't Hooft
lines as $\text{D3}^\uarrow_\gamma$ and
$\text{D3}^\darrow_\gamma$, respectively.

The brane configuration realizing the integrable lattice model is
summarized in Table~\ref{tab:BC-ILM}.

\begin{table}
  \centering
  \begin{tabular}{r@{:\quad}c@{\ $\times$ \ }c@{\ $\times$ \ }c@{\ $\times$ \ }c@{\ $\times$ \ }c}
    Spacetime & $\R^2$ & $T^2$ & $E$ & $\R^2_{67}$ & $\R^2_{89}$
    \\
    $\text{D5}_i$ & $\R^2$ & $T^2$ & $E$ &
    $\{0\}$ & $\{(\phi_x^i,\phi_y^i)\}$
    \\
    $\text{F1}^\dashrarrow_\alpha$ & $\{0\}$ & $\{y = y_\alpha\}$ &
    $\{z_\alpha\}$ & $\{0\}$ & $\{(x^8,\phi_y^{i_\alpha}) \mid x^8 \leq \phi_x^{i_\alpha}\}$
    \\
    $\text{F1}^\dashuarrow_\beta$ & $\{0\}$ & $\{x = x_\beta\}$ &
    $\{z_\beta\}$ & $\{0\}$ & $\{(\phi_x^{i_\beta},x^9) \mid x^9 \leq \phi_y^{i_\beta}\}$
    \\
    $\text{D3}^\uarrow_\gamma$ & $\R^2$ & $\{x = x_\gamma^\uarrow\}$ & $\{z_\gamma^\uarrow\}$ &
    $\{0\}$ & $\{(x^8,\phi_y^{i_\gamma}) \mid x^8 \leq \phi_x^{i_\gamma}\}$
    \\
    $\text{D3}^\darrow_\gamma$ & $\R^2$ & $\{x = x_\gamma^\darrow\}$ & $\{z_\gamma^\darrow\}$ &
    $\{0\}$ & $\{(x^8,\phi_y^{i_\gamma}) \mid x^8 \geq \phi_x^{i_\gamma}\}$
  \end{tabular}
  \caption{A brane configuration for an integrable lattice model.  The
    branes are placed in a background with nonzero RR two-form.
    $\text{D3}_\gamma$ forms a bound state with the D5-branes in the
    region $\{x_\gamma^\uarrow \leq x \leq x_\gamma^\darrow\}$ on $T^2$.}
  \label{tab:BC-ILM}
\end{table}

\subsection{\texorpdfstring{Brane tilings and class-$\CS_k$ theories}{Brane tilings and class-Sk theories}}

Tracing back the chain of dualities, we obtain another realization of
the same integrable lattice model.  By the application of T-duality on
$E$, S-duality, and T-duality on $E$ again,
$\text{F1}^\dashrarrow_\alpha$ and $\text{F1}^\dashuarrow_\beta$ are
converted to D3-branes $\text{D3}^\dashrarrow_\alpha$ and
$\text{D3}^\dashuarrow_\beta$, while $\text{D3}^\uarrow_\gamma$ and
$\text{D3}^\darrow_\gamma$ are converted to NS5-branes
$\text{NS5}^\uarrow_\gamma$ and $\text{NS5}^\darrow_\gamma$.  The dual brane configuration is summarized in
Table~\ref{tab:BC-BT}.

\begin{table}
  \centering
  \begin{tabular}{r@{:\quad}c@{\ $\times$ \ }c@{\ $\times$ \ }c@{\
        $\times$ \ }c@{\ $\times$ \ }c}
    Spacetime & $\R^2$ & $T^2$ & $E$ & $\R^2_{67}$ & $\R^2_{89}$
    \\
    $\text{D5}_i$ & $\R^2$ & $T^2$ & $E$ &
    $\{0\}$ & $\{(\phi_x^i,\phi_y^i)\}$
    \\
    $\text{D3}^\dashrarrow_\alpha$ & $\{0\}$ & $\{y = y_\alpha\}$ &
    $E$ & $\{0\}$ & $\{(x^8,\phi_y^{i_\alpha}) \mid x^8 \leq \phi_x^{i_\alpha}\}$
    \\
    $\text{D3}^\dashuarrow_\beta$ & $\{0\}$ & $\{x = x_\beta\}$ &
    $E$ & $\{0\}$ & $\{(\phi_x^{i_\beta},x^9) \mid x^9 \leq \phi_y^{i_\beta}\}$
    \\
    $\text{NS5}^\uarrow_\gamma$ & $\R^2$ & $\{x = x_\gamma^\uarrow\}$ &
    $E$ & $\{0\}$ & $\{(x^8,\phi_y^{i_\gamma}) \mid x^8 \leq \phi_x^{i_\gamma}\}$
    \\
    $\text{NS5}^\darrow_\gamma$ & $\R^2$ & $\{x = x_\gamma^\darrow\}$ &
    $E$ & $\{0\}$ & $\{(x^8,\phi_y^{i_\gamma}) \mid x^8 \geq \phi_x^{i_\gamma}\}$
  \end{tabular}
  \caption{A brane tiling configuration for an integrable lattice
    model.  The product between $\R^2 \times \R^2_{67}$ and $E$ is
    twisted by rotations of $\R^2$ and $\R^2_{67}$ in opposite
    directions.}
  \label{tab:BC-BT}
\end{table}

Each NS5-brane forms a bound state with the D5-branes over a colored
region in Figure~\ref{fig:lattice-so}.  This bound state of $N$
D5-branes and one NS5-brane is called an $(N,1)$ 5-brane.  In this
terminology, a stack of $N$ D5-branes may be referred to as an $(N,0)$
5-brane.  Our choice of the names for various backgrounds was
motivated by this 5-brane interpretation.

The 5-brane system realizes a five-dimensional $\CN = 1$
supersymmetric gauge theory on
$\R^2 \times S^1 \times E$~\cite{Aharony:1997ju, Aharony:1997bh}, with
the product between $\R^2$ and $E$ being a twisted one.  The D3-branes
create three-dimensional defects supported on
$\{0\} \times S^1 \times E$ in this theory.  Thus, the partition
function of the lattice model translates to the correlation function
of these defects in this theory, also known as the
\emph{supersymmetric index} of the theory on
$\R^2 \times S^1 \times E$ in the presence of the defects.

If we wish, we can introduce additional 't Hooft lines in the
horizontal direction and make a tricolor checkerboard pattern on
$T^2$, as in Figure~\ref{fig:BT-D}; or for that matter, we can
consider entirely different patterns of $(N,0)$ and $(N,\pm 1)$
background regions, such as the one shown in Figure~\ref{fig:BT-T}.
Such configurations of 5-branes, called \emph{brane tilings}, realize
four-dimensional $\CN = 1$ supersymmetric gauge theories on
$\R^2 \times E$~\cite{Hanany:2005ve, Franco:2005rj}.  These theories
have multiple $\SU(N)$ gauge (and flavor) groups, one for each $(N,0)$
background region, and chiral multiplets in the bifundamental
representations under two $\SU(N)$ gauge groups associated to $(N,0)$
background regions sharing a vertex.

\begin{figure}
  \centering
  \subfloat[\label{fig:BT-D}]{
    \begin{tikzpicture}[scale=1.2]
      \fill[lshaded] (0,0) rectangle (3, 2);

      \begin{scope}[shift={(0.25,0)}]
        \fill[ws] (0,0) rectangle (0.5,2);
        \fill[ws] (1,0) rectangle (1.5,2);
        \fill[ws] (2,0) rectangle (2.5,2);
      \end{scope}

      \begin{scope}[shift={(0,0.25)}]
        \fill[ws] (0,0) rectangle (3,0.5);
        \fill[ws] (0,1) rectangle (3,1.5);
      \end{scope}

      \begin{scope}[shift={(0.25,0.25)}]
        \fill[dshaded] (0,0) rectangle (0.5,0.5);
        \fill[dshaded] (1,0) rectangle (1.5,0.5);
        \fill[dshaded] (2,0) rectangle (2.5,0.5);
        \fill[dshaded] (0,1) rectangle (0.5,1.5);
        \fill[dshaded] (1,1) rectangle (1.5,1.5);
        \fill[dshaded] (2,1) rectangle (2.5,1.5);
      \end{scope}

      \begin{scope}[shift={(0.25,0)}]
        \draw[z->] (0,0) -- (0,2);
        \draw[z->] (0.5,2) -- (0.5,0);
        \draw[z->] (1,0) -- (1,2);
        \draw[z->] (1.5,2) -- (1.5,0);
        \draw[z->] (2,0) -- (2,2);
        \draw[z->] (2.5,2) -- (2.5,0);
      \end{scope}

      \begin{scope}[shift={(0,0.25)}]
        \draw[z->] (3,0) -- (0,0);
        \draw[z->] (0,0.5) -- (3,0.5);
        \draw[z->] (3,1) -- (0,1);
        \draw[z->] (0,1.5) -- (3,1.5);
      \end{scope}

      \draw[frame] (0,0) rectangle (3, 2);
    \end{tikzpicture}
  }
  \qquad
  \subfloat[\label{fig:BT-T}]{
    \begin{tikzpicture}[scale=1.2]
      \fill[ws] (0,0) rectangle (3, 2);

      \fill[lshaded] (0.5,0.25) rectangle (0.75,0.5);
      \fill[lshaded] (0.5,1.25) rectangle (0.75,1.5);
      \fill[lshaded] (1.5,0.25) rectangle (1.75,0.5);
      \fill[lshaded] (1.5,1.25) rectangle (1.75,1.5);
      \fill[lshaded] (2.5,0.25) rectangle (2.75,0.5);
      \fill[lshaded] (2.5,1.25) rectangle (2.75,1.5);

      \fill[dshaded] (0.25,0) rectangle (0.5,0.25);
      \fill[dshaded] (1.25,0) rectangle (1.5,0.25);
      \fill[dshaded] (2.25,0) rectangle (2.5,0.25);

      \fill[dshaded] (2.75,0.5) rectangle (3,0.75);
      \fill[dshaded] (2.75,1.5) rectangle (3,1.75);

      \fill[dshaded] (0,0.5) -- (0.5,0.5) -- (0.5,1.25) -- (0.25,1.25)
      -- (0.25,0.75) -- (0,0.75) -- cycle;

      \fill[dshaded] (0.75,0.5) -- (1.5,0.5) -- (1.5,1.25) -- (1.25,1.25)
      -- (1.25,0.75) -- (0.75,0.75) -- cycle;

      \fill[dshaded] (1.75,0.5) -- (2.5,0.5) -- (2.5,1.25) -- (2.25,1.25)
      -- (2.25,0.75) -- (1.75,0.75) -- cycle;
      
      \fill[dshaded] (0,1.5) -- (0.5,1.5) -- (0.5,2) -- (0.25,2)
      -- (0.25,1.75) -- (0,1.75) -- cycle;

      \fill[dshaded] (0.75,1.5) -- (1.5,1.5) -- (1.5,2) -- (1.25,2)
      -- (1.25,1.75) -- (0.75,1.75) -- cycle;

      \fill[dshaded] (1.75,1.5) -- (2.5,1.5) -- (2.5,2) -- (2.25,2)
      -- (2.25,1.75) -- (1.75,1.75) -- cycle;

      \draw[z->] (0,0.75) -- (0.25,0.75) -- (0.25,1.25) -- (0.75,1.25)
      -- (0.75,1.75) -- (1.25,1.75) -- (1.25,2);

      \draw[z->] (1.25,0) -- (1.25,0.25) -- (1.75,0.25)
      -- (1.75,0.75) -- (2.25,0.75) -- (2.25,1.25)
      -- (2.75,1.25) -- (2.75,1.75) -- (3,1.75);
      
      \draw[z->] (0,1.75) -- (0.25,1.75) -- (0.25,2);
      
      \draw[z->] (0.25,0) -- (0.25,0.25) -- (0.75,0.25)
      -- (0.75,0.75) -- (1.25,0.75) -- (1.25,1.25)
      -- (1.75,1.25) -- (1.75,1.75) -- (2.25,1.75)
      -- (2.25,2);
      
      \draw[z->] (2.25,0) -- (2.25,0.25) -- (2.75,0.25)
      -- (2.75,0.75) -- (3,0.75);

      \draw[z->] (0.5,2) -- (0.5,0);
      \draw[z->] (1.5,2) -- (1.5,0);
      \draw[z->] (2.5,2) -- (2.5,0);

      \draw[z->] (3,0.5) -- (0,0.5);
      \draw[z->] (3,1.5) -- (0,1.5);

      \draw[frame] (0,0) rectangle (3, 2);
    \end{tikzpicture}
  }
  \caption{(a) A tricolor checkerboard brane tiling.  (b) Another
    brane tiling.}
  \label{fig:BT}
\end{figure}
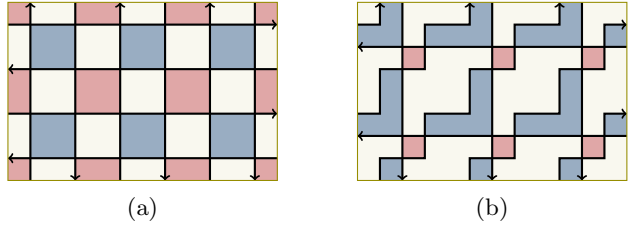

The theories realized by the brane tilings in Figure~\ref{fig:BT} are
also examples of $\CN = 1$ supersymmetric theories of \emph{class
  $\CS_k$}~\cite{Gaiotto:2009we, Gaiotto:2009hg, Gaiotto:2015usa}.
Theories of class $\CS_k$ describe the dynamics of M5-branes probing a
transverse $\C^2/\Z_k$ singularity, compactified on punctured Riemann
surfaces which in our case are tori.  This brane setup is obtained by
T-duality in the horizontal direction of $T^2$, followed by a lift to
M-theory.  The D3-branes are lifted to M2- and M5-branes, producing
surface operators in the class-$\CS_k$ theories.

It is known that surface operators act on the supersymmetric indices
of brane tiling and class-$\CS_k$ theories as difference
operators~\cite{Gaiotto:2012xa, Gadde:2013ftv, Alday:2013kda,
  Gaiotto:2015usa, Maruyoshi:2016caf, Yagi:2017hmj, Ito:2016fpl}.  Our
construction shows that these difference operators are nothing but
transfer matrices of L-operators.  This result, obtained
in~\cite{Maruyoshi:2016caf, Yagi:2017hmj} from the perspective of
brane tilings, was a primary motivation for us to study surface
operators in four-dimensional Chern--Simons theory.

\subsection{Linear quiver theories}

Another interesting chain of dualities we can apply to the brane
configuration in Table~\ref{tab:BC-ILM} is S-duality and T-duality in
the horizontal direction of $T^2$.  This turns $\text{D5}_i$ into
NS5-branes $\text{NS5}_i$, $\text{F1}^\dashrarrow_\alpha$ into
D0-branes $\text{D0}^\dashrarrow_\alpha$,
$\text{F1}^\dashuarrow_\beta$ into D2-branes
$\text{D2}^\dashuarrow_\beta$, and $\text{D3}^\uarrow_\gamma$ and
$\text{D3}^\darrow_\gamma$ into D4-branes $\text{D4}^\uarrow_\gamma$
and $\text{D4}^\darrow_\gamma$, as summarized in
Table~\ref{tab:BC-LQ}.  A schematic picture of this brane setup is
shown in Figure~\ref{fig:HW}.  These branes are placed in a background
with a nonzero B-field.

\begin{table}
  \centering
  \begin{tabular}{r@{:\quad}c@{\ $\times$ \ }c@{\ $\times$ \ }c@{\ $\times$ \ }c@{\ $\times$ \ }c}
    Spacetime & $\R^2$ & $T^2$ & $E$ & $\R^2_{67}$ & $\R^2_{89}$
    \\
    $\text{NS5}_i$ & $\R^2$ & $T^2$ & $E$ &
    $\{0\}$ & $\{(\phi_x^i,\phi_y^i)\}$
    \\
    $\text{D0}^\dashrarrow_\alpha$ & $\{0\}$ & $\{(x_\alpha,y_\alpha)\}$ &
    $\{z_\alpha\}$ & $\{0\}$ & $\{(x^8,\phi_y^{i_\alpha}) \mid x^8 \leq \phi_x^{i_\alpha}\}$
    \\
    $\text{D2}^\dashuarrow_\beta$ & $\{0\}$ & $T^2$ &
    $\{z_\beta\}$ & $\{0\}$ & $\{(\phi_x^{i_\beta},x^9) \mid x^9 \leq \phi_y^{i_\beta}\}$
    \\
    $\text{D4}^\uarrow_\gamma$ & $\R^2$ & $T^2$ & $\{z^\uarrow_\gamma\}$ & $\{0\}$ &
    $\{(x^8,\phi_y^{i_\gamma}) \mid x^8 \leq \phi_x^{i_\gamma}\}$
    \\
    $\text{D4}^\darrow_\gamma$ & $\R^2$ & $T^2$ & $\{z^\darrow_\gamma\}$ & $\{0\}$ &
    $\{(x^8,\phi_y^{i_\gamma}) \mid x^8 \geq \phi_x^{i_\gamma}\}$
  \end{tabular}
  \caption{A brane configuration of Hanany--Witten type for an
    integrable lattice model.}
  \label{tab:BC-LQ}
\end{table}

\begin{figure}
  \centering
  \subfloat[\label{fig:HW}]{
  \begin{tikzpicture}[align at bottom]
    \fill[brown!8] (-1,0) -- ++(-135:0.4) -- ++(0:4.7) -- ++(45:0.8)
    -- ++(0:-4.7) -- cycle;

    \begin{scope}[shift={($(-2.5,0)+(-135:0.4)$)}]
      \draw[->,>=stealth] (0,0) -- ++(90:0.5) node[above] {$z$};
      \draw[->,>=stealth] (0,0) -- ++(0:0.5) node[right, yshift=2pt] {$x^8$};
      \draw[->,>=stealth] (0,0) -- ++(45:0.5) node[above right=-2pt] {$x^9$};
    \end{scope}

    \draw[thick, midnightblue, shift={(0:{0.2/sqrt(2)})}] (-1,1) -- ++(0:2);
    \draw[thick, firebrick] ($(1,1.1)+(0:{0.2/sqrt(2)})$)
    -- ++($(0:0.5)-(0:{0.2/sqrt(2)})$);
    \node at (1.9,1.1) {$\dots$};
    \draw[thick, firebrick] (2.2,1.1)
    -- ++($(0:1.5)+(0:{0.2/sqrt(2)})$);
    \draw[thick, shift={(45:0.2)}, olive] (1,0)
    -- ++(90:2.5) node[above, black] {$\text{NS5}_2$};

    \draw[thick, midnightblue] (-1,0.5) -- ++(0:1);
    \draw[line width=3pt, white] (0,0.4) -- ++(0:1.5);
    \draw[thick, firebrick] (0,0.4) -- ++(0:1.5);
    \node at (1.9,0.4) {$\dots$};
    \draw[thick, firebrick] (2.2,0.4) -- ++(0:1.5);
    \draw[line width=3pt, white] (0,0.9) -- ++(90:0.2);
    \draw[line width=3pt, white] (0,2.1) -- ++(90:0.2);
    \draw[thick, olive] (0,0) -- ++(90:2.5)
    node[above, black] {$\text{NS5}_1$};

    \draw[line width=3pt, white] ($(-1,1.8)-(0:{0.2/sqrt(2)})$)
    -- ++($(0:2.5)+(0:{0.2/sqrt(2)})$);
    \draw[thick, midnightblue] ($(-1,1.8)-(0:{0.2/sqrt(2)})$)
    -- ++($(0:2.5)+(0:{0.2/sqrt(2)})$);
    \node at (1.9,1.8) {$\dots$};
    \draw[thick, midnightblue] (2.2,1.8)
    -- ++($(0:0.5)-(0:{0.2/sqrt(2)})$);
    \draw[thick, firebrick] ($(2.7,1.9)-(0:{0.2/sqrt(2)})$)
    -- ++($(0:1)+(0:{0/sqrt(2)})$);
    \draw[line width=3pt, shift={(45:-0.2)}, white] (2.7,0.5)
    -- ++(90:1.2);
    \draw[thick, shift={(45:-0.2)}, olive] (2.7,0)
    -- ++(90:2.5) node[above, black] {$\text{NS5}_N$};

    \node at (-0.5,1.5) {$\vdots$};
    \node at (3.2,1.6) {$\vdots$};
    
    \draw[decorate, decoration={brace}, xshift=-4pt]
    ($(-1,0.45)-(0:{0.2/sqrt(2)})$) -- ++(90:1.4)
    node[left=4pt, pos=0.5] {$k$ $\text{D4}^\suarrow$};

    \draw[decorate, decoration={brace, mirror}, xshift=4pt]
    ($(3.7,0.35)+(0:{0.2/sqrt(2)})$) -- ++(90:1.6)
    node[right=4pt, pos=0.5] {$k$ $\text{D4}^\sdarrow$};

    \draw[line width=3pt, white, shift={(45:0.2)}] ($(1,0.6)+(45:-0.4)$)
    -- ++(-135:0.218);
    \draw[densely dashed, thick, shift={(45:0.2)}] ($(1,0.6)$)
    -- ++(-135:1.2) node[below, shift={(-0.1,0)}] {$\text{D2}^\sdashuarrow$};

    \draw[densely dashed,thick] (-1,2.2)
    node[left] {$\text{D0}^\sdashrarrow$} -- ++($(0:1)$);
  \end{tikzpicture}
  }
  \qquad
  \subfloat[\label{fig:LQ}]{
    \begin{tikzpicture}[align at bottom]
      \draw (0,0) node [fnode, minimum size=12pt] {$k$}
      -- (1,0) node[gnode, minimum size=12pt] {$k$}
      -- (2,0) node[gnode, minimum size=12pt] {$k$}
      -- (2.5,0);
      \node at (2.75,0) {$\,\dots$};
      \draw (3,0) -- (3.5,0) node[gnode, minimum size=12pt] {$k$}
      -- (4.5,0) node [fnode, minimum size=12pt] {$k$};
      
      \draw[decoration={brace, mirror}, decorate, yshift=-10pt] (-0.25,0)
      -- node[pos=0.5, yshift=-10pt] {$N+1$ nodes} (4.75,0); 

      \node at (0,-1.2) {};
    \end{tikzpicture}
  }
  \caption{(a) A D4--NS5 brane configuration of Hanany--Witten type
    with additional D0- and D2-branes. (b)~The linear quiver for the
    theory realized by the D4--NS5 brane configuration.}
  \label{tab:BC-HW}
\end{figure}
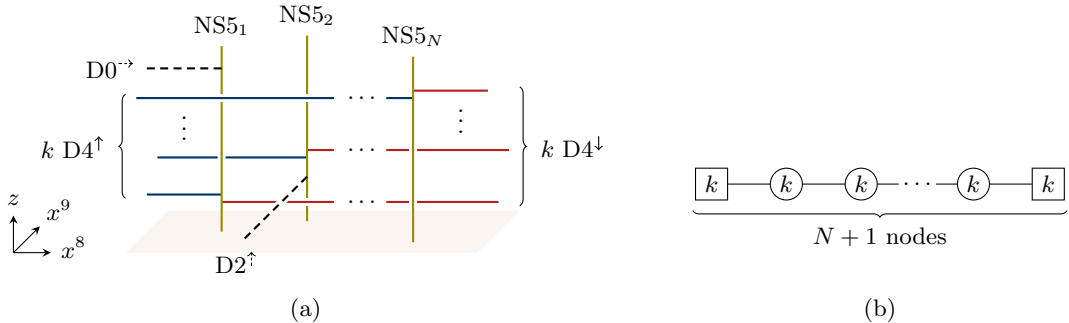

Let us decompactify the holomorphic surface $C = E$ to $\C$.  Then,
the part of the system consisting of the D4- and NS5-branes is a
well-known brane configuration studied in Witten's classic
paper~\cite{Witten:1997sc}, which builds on his earlier
work~\cite{Hanany:1996ie} with Hanany.

The D4--NS5 brane configuration realizes a four-dimensional $\CN = 2$
supersymmetric gauge theory on $\R^2 \times T^2$.  This theory is
described by a linear quiver shown in Figure~\ref{fig:LQ}.  A circle
node represents a vector multiplet for an $\SU(k)$ gauge group, a
square node an $\SU(k)$ flavor group, and an edge a bifundamental
hypermultiplet.

The value $\phi_x^{i+1} - \phi_x^i$ determines the gauge coupling of
the $i$th $\SU(k)$ gauge group, while the difference of the periodic
scalars on $\text{NS5}_i$ and $\text{NS5}_{i+1}$ gives the
$\theta$-angle for this group; together they form a complexified gauge
coupling.  The positions $z^\uarrow_\gamma$ and $z^\darrow_\gamma$
of the D4-branes in $C$ determine the masses of the hypermultiplets
charged under the left and right $\SU(k)$ flavor groups, respectively.
For generic values of $\phi_y^i$, the theory is in the Higgs phase in
which the gauge symmetry is completely broken.

The topological twist used in the construction of the six-dimensional
topological--holomorphic theory becomes the Donaldson--Witten twist of
the linear quiver theory, as can be seen as follows.  If there are
only the NS5- and D4-branes, the dualities used above can be applied
to a more general setup where $M$ is the product of a three-manifold
$W$ and $S^1$, instead of $\R^2 \times T^2$.  By dimensional reduction
on $S^1$, the linear quiver theory reduces to a three-dimensional
$\CN = 4$ supersymmetric gauge theory on $W$.  There are two
topological twists for a general $\CN = 4$ supersymmetric gauge
theory~\cite{Blau:1996bx}, and what we get here is the one using the
$\SU(2)$ R-symmetry coming from the rotation symmetry of $\R^3_{679}$.
This is known to be the dimensional reduction of the Donaldson--Witten
twist.

The presence of the B-field and other $\eps$-dependent part of the
background has the effect of introducing the standard
$\Omega$-deformation.  A quick way to see this is to note that if we
apply S-duality, T-duality in the horizontal direction of $T^2$, and
T-duality on $E$ to the brane configuration in Table~\ref{tab:BC-BT},
we arrive at an almost identical Hanany--Witten configuration, in
which $E$ is replaced with the dual elliptic curve~$E^\vee$.  The
linear quiver theory realized by this brane configuration is clearly
subjected to the standard $\Omega$-deformation because the last
T-duality is applied to a twisted product of $\R^2$ and $E$ and, as
discussed earlier, this is how the standard $\Omega$-deformation is
constructed.  The theories realized by the two Hanany--Witten
configurations are related by a diffeomorphism between the elliptic
curves, so the deformations they receive are the same.

The D0-branes insert local operators in the linear quiver theory,
while the D2-branes create surface operators supported on
$\{0\} \times T^2$.  In particular, each D0-brane acts on the
partition function of the $\Omega$-deformed linear quiver theory as a
transfer matrix.

Let us consider the situation where all $\text{D4}^\uarrow_\gamma$
and $\text{D4}^\darrow_\gamma$ end on the same NS5-brane, say
$\text{NS5}_1$.  In this case, this transfer matrix is constructed
from $k$ copies of a rational version of $L^{(N,1)}$ corresponding to
the decompactification of $E$ to $\C$.%
\footnote{The dynamical parameter is absent for $C = \C$ as we explain
  in section~\ref{sec:2dNS}, so the decompactification acts as if
  wrapping $T^2$ with a surface operator and then taking the rational
  limit.  The transfer matrix still consists of $L^{(N,1)}$ if the
  positions of the 't Hooft lines are pairwise interchanged.  To be
  precise, the L-operator that enters the transfer matrix is not equal
  but gauge equivalent to $L^{(N,1)}$ because we have defined
  $L^{(N,1)}$ as the L-operator in the background with
  $\CA_x = \CA_y = 0$.}

If we further specialize to the case $N = 2$, these L-operators are
R-matrices for the rational six-vertex model (the rational limit of
the eight-vertex model) whose vertical lines carry Verma modules of
$\slf_2$.  The module structure comes from dynamical creation and
annihilation of D2-branes stretched between
$\text{D4}^\darrow_\gamma$ and $\text{NS5}_2$~\cite{Dorey:2011pa}.

A transfer matrix of the rational six-vertex model is a generating
function of the conserved charges of the XXX spin chain.  Thus, our
brane construction naturally explains the appearance of the
``noncompact'' XXX spin chain of length $k$, whose spins take values
in Verma modules of $\slf_2$, from the $\Omega$-deformed $\CN = 2$
supersymmetric gauge theory with a single $\SU(k)$ gauge group and two
fundamental hypermultiplets \cite{Nekrasov:2009rc, Dorey:2011pa}.
This phenomenon generalizes to any $N \geq 2$, for which an $\slf_N$
spin chain arises~\cite{Chen:2011sj, Nekrasov:2012xe,
  Nekrasov:2013xda}.

Now let us make $C$ compact again, taking $C = E$.  Then, the D4--NS5
brane configuration realizes a six-dimensional lift of the linear
quiver theory compactified on $E$, as one can see by applying
T-duality on $E$.  Correspondingly, the six-vertex model is promoted
to the eight-vertex model, whose transfer matrix generates the
conserved charges of the XYZ spin chain.  If we compactify only one
direction so that $C = \C^\times$, the brane configuration produces a
five-dimensional gauge theory and the XXZ spin chain.

\subsection{Nekrasov--Shatashvili realization of compact spin chains}
\label{sec:2dNS}

In the same brane configuration, the crossings of the D0- and
D2-branes create transfer matrices constructed from R-matrices in the
vector representation of $\slf_N$.  Therefore, the $\slf_N$ spin
chains with spins in the vector representation also appear in this
setup.  It is interesting to look at these spin chains from the point
of view of the D2-branes.

For the moment let us take $N = 2$, so there are two NS5-branes.  The
possible configurations of $n$ D2-branes ending on either NS5-brane
are classified by an integer $M$ such that $0 \leq M \leq n$, namely
the number of D2-branes ending on $\text{NS5}_2$.  This is the magnon
number of the spin chain, counting the total number of ``up'' spins in
the chain.  A case with $M = 2$ is illustrated in
Figure~\ref{fig:D2-NS5}.

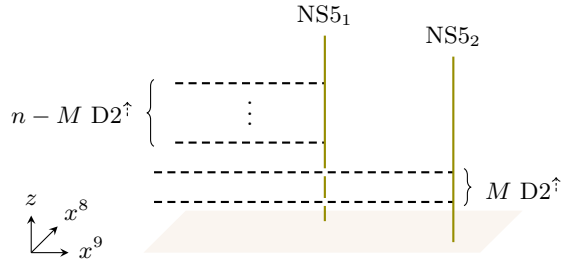
\begin{figure}
  \centering
  \begin{tikzpicture}[align at bottom]
    \fill[brown!8] (-2,0) -- ++(-135:0.4) -- ++(0:4.5) -- ++(45:0.8)
    -- ++(0:-4.5) -- cycle;

    \begin{scope}[shift={($(-3.5,0)+(-135:0.4)$)}]
      \draw[->,>=stealth] (0,0) -- ++(90:0.5) node[above] {$z$};
      \draw[->,>=stealth] (0,0) -- ++(0:0.5) node[right, yshift=2pt] {$x^9$};
      \draw[->,>=stealth] (0,0) -- ++(45:0.5) node[above right=-2pt] {$x^8$};
    \end{scope}

    \draw[densely dashed,thick] ($(-2,1.2)+(0:{0.2/sqrt(2)})$) -- ++(0:2);
    \draw[densely dashed,thick] ($(-2,2)+(0:{0.2/sqrt(2)})$) -- ++(0:2);
    \draw[-,thick, shift={(45:0.2)}, olive] (0,0) -- ++(90:2.5)
    node[above, black] {$\text{NS5}_1$};

    \draw[line width=3pt, white] ($(-2,0.4)-(0:{0.2/sqrt(2)})$) -- ++(0:3);
    \draw[line width=3pt, white] ($(-2,0.8)-(0:{0.2/sqrt(2)})$) -- ++(0:3);
    \draw[densely dashed,thick] ($(-2,0.4)-(0:{0.2/sqrt(2)})$) -- ++(0:4);
    \draw[densely dashed,thick] ($(-2,0.8)-(0:{0.2/sqrt(2)})$) -- ++(0:4);
    \draw[-,thick, shift={(45:-0.2)}, olive] (2,0) -- ++(90:2.5)
    node[above, black] {$\text{NS5}_2$};

    \node[shift={(0:{0.2/sqrt(2)})}] at (-1,1.7) {$\vdots$};

    \draw[decorate, decoration={brace,mirror}, xshift=4pt]
    ($(2,0.35)-(0:{0.2/sqrt(2)})$) -- ++(90:0.5)
    node[right=4pt, pos=0.5] {$M$ $\text{D2}^\sdashuarrow$};

    \draw[decorate, decoration={brace}, xshift=-4pt]
    (-2,1.15) -- (-2,2.05)
    node[left=4pt, pos=0.5] {$n - M$ $\text{D2}^\sdashuarrow$};
  \end{tikzpicture}
  \caption{A brane configuration for a two-dimensional $\CN = (4,4)$
    supersymmetric gauge theory.}
  \label{fig:D2-NS5}
\end{figure}

In the case when $C = \C$ and the $\Omega$-deformation is absent, the
D2--NS5 brane configuration with fixed $M$ realizes an $\CN = (4,4)$
supersymmetric gauge theory on~$T^2$.  This theory has a $\U(M)$ gauge
group and a hypermultiplet in the bifundamental representation of the
gauge group and a $\U(n)$ flavor symmetry.

The separation $\phi_y^2 - \phi_y^1$ of the NS5-branes in the
$x^9$-direction determines the gauge coupling.  The separation
$\phi_x^2 - \phi_x^1$ in the $x^8$-direction is proportional to the
Fayet--Iliopoulos (FI) parameter $r$ for the $\U(1)$ part of the gauge
group, and it combines with the two-dimensional $\theta$-angle
$\vtheta$ to form the complexified FI parameter
\begin{equation}
  t = \frac{\vtheta}{2\pi} + \iu r \,.
\end{equation}
The positions $z_\beta$ of $\text{D2}^\dashuarrow_\beta$ in $C$ are
twisted masses of the hypermultiplet.  These complex mass parameters
may be thought of as the eigenvalues of the scalar field in the
nondynamical vector multiplet for the $\U(n)$ flavor symmetry.

To this theory the $\Omega$-deformation is applied.  This makes use of
the $\U(1)$ isometry of a plane transverse to the D2-branes, and breaks
the $\CN = (4,4)$ supersymmetry to $\CN = (2,2)$ supersymmetry.  In
the language of $\CN = (2,2)$ supersymmetry, the $\CN = (4,4)$ vector
multiplet consists of a vector multiplet and a chiral multiplet in the
adjoint representation, whereas the $\CN = (4,4)$ fundamental
hypermultiplet splits into a pair of fundamental and antifundamental
chiral multiplets.  The $\Omega$-deformation gives the adjoint chiral
multiplet a twisted mass $u$ proportional to $\eps$, and the
fundamental and antifundamental chiral multiplets twisted masses
$-u/2$ and $u/2$, respectively~\cite{Hellerman:2011mv}.

The topological twist is the A-twist here.  We can see this from the
fact that the scalar field $\sigma$ of the vector multiplet for the
gauge symmetry, whose eigenvalues parametrize the positions of the
D2-branes on $C$, is unaffected by the twist.  Alternatively, we may
note that the D2-branes are surface operators in the theory on the
D4-branes, and the Donaldson--Witten twist reduces to the A-twist in
two dimensions.

Suppose that $r \neq 0$ and the twisted masses are vanishing,
including those induced by the $\Omega$-deformation.  Then, the theory
is in the Higgs phase and flows in the infrared to a topological sigma
model whose target space is the cotangent bundle $T^*\Gr(M,n)$ of the
Grassmannian $\Gr(M,n)$, endowed with a hyperk\"ahler metric.  This is
the A-model~\cite{Witten:1988xj}, and its algebra of local operators
is given by the quantum cohomology ring $QH^\bullet(T^*\Gr(M,n))$.  By
the state--operator correspondence this is isomorphic as a vector
space to the Hilbert space of states.

Now we turn on all the twisted masses.  As the supercharge of an
A-twisted gauge theory squares to a gauge transformation generated by
the adjoint scalar in the vector multiplet, this amounts to working
equivariantly with respect to the $\U(n)$ flavor symmetry as well as
the $\U(1)$ isometry used in the $\Omega$-deformation.  The algebra of
local operators is therefore deformed to the equivariant quantum
cohomology $QH^\bullet_{(\C^\times)^n \times \C^\times}(T^*\Gr(M,n))$,
where $(\C^\times)^n$ is the diagonal torus of the complexification of
the $\U(n)$ flavor symmetry and the last $\C^\times$ is the
complexification of the $\U(1)$ isometry.

The D0-branes create local operators in the theory.  According to our
brane construction, these operators can be understand as transfer
matrices, constructed from a rational version of $R^{(N,0)}$.
Although there is a rational solution of the dynamical Yang--Baxter
equation~\cite{MR1645196}, what we get here is a \emph{nondynamical}
one: flat connections on $C = \C$ are all gauge equivalent to zero, so
the relevant moduli space has no directions that would correspond to a
dynamical parameter.  (In our brane setup $\phi$ goes to a constant
value at the infinity of $C$, and with this boundary condition the
argument in section~\ref{sec:4dCS-from-6d} applies.)  The transfer
matrices of this rational R-matrix are those of the XXX spin chain
whose spins are in the vector representation.

By integrability, these transfer matrices generate a commutative
algebra of operators, called a \emph{Bethe algebra}, which has the
same dimension as the Hilbert space of the spin chain.  Since the
total spin is a conserved quantity in the XXX spin chain, this is the
direct sum of $n+1$ commutative algebras, each acting on a subspace of
a fixed magnon number.  In the present setup, the local operators
created by the D0-branes generate the summand corresponding to the
$M$-magnon sector.  The dimension of this summand is actually equal to
the dimension of
$QH^\bullet_{(\C^\times)^n \times \C^\times}(T^*\Gr(M,n))$, so the
D0-branes generate the whole algebra of local operators of the
A-model.  Hence, the Bethe algebra for the $M$-magnon sector of the
XXX spin chain of length $n$ is isomorphic to
$QH^\bullet_{(\C^\times)^n \times \C^\times}(T^*\Gr(M,n))$.

Our brane construction thus explains the correspondence between the
XXX spin chain and the equivariant cohomology of the cotangent bundles
of Grassmannians, discovered by Nekrasov and
Shatashvili~\cite{Nekrasov:2009uh, Nekrasov:2009ui} and mathematically
developed in~\cite{MR3118573, Maulik:2012wi, MR3354333}.  The above
brane configuration has been studied in this context
in~\cite{Orlando:2011nc}.

If we take $C = \C^\times$, the brane configuration realizes a
three-dimensional lift of the above theory, and the rational R-matrix
is replaced with a trigonometric one.  This is again a nondynamical
R-matrix for the following reason.  Physically, we expect that the
trigonometric case is equivalent to the limit $\tau \to \iu\infty$ of
the elliptic case where $E$ degenerates to a cylinder.  If the
dynamical parameter $\uplambdab$ is fixed in this limit, the dynamical
elliptic R-matrix becomes a dynamical trigonometric R-matrix.  In our
case, however, $\uplambdab$ is determined by the background gauge
field according to formula~\eqref{eq:uplambda}.  Provided that the
holonomy of $A$ around the one-cycle $\CC_a$ is generic and fixed,
taking $\tau \to \iu\infty$ entails the limit
$|\uplambda_i| \to \infty$.  In this limit of infinite dynamical
parameter, the dynamical trigonometric R-matrix reduces to a
nondynamical one.

Hence, the three-dimensional theory corresponds to the $M$-magnon
sector of the XXZ spin chain, whose transfer matrices coincide with
those of the nondynamical trigonometric R-matrix.  The trigonometric
case of the Nekrasov--Shatashvili correspondence has been
mathematically established~\cite{MR3350271}.

For the elliptic case $C = E$, one may be tempted to say that the
four-dimensional lift of the theory should correspond to the
``$M$-magnon sector'' of the XYZ spin chain.  However, such a
statement does not make sense since the total spin is not a conserved
quantity in the XYZ spin chain.  This is not a contradiction.  The
point is that the R-matrix for the XYZ spin chain is Baxter's
nondynamical elliptic R-matrix, while what the theory gives is the
dynamical elliptic R-matrix $R^{(N,0)}$.  The transfer matrices of
$R^{(N,0)}$ do preserve the total spin.  The correct statement is
therefore that the Higgs branch of the four-dimensional theory
corresponds to the Bethe algebra for the $M$-magnon sector of the spin
chain defined by $R^{(N,0)}$.

For general $N \geq 2$, the configurations of $n$ D2-branes ending on
$N$ NS5-branes are classified by integers
$\mathbf{M} = (M_0, \dotsc, M_N)$ such that
\begin{equation}
  0 = M_0 \leq M_1 \leq \dotsb \leq M_N = n \,.
\end{equation}
The gauge theory on the D2-branes has gauge group
$\U(M_1) \times \dotsb \times \U(M_{N-1})$ and flavor group $\U(M_N)$,
and a bifundamental hypermultiplet of $\U(M_i) \times \U(M_{i+1})$ for
each $i = 1$, $\dotsc$, $N-1$.  For generic values of the FI parameters,
it flows to the A-model (or its three- or four-dimensional lift) whose
target space is the cotangent bundle $T^*\CF_{\mathbf{M}}$ of the
partial flag manifold $\CF_{\mathbf{M}}$ parametrizing chains of
subspaces
\begin{equation}
  0 = F_0 \subset F_1 \subset \dotsb \subset F_N = \C^n \,,
  \qquad
  \dim F_i = M_i \,.
\end{equation}
Everything we have said about $T^*\Gr(M,n)$ generalizes
straightforwardly to $T^*\CF_{\mathbf{M}}$, and we find that the
subsector of an $\slf_N$ spin chain with total $\slf_N$ weight
$\sum_{i=1}^N (M_i - M_{i-1}) \upomegab_{N-i+1}$ arises from this
theory.

\subsection{Q-operators}

In all of these spin chains there are important operators called
\emph{Q-operators}, which are of great help in solving the spectra.
One of the main results of~\cite{Nekrasov:2009uh, Nekrasov:2009ui} is
that the Q-operator $Q(z)$ for the XXX spin chain is identified with
the local operator
\begin{equation}
  \label{eq:det(z-sigma)}
  \det(z - \sigma)
\end{equation}
in the gauge theory.  (Similar results have been obtained for the
trigonometric case in~\cite{Pushkar:2016qvw, Koroteev:2017nab}.)  We
can understand this identification as follows.

Let us enrich the system by introducing an additional 't Hooft line
along the horizontal direction of the lattice, with the Dirac string
extending along $C = \C$ and going off to $\infty$.  By following the
chain of dualities we see that this is another kind of D2-brane, which
covers the $\Omega$-deformation plane $\R^2$ and ends on one of the
two NS5-branes, say $\text{NS5}_2$, from the positive $x^9$-direction.
(Which NS5-brane it ends on is immaterial due to the symmetry under
flipping of all spins, or the isomorphism $\Gr(M,n) \iso \Gr(n-M,n)$,
or Hanany--Witten transitions involving D6-branes.)  This D2-brane is
supported at a point on $T^2$, hence creates a local operator in the
theory.

From the point of view of the linear quiver theory on
$\R^2 \times T^2$, the original D2-branes and the additional one
represent two kinds of surface operators, one extending along $T^2$
and the other along $\R^2$.  Such intersecting surface operators were
studied in~\cite{Gomis:2016ljm} by means of the correspondence to
Liouville theory~\cite{Alday:2009aq}.  There it was found that open
strings stretched between intersecting D2-branes give rise to a
zero-dimensional $\CN = (0,2)$ Fermi multiplet at the intersection.
In the present case, this multiplet takes values in the bifundamental
representation of $\U(M) \times \U(1)$, where $\U(M)$ is the flavor
symmetry on the original $M$ D2-branes attached to $\text{NS5}_2$
(which is the gauge symmetry of the two-dimensional gauge theory) and
$\U(1)$ is the global symmetry on the additional D2-brane.  The
partition function of this multiplet turns out to be given precisely
by the operator~\eqref{eq:det(z-sigma)}, with $z$ being the value of
the scalar field in the $\U(1)$ vector multiplet, or the position of
the additional D2-brane on $C$.

Thus, we identify $Q(z)$ with a horizontal 't Hooft line with spectral
parameter $z$ crossing the vertical Wilson lines.  In our forthcoming
paper, we will present a more explicit derivation of this
identification using a description of surface operators in
four-dimensional Chern--Simons theory in terms of two-dimensional
degrees of freedom.

\subsection{Theories for open spin chains}

In the above discussions on the appearances of spin chains from linear
quiver theories and theories related to the cotangent bundles of
partial flag manifolds, it is crucial that the horizontal direction of
$T^2$ is periodic because we need to use T-duality in this direction
to arrive at the relevant brane configurations.  Consequently, the
spin chains appearing in these theories are closed ones with periodic
boundary conditions.

In four-dimensional Chern--Simons theory, however, there is nothing
that stops us from considering lattices on a noncompact surface such
as $\R^2$.  Even though we can no longer apply the T-duality then,
S-duality still leads to an interesting configuration consisting of
NS5-branes $\text{NS5}_i$, D1-branes $\text{D1}^\dashrarrow_\alpha$
and $\text{D1}^\dashuarrow_\beta$, and D3-branes
$\text{D3}^\uarrow_\gamma$ and $\text{D3}^\darrow_\gamma$.

For $C = \C$, the part of the system comprised of $\text{NS5}_i$ and
$\text{D1}^\dashuarrow_\beta$ realizes a one-dimensional gauge theory
which is the dimensional reduction of the two-dimensional
$\CN = (2,2)$ supersymmetric gauge theory discussed above.  Our
construction therefore implies that an \emph{open} $\slf_N$ spin chain
arises from this one-dimensional theory.

Since the horizontal Wilson lines now extend indefinitely, they do not
represent transfer matrices anymore.  Rather, these Wilson lines
crossing the vertical ones are monodromy matrices $T_\alpha$
constructed from the rational version of $R^{(N,0)}$, which is a
nondynamical R-matrix.  The monodromy matrices satisfy the RLL
relation~\eqref{eq:RLL(N,0)}, with $T_\alpha$ taking the place of
$L^{(N,0)}_\alpha$.  As such, they provide a representation of the
corresponding quantum algebra, namely the Yangian of $\slf_N$.

Again, the theory flows to a sigma model on $T^*\CF_{\mathbf{M}}$ for
generic values of the FI parameters.  Due to the topological twist
this is topological quantum mechanics on $T^*\CF_{\mathbf{M}}$, whose
algebra of local operators is the equivariant cohomology
$H^\bullet_{(\C^\times)^n \times \C^\times}(T^*\CF_{\mathbf{M}})$.
Thus, the action of monodromy matrices on the Hilbert space defines an
action of the Yangian on
$H^\bullet_{(\C^\times)^n \times \C^\times}(T^*\CF_{\mathbf{M}})$.
This statement was proved in~\cite{Maulik:2012wi, MR3118573,
  MR3354333}.

Similarly, we obtain in the trigonometric case an action of the
quantum loop algebra of $\slf_N$ on the equivariant K-theory of
$T^*\CF_{\mathbf{M}}$~\cite{MR3350271}, and in the elliptic case an
action of the elliptic quantum group of $\slf_N$ on the equivariant
elliptic cohomology of $T^*\CF_{\mathbf{M}}$~\cite{Aganagic:2016jmx,
  Felder:2017, MR3825878}.

\subsection{Yangians in three-dimensional linear quiver theories}

The D3--NS5 part of the above system, for $C = \C$, is the original
Hanany--Witten configuration for a three-dimensional $\CN = 4$
supersymmetric gauge theory, described by a linear quiver with $\U(k)$
gauge and flavor groups~\cite{Hanany:1996ie}.  In this theory
$\text{D1}_\alpha^\dashrarrow$ create local operators which involve monopoles,
and they represent monodromy matrices with $k$ vertical lines carrying
infinite-dimensional representations of~$\slf_N$.

For generic values of the FI parameters, the theory is in the Higgs
phase, and the topological twist and the $\Omega$-deformation reduce
it to topological quantum mechanics on the moduli space of
vortices~\cite{Bullimore:2016hdc}.  Therefore, $\text{D1}_\alpha^\dashrarrow$
act as the Yangian on the equivariant cohomology of this space.  This
conclusion fits nicely with results obtained
in~\cite{Bullimore:2015lsa, Braverman:2016pwk}, where it was found
that the algebra of local operators of the topologically twisted and
$\Omega$-deformed linear quiver theory is a certain quotient of the
Yangian.

\section*{Acknowledgments}

We would like to thank Davide Gaiotto for valuable discussions.  This
work is supported by the NSERC Discovery Grant program and by the
Perimeter Institute for Theoretical Physics.  Part of the research was
conducted while JY was supported by the ERC Starting Grant No.\ 335739
``Quantum fields and knot homologies'' funded by the European Research
Council under the European Union's Seventh Framework Programme and by
the National Key Research and Development Program of China
(No.~2020YFA0713000).  Research at Perimeter Institute is supported by
the Government of Canada through Industry Canada and by the Province
of Ontario through the Ministry of Research and Innovation.

\end{document}